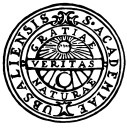

# The Complex World of Superstrings

*On Semichiral Sigma Models and N=(4,4) Supersymmetry*

MALIN GÖTEMAN

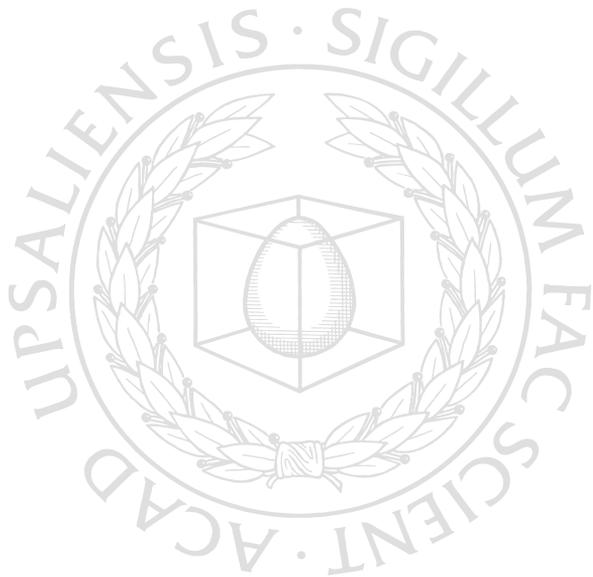

**Abstract**
Göteman, M. 2012. The Complex World of Superstrings: On Semichiral Sigma Models and N=(4,4) Supersymmetry. Acta Universitatis Upsaliensis. *Digital Comprehensive Summaries of Uppsala Dissertations from the Faculty of Science and Technology* 989. 139 pp. Uppsala. ISBN 978-91-554-8519-1.

Non-linear sigma models with extended supersymmetry have constrained target space geometries, and can serve as effective tools for investigating and constructing new geometries. Analyzing the geometrical and topological properties of sigma models is necessary to understand the underlying structures of string theory.

The most general two-dimensional sigma model with manifest N=(2,2) supersymmetry can be parametrized by chiral, twisted chiral and semichiral superfields. In the research presented in this thesis, N=(4,4) (twisted) supersymmetry is constructed for a semichiral sigma model. It is found that the model can only have additional supersymmetry off-shell if the target space has a dimension larger than four. For four-dimensional target manifolds, supersymmetry can be introduced on-shell, leading to a hyperkähler manifold, or pseudo-supersymmetry can be imposed off-shell, implying a target space which is neutral hyperkähler.

Different sigma models and corresponding geometries can be related to each other by T-duality, obtained by gauging isometries of the Lagrangian. The semichiral vector multiplet and the large vector multiplet are needed for gauging isometries mixing semichiral superfields, and chiral and twisted chiral superfields, respectively. We find transformations that close off-shell to a N=(4,4) supersymmetry on the field strengths and gauge potentials of the semichiral vector multiplet, and show that this is not possible for the large vector multiplet.

A sigma model parametrized by chiral and twisted chiral superfields can be related to a semichiral sigma model by T-duality. The N=(4,4) supersymmetry transformations of the former model are linear and close off-shell, whereas those of the latter are non-linear and close only on-shell. We show that this discrepancy can be understood from T-duality, and find the origin of the non-linear terms in the transformations.

*Malin Göteman, Uppsala University, Department of Physics and Astronomy, Theoretical Physics, Box 516, SE-751 20 Uppsala, Sweden.*

*It was the secrets of heaven and earth that I desired to learn.*

Mary Shelley, Frankenstein: Or, The Modern Prometheus (1897).

# List of papers

This thesis is based on the following papers, which are referred to in the text by their Roman numerals.

I  M. Göteman and U. Lindström, *Pseudo-hyperkähler geometry and generalized Kähler geometry*, Lett. Math. Phys. **95** (2011) 211 [arXiv:0903.2376 [hep-th]].

II  M. Göteman, U. Lindström, M. Roček and I. Ryb, *Sigma models with off-shell N=(4,4) supersymmetry and non-commuting complex structures*, JHEP **1009** (2010) 055 [arXiv:0912.4724 [hep-th]].

III  M. Göteman, U. Lindström, M. Roček and I. Ryb, *Off-shell N=(4,4) supersymmetry for new (2,2) vector multiplets*, JHEP **1103** (2011) 088 [arXiv:1008.3186 [hep-th]].

IV  M. Göteman, U. Lindström and M. Roček, *Semichiral sigma models with 4D hyperkähler geometry*, accepted for publication in Journal of High Energy Physics, arXiv:1207.4753 [hep-th].

V  M. Göteman, *N=(4,4) supersymmetry and T-duality*, Symmetry **4** (2012) 603, arXiv:1208.2166 [hep-th].

Reprints were made with permission from the publishers.

# Contents

# 1. Introduction

> We have a hunger of the mind which asks for knowledge of all around us, and the more we gain, the more is our desire; the more we see, the more we are capable of seeing.
> *Maria Mitchell, astronomer (1818-1889)*

What motivates a person to devote years to studying certain mathematical properties of models, that, from a very optimistic perspective, may be described as distant relatives to a physical system?

From my point of view, it is part of a bigger quest originating from the desire to understand the physical world we inhabit. Throughout history, mankind has always been curious and has felt a need to explain the phenomena we encounter in life. What's on the other side of the ocean? What is thunder? What are the stars? Does the universe have an end?

Stories of creation and religion have provided their attempts to answer these questions. Another approach, and the one I am exerting, is the scientific one.

The scientific principle is simple. Acceptable research has to follow certain rules: it should be objective, independent of who is performing the research, and repeatable. That is, if one experiment hints towards a new result, then other experiments, performed in other laboratories and by other people, should produce the same result. Independent readers should be able to follow the steps of a proof of a theorem in order for it to earn its validity.

The scientific method is empirical, methodical and based on logic. But the empirical observations do not always come first. Sometimes, theoretical considerations based on logic, symmetries or mathematical structures lay out the directions and make predictions for possible future observations. Theoretical physics has celebrated many great triumphs when hypotheses have finally been confirmed by experiments. Einstein's theory of general relativity from 1915 was not tested in accuracy until in 1959, when gravitational redshift could be measured to a great precision. Another example is the prediction of the top quark, whose existence was suggested in 1973, but not confirmed until 20 years later. And at the time of writing, it seems that the long foreseen Higgs particle has finally been observed, and pictures of Peter Higgs, shed-ding a tear of happiness at the press conference at CERN, have been cabled out around the world.†

However, theoretical basic research doesn't always come with predictions for experiments. History has taught us, that in many important examples, the application of a research result didn't come until much later, and sometimes from unexpected directions. Modern cryptography is to a large extent based on number theory concerning prime numbers, an area developed by mathematicians who did not have a clue about the major impact their research would have in today's computer society; nevertheless, they performed their research, perhaps for the beauty of science itself, or for the intellectual challenge, or simply out of curiosity.

Although the aim of science is to be objective, of course, research is not independent of the cultural context. Even if a researcher believes that her or his conclusions are objective, biased predictions or prejudices sometimes cloud the interpretation of the results. A clear example is the research performed at the State Institute for Racial Biology in Uppsala, Sweden, which was the first of its kind when it opened in 1922. Even though the researchers claimed to exercise good science and to explain the effect biological heritage and the environment has on people, today we see that they were heavily influenced by racist prejudices, and so were their "scientific" conclusions. Another, perhaps less drastic example is when researchers tend to see the data that supports their hypothesis, but overlook the data that contradicts it. The risk of subjective and false conclusions is minimized when researchers are aware of the possibility of biases, and when research is replicated and reviewed by different people around the world.

There is a myth that science is driven forward by an exclusive group of geniuses. True, not all scientists contribute to the major breakthroughs, sometimes new perspectives and methods are needed to solve a long-standing problem. But in general, doing science is a collective effort. Many detours have to be taken before the right path is found, much data has to be collected for patterns to arise, a lot of calculations have to be performed; for every successful experiment there are a number of failed ones. Albert Einstein would not have been able to develop his theory of general relativity without the, at the time, newly developed tools in tensor calculus. James Watson and Francis Crick would not have been able to determine the structure of DNA, for which they were awarded the Nobel Prize, if it had not been for the X-ray images developed by Rosalind Franklin. To take a more recent example that has already been mentioned, thousands and yet thousands of people have been involved in the search for, and the recent discovery of, the Higgs-like particle of the Standard Model. Theoreticians, experimentalists and engineers all depend on mutual teamwork. They carry out their tasks in a methodic, controlled way, and the hard work of each individual form a greater picture, that may be broader than the sum of the separate parts.

---

† Although the Higgs particle bears the name after Peter Higgs, his paper [Hig64] was not the first one to suggest the existence of a Higgs-like particle. The same mechanism was presented before by Englert and Brout in [EB64] and was further developed by Guralnik, Hagen and Kibble in [GHK64]. It has been proposed that the Higgs mechanism should more accurately be called the Englert-Brout-Higgs-Guralnik-Hagen-Kibble mechanism.

But, despite the frames and rules for the scientific methods being fixed, research is extremely creative. As the artist, the researcher starts from a white sheet of paper; she then performs her experiment, collects her data or completes her calculation, and creates something new, a result that no one before her has ever seen.

This, I believe, is what drives a person to spend several years studying geometrical properties of supersymmetric sigma models.

## 1.1 A physics conception of the world

Physics is a natural science that tries to explain the fundamental properties of nature, such as matter, dynamics, energy, forces.

The foundations for modern physics were laid during the Scientific Revolution in the 16th and 17th centuries. Natural philosophers studied and found explanations for the dynamics of mechanical bodies, the motion of astronomical objects, optics, thermodynamics and other phenomena. During the 18th and 19th century, the knowledge of physics was broadened with the theory of electromagnetism and the theory of classical mechanics, together with important advancements in the language of physics: mathematics. These laws govern what is called *classical* physics, that is, physics on macroscopic (non-atomic) scales and for non-relativistic velocities, i.e., for bodies moving much slower than the speed of light.

However, the laws of classical physics cannot explain how atoms behave, or how the gravitational redshift of light occurs. New concepts were needed to explain these and other phenomena. Two ideas that approached these problems were developed around a century ago and are now the basic frameworks for modern physics: quantum physics and relativity. Both concepts have had a major impact on our understanding of the physical world we inhabit. Of course, the classical theories of physics are incorporated in the new theories and are obtained in the classical limits.

Quantum theory explains physics at atomic scales and reveals that the quantum world behaves according to rules that might feel counter-intuitive, like the famous *Schrödinger's cat*, that is both dead and alive at the same time. The theory has been further developed and refined into relativistic quantum field theories, which form the basis for the Standard Model that governs our understanding of particle physics. The three fundamental forces that dictate the behavior of the elementary particles (like electrons and quarks) are the electromagnetic force, the weak force and the strong force. The dynamics of quantum particles can be understood and predicted to an incredible precision with the Standard Model.

The fourth fundamental force is gravity, which by the theory of general relativity is understood in geometrical terms as the curvature of space-time due to the mass or energy present. The theory makes predictions that differ significantly from those of classical physics; for example, gravitational time dilation in the gravitational field around the Earth, which has been measured and confirmed numerous times using atomic clocks.

So, do we now have a perfect understanding of the laws of physics? Can we relax, and go home? No, far from it. There are still many things we do not understand. Measurements of cosmic microwave background show that the energy density of the universe must be much higher than the observable matter; around 22% must be made up of dark matter, and 74% of dark energy, both of which are unknown to us. The Standard Model leaves many open questions. Why are there so many parameters, and why do they have their specific values? Why are there three generations of matter? Why do we observe an asymmetry between matter and antimatter? Can we fully explain black holes, and what about the Big Bang?

Theoretical physicists still have a lot to work out, together with both experimentalists and mathematicians. One approach that suggests a new way of looking at both particle physics and gravity at the same time is string theory.

## 1.2   String theory and sigma models

The basic assumption of string theory is that the fundamental objects in nature are not zero-dimensional point-particles, but instead higher-dimensional objects, such as one-dimensional strings.

The theory was first developed in the late 1960's, not as a self-contained fundamental physical theory, but as an attempt to model the strong interaction. It had been found that a certain behavior of the hadron masses, the so called Regge trajectories, could be explained if the hadrons were modeled not as point-like particles, but as one-dimensional vibrating strings [Ven68]. But this stringy description of the hadrons suffered from some technical problems, and when a new promising theory for the strong force, quantum chromodynamics, was developed, the strings were soon abandoned as a description for the strong interaction.

Instead, with the discovery that the theory includes a particle that could be interpreted as the graviton, the proposed quantum particle for the gravita-tional force, the awareness grew that string theory could perhaps be used for something much more profound: a quantum theory of gravity.

The field has since grown rapidly and evolved in many directions. Whereas the original theory included only bosons, fermions were soon included in the theory by supersymmetry. Several different consistent string theories could be defined, and were later unified by dualities into one single theory. The more recent AdS/CFT duality [Mal98] relates string theory in a certain space to a lower-dimensional field theory, connecting back to particle physics.

String theory predicts that space-time is ten-dimensional, instead of the ordinary four dimensions one would expect for a physical theory. Compactification of the extra dimensions has significant implications for the geometry and topology of the space; this is one of many examples of where rich mathematical structures appear in string theory, and the superspace formalism provides powerful calculational tools, useful also outside string theory. As a summary, string theory needs very little input, but has a huge output. Starting from vibrating strings, the theory implies both gravity and Yang-Mills theory, gives insights concerning many other areas of physics and new results in mathematics.

But it should also be said that string theory is still a developing theory and not yet fully understood. It is far from being a *theory for everything*, as it has sometimes been proclaimed to be, and there are no experimental evidences. If string theory is indeed the correct description of nature, or if it is only an extremely powerful tool for understanding physics and mathematics, remains to be seen.

## 1.3 Goals and research questions

The aim of the research presented in this thesis is to understand a branch that is studied both in string theory and in mathematics, namely supersymmetric sigma models and their intimate connection to geometry and topology. This, of course, is a subject far too vast to be covered in one single thesis, and the focus is therefore sharpened to a much more narrow area: two-dimensional sigma models, to a large extent described by semichiral superfields, and the implications of $N=(4,4)$ supersymmetry on the target space geometry. Further, the ambition is to understand these models in relation to other known models and geometries, in particular to sigma models parametrized by chiral and twisted chiral fields and bihermitian local product geometry. The understanding of this particular model also sheds light on sigma models and geometry in a broader sense, involving generalized complex geometry, pseudo-supersymmetry, neutral hyperkähler metrics, supersymmetry representations, auxiliary superfields,

T-duality, Legendre transforms and quotient reduction, vector multiplets and much more.

At the beginning of each of the chapters 6 to 8 in the part II, summarizing the developments on the subject in the papers [I-V], the specific research questions belonging to each of the papers in the thesis will be reformulated in a more detailed setting.

## 1.4 Outline of the thesis

The main part of the thesis is divided into two parts. The first part includes chapters 2-5 and is a background and introduction to the research subject. Chapter 2 is an initiation into the mathematical preliminaries: complex geometry, generalized complex geometry and related issues, such as neutral hyperkähler geometry. Supersymmetry is introduced from an algebraic point of view in chapter 3, and the notation for superspace and superfields is set. Supersymmetric sigma models are treated extensively in chapter 4; special focus is given to manifest $N=2$ sigma models and their different manifestations of generalized Kähler geometry. In chapter 5, the concept of T-duality is developed as a tool for relating and constructing new geometries, machinery that will be necessary for later chapters.

The second part is devoted to the research results of the papers [I-V]. Papers [I], [II] and [IV] all discuss different aspects of semichiral sigma models and $N=(4,4)$ (pseudo-) supersymmetry, and the results will be presented in a more coherent setting in chapter 6. Chapter 7 is based on paper [III] and deals with the semichiral and large vector multiplets and extended supersymmetry. Finally, the subjects of the preceding chapters meet in chapter 8, where semichiral sigma models, extended supersymmetry and T-duality is discussed, based on the results of paper [V].

Excerpts of the text have appeared also in the author's licentiate thesis, successfully defended at Uppsala University in March 2011. This concerns in particular parts of the background chapters 3 and 4. The five papers that this thesis is based on are reprinted at the end, but the thesis can be read independently, as a comprehensive summary and discussion of the research and results of the papers [I-V].

# Part I:

# Background

# 2. Geometry

> No other field can offer, to such an extent as mathematics, the joy of discovery, which is perhaps the greatest human joy.
> *Rózsa Péter, mathematician (1905-1977)*

The first observations of the deep connections between supersymmetric sigma models and geometry [Zum79, AGF80] sparked a great interest in further investigating the correlation between the subjects. As will be reviewed in detail in chapter 4, the target space of a two-dimensional sigma model with $N=(2,2)$ supersymmetry is Kähler in the absence of torsion [Zum79], and bihermitian if torsion is present [GHR84]. More supersymmetry requires more structures on the target manifold; $N=(4,4)$ supersymmetry implies hyperkähler geometry in the torsion-free case [AGF80] and bihyperhermitian geometry in the case with torsion.

The ambition to unify complex and symplectic geometry, two seemingly different geometries, led mathematicians to study and develop the subject of generalized complex geometry [Gua03, Hit03, Cav05, Wit05]. Independently, a connection between the two geometries had been studied from the viewpoint of mirror symmetry in string theory [LVW89, GP90]. A special case of the generalized geometry was shown to be equivalent with bihermitian geometry [Gua03], which arises naturally in string theory when studying sigma models with extended supersymmetry. This inspired both physicists and mathematicians to study generalized complex geometry and supersymmetric sigma models. More recently, with the advancements of flux compactifications and supergravity, the subject has continued to grow and now covers a wide spectrum including non-geometries, double field theory, projective superspace, gerbes, dualities, topology changes and much more.

In this chapter, the preliminaries of (generalized) complex geometry will be given. Section 2.1 covers complex geometry, with particular focus on the special case of hyperkähler geometry. Some subtleties will be omitted but can be found in standard textbooks, such as [Nak03]. The formalism of generalized complex geometry is introduced in section 2.2, and the equivalence of generalized Kähler geometry and bihermitian geometry is explained.

## 2.1 Complex geometry

### 2.1.1 Complex manifolds

A *topological space* is a set $X$ and a collection of subsets $U = \{U_i\}$ with $U_i \subset X$, such that the empty set and $X$ are both elements of $U$, and $U$ is closed under finite intersections and arbitrary unions. A topological manifold is a topological space that is locally Euclidean and Hausdorff, i.e., points are separable.

A *smooth manifold* is a topological manifold together with an atlas of charts $\{U_i, \varphi_i\}$, where $\{U_i\}$ is an open covering $M = \bigcup U_i$ and $\varphi_i$ are homeomorphisms $\varphi_i : U_i \to \mathbb{R}^n$, such that the transition functions

$$\varphi_j \circ \varphi_i^{-1} : \varphi_i(U_i \cap U_j) \to \varphi_j(U_i \cap U_j) \tag{2.1}$$

are infinitely differentiable on all non-empty intersections $U_i \cap U_j \neq \emptyset$.

Consider a topological manifold $M$ with an open covering $\{U_i\}$ and an atlas of charts $\{U_i, \varphi_i\}$ to $\mathbb{C}^n$, assigning complex coordinates $(z_1^i, \ldots, z_n^i)$ to all points in $U_i$. The space $M$ is called a *complex manifold* if, for all non-empty intersections, the change of coordinates

$$\begin{aligned}
\varphi_j \circ \varphi_i^{-1} : \varphi_i(U_i \cap U_j) &\longrightarrow \varphi_j(U_i \cap U_j) \\
(z_i^1, \ldots, z_i^n) &\longmapsto (z_j^1(z_i^1, \ldots, z_i^n), \ldots, z_j^n(z_i^1, \ldots, z_i^n))
\end{aligned} \tag{2.2}$$

are holomorphic. In other words, every neighborhood of the manifold looks like the complex space $\mathbb{C}^n$ in a consistent way. A complex manifold necessarily has an even number of real dimensions, and all complex manifolds are also real differentiable manifolds, but not the other way around. The two-sphere $S^2$ is a complex manifold, for example, whereas the four-sphere $S^4$ is not. For the six-sphere $S^6$ it is not yet known if the manifold is complex, showing that the classification of complex manifolds is indeed not trivial.

A complex $n$-dimensional manifold can be viewed as a real $2n$-dimensional manifold together with a complex structure $J$ containing information about how the real and imaginary parts of the complex vector fields relate to one another and which differential equations they have to fulfill in order for the change of coordinates between the vector fields to be holomorphic.

Consider a real $2n$-dimensional differential manifold $M$, with coordinates $(x^\mu, y^\mu)$, where $\mu = 1, \ldots n$. The tangent space and cotangent space are spanned by

$$T_p M = \mathrm{span}_\mathbb{R} \left( \frac{\partial}{\partial x^\mu}, \frac{\partial}{\partial y^\mu} \right), \quad T_p^* M = \mathrm{span}_\mathbb{R} (dx^\mu, dy^\mu). \tag{2.3}$$

The manifold can be complexified by introducing complex coordinates defined as $z^\mu = x^\mu + iy^\mu$. The basis of the complexified tangent space and the dual basis of the cotangent space are now

$$T_p M^{\mathbb{C}} = \text{span}_{\mathbb{C}} \left( \frac{\partial}{\partial z^\mu}, \frac{\partial}{\partial \bar{z}^\mu} \right), \quad T_p^* M^{\mathbb{C}} = \text{span}_{\mathbb{C}} \left( dz^\mu, d\bar{z}^\mu \right). \tag{2.4}$$

Any complex vector field $Z \in T_p M^{\mathbb{C}}$ can be divided into a real and an imaginary part as $Z = X + iY$. Consider a map $J$ acting as multiplication of the vector field with $\pm i$. Applying this map twice gives $J^2 = -1$. Any map fulfilling this condition is called an *almost complex structure*. Any almost complex structure

$$J : T_p M \to T_p M, \quad J^2 = -1 \tag{2.5}$$

has eigenvalues $\pm i$. This implies that the tangent space of the manifold can be divided into two disjoint vector spaces corresponding to the eigenspaces of $J$,

$$T_p M^{\mathbb{C}} = T_p M^+ \oplus T_p M^-, \quad T_p M^\pm = \left\{ Z \in T_p M^{\mathbb{C}} : JZ = \pm iZ \right\}, \tag{2.6}$$

where $Z \in T_p M^+$ is a holomorphic and $Z \in T_p M^-$ an anti-holomorphic vector. As already mentioned, not all real even dimensional manifolds are complex; the six-sphere, for example, can be complexified and admits an almost complex structure, but this doesn't make it a complex manifold. The condition $J^2 = -1$ is not sufficient for the change of coordinates to be holomorphic; a sufficient and necessary condition for the manifold to be complex is that the almost complex structure $J$ is integrable [NN57], that is, that the almost complex structure defines integrable eigenspaces.

Consider two (anti-) holomorphic complex vectors $X, Y \in T_p M^\pm$. The distribution $T_p M^\pm$ is called *integrable* if and only if it is closed under the Lie bracket,

$$X, Y \in T_p M^\pm \quad \Rightarrow \quad [X, Y] \in T_p M^\pm. \tag{2.7}$$

Using projection operators $P^\pm = \frac{1}{2}(1 \mp iJ)$, this condition for integrability can be rewritten as $P^\mp [P^\pm X, P^\pm Y] = 0$ for $X, Y \in T_p M^{\mathbb{C}}$. The Nijenhuis tensor for any tensor $J$ of rank $(1,1)$ is defined as [Nij51]

$$\mathcal{N}_J(X, Y) = J^2[X, Y] - J[JX, Y] - J[X, JY] + [JX, JY], \tag{2.8}$$

or, in components,

$$\mathcal{N}(J)^i_{jk} = J^l_{[j} J^i_{k],l} + J^i_l J^l_{[j,k]}. \tag{2.9}$$

The integrability condition written in terms of the projection operators is proportional to the Nijenhuis tensor. The integrability condition for the almost complex structure $J$ can thus be rewritten in terms of the vanishing of the Nijenhuis tensor,

$$\mathcal{N}_J(X, Y) = 0. \tag{2.10}$$

A structure $J$ fulfilling the two conditions $J^2 = -1$ and $\mathcal{N}_J(X,Y) = 0$ is called a *complex structure*, and a real differentiable manifold with a complex structure is called a complex manifold.

For a complex manifold with corresponding complex structure, one can always find a change to an (anti-) holomorphic coordinate system $(z,\bar{z})$ in which the complex structure takes the canonical form with constant entries

$$J = \begin{pmatrix} i & 0 \\ 0 & -i \end{pmatrix}. \tag{2.11}$$

For a manifold with more than one complex structure, the structures are said to be *simultaneously integrable* when there exists an atlas on the manifold such that on every patch, coordinates can be found in which all structures are constant. The simultaneous integrability can be defined in terms of the Magri-Morosi concomitant, defined in components as [YA68]

$$\mathcal{M}(I,J)^i_{jk} = I^l_j J^i_{k,l} - J^l_k I^i_{j,l} - I^i_l J^l_{k,j} + J^i_l I^l_{j,k}. \tag{2.12}$$

For $I = J$, the Magri-Morosi concomitant reduces to the ordinary Nijenhuis tensor, $\mathcal{M}(J,J) = \mathcal{N}(J)$. The sum of two structures is integrable if the structures are separately integrable and their Magri-Morosi concomitant vanishes,

$$\mathcal{N}(I+J)^i_{jk} = [\mathcal{N}(I) + \mathcal{N}(J) + \mathcal{M}(I,J) + \mathcal{M}(J,I)]^i_{jk}. \tag{2.13}$$

The last two terms are also known as the Nijenhuis concomitant [FN56], $\mathcal{N}(I,J) = \mathcal{M}(I,J) + \mathcal{M}(J,I)$. Two commuting complex structures are simultaneously integrable if and only if their Magri-Morosi concomitant vanishes [HP88]. For two general endomorphisms $I$, $J$ of the target space, the Magri-Morosi concomitant is

$$\mathcal{M}_{I,J}(X,Y) = [IX, JY] - I[X, JY] - J[IX, Y] + IJ[X, Y] + [I,J]XY, \tag{2.14}$$

from which it is clear that the Magri-Morosi concomitant takes the form of the Nijenhuis tensor in (2.8) when $I = J$.

### 2.1.2 Dolbeault cohomology

Consider first real $p$-forms $\omega \in \Omega^p(M)$, where $M$ is an $n$-dimensional manifold with metric $g$. Define the Hodge operator $* : \Omega^p(M) \to \Omega^{n-p}(M)$ by

$$*(dx^{\mu_1} \wedge \cdots \wedge dx^{\mu_p}) = \frac{\sqrt{|g|}}{(n-p)!} \varepsilon^{\mu_1 \ldots \mu_p}{}_{\nu_{p+1} \ldots \nu_n} dx^{\nu_{p+1}} \wedge \cdots \wedge dx^{\nu_n}, \tag{2.15}$$

where $\varepsilon$ is the totally anti-symmetric tensor. The *adjoint* of the exterior derivative $d : \Omega^{p-1}(M) \to \Omega^p(M)$ is an operator $d^\dagger$ defined as

$$d^\dagger : \Omega^p(M) \to \Omega^{p-1}(M), \quad d^\dagger = \pm(-1)^{np+n+1} *d*, \tag{2.16}$$

where the sign in (2.16) depends on the signature of the metric. The *Laplacian* $\Delta : \Omega^p(M) \to \Omega^p(M)$ is defined in terms of the exterior derivative and its adjoint as

$$\Delta = (d+d^\dagger)^2 = dd^\dagger + d^\dagger d. \qquad (2.17)$$

A form satisfying $\Delta\omega = 0$ is *harmonic*. For real manifolds, the *de Rham cohomology* $H^p_{dR}(M,\mathbb{R})$ is defined as all closed $p$-forms modulo the exact $p$-forms,

$$H^p_{dR}(M,\mathbb{R}) = \frac{\ker d : \Omega^p(M) \to \Omega^{p+1}(M)}{\operatorname{im} d : \Omega^{p-1}(M) \to \Omega^p(M)}. \qquad (2.18)$$

The dimension of the vector space $H^p_{dR}(M,\mathbb{R})$ is a topological invariant called the *Betti number* $b^p$. Hodge's theorem for the de Rham cohomology groups says, that on a compact orientable Riemannian manifold, the de Rham cohomology group is isomorphic to the group of harmonic forms,

$$H^p_{dR}(M) \cong \operatorname{Harm}^p(M). \qquad (2.19)$$

Now move over to complex forms. A *form of bidegree* $(r,s)$ has a basis of $r$ holomorphic and $s$ anti-holomorphic forms,

$$\omega^{(r,s)} = \frac{1}{r!s!}\omega_{\mu_1\ldots\mu_r\bar\nu_1\ldots\bar\nu_s}dz^{\mu_1}\wedge\cdots\wedge dz^{\mu_r}\wedge d\bar z^{\bar\nu_1}\wedge\cdots\wedge d\bar z^{\bar\nu_s} \in \Omega^{r,s}(M). \qquad (2.20)$$

Any complex $p$-form can be written as a sum of such forms of bidegree $(r,s)$ with $r+s=p$. For complex manifolds, the exterior derivative $d$ can be split into holomorphic and anti-holomorphic *Dolbeault operators*, $d = \partial + \bar\partial$. A holomorphic $r$-form is defined as a form $\omega \in \Omega^{r,0}(M)$ satisfying $\bar\partial\omega = 0$. This happens precisely when the component function is a holomorphic function, $\omega_{\mu_1\ldots\mu_r,\bar\lambda} = 0$. The Dolbeault operators define the Dolbeault complex as

$$\Omega^{r,s+1}(M) \xleftarrow{\bar\partial} \Omega^{r,s}(M) \xrightarrow{\partial} \Omega^{r+1,s}(M). \qquad (2.21)$$

The Dolbeault cohomology is the complex analogue of the de Rham cohomology; with $Z^{r,s}_{\bar\partial}(M)$ and $B^{r,s}_{\bar\partial}(M)$ denoting the $\bar\partial$-closed and $\bar\partial$-exact forms of bidegree $(r,s)$, the Dolbeault $\bar\partial$-cohomology group is the quotient

$$H^{r,s}_{\bar\partial}(M) = Z^{r,s}_{\bar\partial}(M)/B^{r,s}_{\bar\partial}(M). \qquad (2.22)$$

In other words, the elements in $H^{r,s}_{\bar\partial}(M)$ are equivalence classes, and two $(r,s)$-forms $\omega$ and $\omega'$ belong to the same Dolbeault cohomology equivalence class if they are $\bar\partial$-closed, $\bar\partial\omega = \bar\partial\omega' = 0$, and differ only by $\omega - \omega' = \bar\partial\alpha$ for some form $\alpha \in \Omega^{r,s-1}(M)$. Analogously to the Betti numbers for the de Rham cohomology, the complex dimension of the Dolbeault cohomology vector space $H^{r,s}_{\bar\partial}(M)$ is given by the *Hodge number* $h^{r,s}$. Generally, there is no simple relation between de Rham and Dolbeault cohomology, and the latter carries no topological information. But for Kähler manifolds, the rich geometrical structure enables relations between them, as will soon be reviewed.

## 2.1.3 Connections

To develop the concept of holonomy needed later, a short introduction to principal bundles must first be given. A *fiber bundle* is the set of data $(E, \pi, B, F)$, where $B$ is denoted the base space and $F$ the fiber, and the projection $\pi : E \to B$ is locally trivial, in other words, such that locally, the bundle looks like the trivial bundle $E = B \times F$. The inverse image $\pi^{-1}(b)$ is the fiber at $b \in B$. In a differentiable fiber bundle, the base space, the fiber and the total space are differentiable manifolds, which is from now on assumed. A global *section* is a smooth map $s : B \to E$ that maps points of the base space to unique points on the fiber, $\pi s(b) = b \in B$. In a principal bundle $(P, \pi, B, G)$, the fiber is a (Lie) structure group $G$, acting free and transitive, $\pi(pg) = \pi(p)$ with $g \in G$ and $p \in P$. The base space is isomorphic to the space of orbits, $B \cong P/G$.

A *connection* on a principal bundle is a smooth and unique separation of the tangent space of $P$ into a vertical subspace $V_p P$, tangent to the fiber, and a horizontal space,

$$T_p P = V_p P \oplus H_p P, \qquad (2.23)$$

in such a way that choosing the horizontal subspace at one point, the horizontal subspaces at all other points are uniquely determined, $H_{pg} P = R_{g*} H_p P$, where $R_{g*} : T_p P \to T_{pg} P$ is the push-forward of the right-translation of $G$. An element $A$ in the Lie algebra generates a flow along the fiber, $\sigma_t(p) = p e^{tA}$, satisfying $\pi(\sigma_t(p)) = \pi(p) = m \in B$. Consider an arbitrary smooth function $f : P \to \mathbb{R}$ and define a vector $X_A \in T_p P$ by

$$X_A\big(f(p)\big) = \frac{d}{dt} f\big(\sigma_t(p)\big)\Big|_{t=0}. \qquad (2.24)$$

The vector field is tangent along the flow $\sigma_t(p)$ and defines the vertical subspace. It is called the fundamental vector field associated to $A$.

In a principal bundle, a horizontal lift $\gamma_P$ can be defined that lifts a curve in the base space, $\gamma = [0,1] \subset B$, to the fibers in such a way that all tangent vectors to the lifted curve lie in the horizontal subspace $H_p P$. Given a connection, the *parallel transport* of a point $p \in P$ along a curve $\gamma$ in $B$ can then be uniquely determined by moving it along the horizontal lift $\gamma_P$. For a closed loop $\gamma(0) = \gamma(1) = m$, the parallel transported endpoints lie on the same fiber, $\pi(\gamma_P(0)) = \pi(\gamma_P(1)) = m$, but are not necessarily equal; the loop defines a transformation $\tau_\gamma : \pi^{-1}(m) \to \pi^{-1}(m)$ on the fiber. Varying the closed loops for a fixed point $m \in B$ and denoting all the loops by $\mathcal{C}_m(B)$ generates the *holonomy* of the connection,

$$\text{Hol}_m(P) = \big\{ g \in G \,\big|\, \tau_\gamma(m) = mg,\ \gamma \in \mathcal{C}_m(B) \big\}, \qquad (2.25)$$

measuring to which extent the parallel transport around closed loops fails to preserve the geometrical data being transported. The holonomy depends on both the connection and the principal bundle.

A connection one-form is defined as a projection of the tangent space $T_pP$ onto the vertical subspace, satisfying $\omega(X_A) = A$ and $\omega_{pg}(R_{g*}X) = g^{-1}\omega_p(X)g$, where $R_{g*}$ is the push-forward defined above. The last constraints implies that the horizontal subspace is equivariant; if a vector $X$ lies in the horizontal subspace $H_pP$, then the push-forward vector $R_{g*}X$ lies in the horizontal subspace $H_{pg}P$. With this definition, the horizontal subspace can also be defined as the vectors in the tangent space satisfying $H_pP = \{X \in T_pP | \omega(X) = 0\}$. Let $\{U_i\}$ be an open covering of $B$ and $s_i$ be a local section defined on every subspace. The *local connection one-form* is defined as the pullback of the local section of $\omega$,

$$A_i = s_i^*\omega \in \mathbf{g} \otimes \Omega^1(U_i). \qquad (2.26)$$

This is the Lie-algebra valued gauge potential that arises in physics and will be used in later chapters; in particular the discussion presented here will become useful when discussing the gauging of sigma model isometries in chapter 5. On the non-empty intersections $U_i \cap U_j$, the gauge potentials relate to each other by the gauge transformations $A_j = g_{ij}^{-1}A_ig_{ij} + g_{ij}^{-1}dg_{ij}$, where $g_{ij}$ are the transition functions, see, e.g., [Nak03].

A *vector bundle* is a fiber bundle whose fiber is a vector space. The prototype of a vector bundle is the tangent bundle $TM$ over an $n$-dimensional manifold $M$, whose fiber is the tangent spaces $T_pM \cong \mathbb{R}^n$ at each point $p \in M$. The sections of a tangent bundle are vector fields over $M$. On a tangent bundle, each fiber has a natural basis $\{\partial/\partial x^\mu\}$ given by the coordinate system on $U_i \subset M$. The basis vectors of the tangent spaces form a local frame over $U_i$, and the set of frames $L_pM$ at each point $p \in M$ defines the *frame bundle*. This frame bundle is actually a principal bundle with the structure group being the set of non-singular linear transformations, $GL(n, \mathbb{R})$, and a local connection one-form can be defined as in (2.26), where $\mathbf{g}$ is the Lie algebra of all invertible $n \times n$ matrices. These matrix-valued forms define the Christoffel symbols for a covariant derivative; hence any connection on the frame bundle defines a covariant derivative on the tangent bundle. If the connection is torsion-free and the covariant derivative preserves the metric, it is the *Levi-Civita* connection.

### 2.1.4 Kähler geometry

A manifold endowed with a complex structure $J$ always admits a *hermitian metric g* satisfying

$$g(JX, JY) = g(X, Y), \qquad (2.27)$$

or differently expressed, $J^t g J = g$. Explicitly, given a Riemannian metric $\tilde{g}$, the hermitian metric can be defined as $g(X,Y) = \frac{1}{2}(\tilde{g}(X,Y) + \tilde{g}(JX,JY))$, obviously satisfying (2.27). In the (anti-) holomorphic coordinates in which the complex structure takes the canonical form in (2.11), the hermitian metric has only off-diagonal entries,

$$g = g_{\mu\bar{\nu}}dz^\mu \otimes d\bar{z}^{\bar{\nu}} + g_{\bar{\mu}\nu}d\bar{z}^{\bar{\mu}} \otimes dz^\nu. \tag{2.28}$$

The *Kähler form* is a two-form defined in terms of the metric and the complex structure as

$$\omega(X,Y) = g(JX,Y), \tag{2.29}$$

with $X,Y \in T_pM$. When the Kähler form is closed, the manifold is said to be a *Kähler manifold* and the metric a Kähler metric. This condition is equivalent to the complex structure being covariantly constant with respect to the Levi-Civita connection,

$$d\omega = 0 \iff \nabla J = 0. \tag{2.30}$$

Despite the fact that a complex structure is locally trivial in every coordinate patch, in general, the integrability does not imply that it is invariant when parallel transported along a closed curve that traverses several patches. A necessary and sufficient condition for the complex structure to be covariantly constant is that the Kähler form is closed.

Since the Kähler form is closed, non-degenerate and anti-symmetric, it is a symplectic form. Writing the Kähler form in components as $\omega = ig_{\mu\bar{\nu}}dz^\mu \wedge d\bar{z}^{\bar{\nu}}$, the closeness constraint (2.30) implies that the Kähler metric satisfies the two relations $g_{\mu\bar{\nu},\lambda} = g_{\lambda\bar{\nu},\mu}$ and $g_{\mu\bar{\nu},\bar{\lambda}} = g_{\mu\bar{\lambda},\bar{\nu}}$. A metric written in terms of a *Kähler potential* as

$$g_{\mu\bar{\nu}} = \frac{\partial^2 \tilde{K}}{\partial z^\mu \partial \bar{z}^{\bar{\nu}}} \tag{2.31}$$

clearly satisfies these conditions. The converse is also true; if $\{U_i\}$ is an open covering on a Kähler manifold, then locally on a chart $U_i$ the metric can be written as second derivatives of a Kähler potential $\tilde{K}_i$ as in (2.31). On overlapping charts, two Kähler potentials may differ up to a Kähler transformation, $\tilde{K}_i(z,\bar{z}) - \tilde{K}_j(z,\bar{z}) = f_{ij}(z) + \bar{f}_{ij}(\bar{z})$, where $f_{ij}(z)$ is a holomorphic function. The constraints for the Kähler metric also imply that all Christoffel symbols in the Levi-Civita connection with mixed holomorphic and anti-holomorphic indices vanish, e.g.,

$$\Gamma^{(0)\mu}_{\nu\bar{\rho}} = g^{\mu\bar{\lambda}}\left(g_{\nu\bar{\lambda},\bar{\rho}} + 0 - g_{\nu\bar{\rho},\bar{\lambda}}\right) = 0. \tag{2.32}$$

Since the Levi-Civita connection does not mix holomorphic with antiholomorphic indices, it preserves holomorphicity. In other words, a holomorphic vector remains holomorphic after parallel transport. This in particular implies that the holonomy of a Kähler manifold is contained in $U(n)$.

The complex projective space $\mathbb{C}P^n$ with the well-known Fubini-Study metric is an example of a Kähler manifold. An important special case of Kähler manifolds are Calabi-Yau spaces, defined as compact Kähler manifolds that are Ricci-flat. This restricts the holonomy of the manifold to $SU(n) \subset U(n)$.

Laplacians for hermitian manifolds can be defined analogously to (2.17) as

$$\Delta_\partial = (\partial + \partial^\dagger)^2 = \partial\partial^\dagger + \partial^\dagger\partial,$$
$$\Delta_{\bar\partial} = (\bar\partial + \bar\partial^\dagger)^2 = \bar\partial\bar\partial^\dagger + \bar\partial^\dagger\bar\partial. \quad (2.33)$$

A form satisfying $\Delta_\partial \omega = 0$ is called $\partial$-harmonic. Generally, there is no simple relationship between the Laplacians in (2.17) and (2.33). But for Kähler manifolds, they are related simply by $\Delta_\partial = \Delta_{\bar\partial} = \frac{1}{2}\Delta$. This means, that a holomorphic $r$-form not only satisfies $\Delta_{\bar\partial}\omega = 0$, but also $\Delta\omega = 0$, in other words, that all holomorphic forms are harmonic with respect to the Kähler metric. Moreover, for Kähler manifolds, but not for hermitian manifolds in general, de Rham and Dolbeault cohomology are related by the Hodge decomposition

$$H^p_{dR}(M)^{\mathbb{C}} = \oplus_{r+s=p} H^{r,s}(M), \quad (2.34)$$

implying that the Betti numbers $b^p$ can be computed as the sum of all Hodge numbers $h^{r,s}$ with $r+s = p$.

### 2.1.5 Bihermitian geometry

If torsion is included in the connection via $H = db$,

$$\nabla^{(\pm)} = \nabla \pm \tfrac{1}{2} g^{-1} H, \quad (2.35)$$

and the manifold is equipped with two complex structures which are covariantly constant with respect to these connections,

$$\nabla^{(\pm)} J^{(\pm)} = 0, \quad (2.36)$$

and further the metric is hermitian with respect to both complex structures, the geometry of the manifold is called *bihermitian*. The corresponding Kähler forms are defined as in (2.29) as $\omega^{(\pm)} = gJ^{(\pm)}$. For only one complex structure, the geometry defined by the constraint $\nabla^{(+)}J^{(+)} = 0$ is also known as *strong Kähler with torsion* (SKT) [HP96], so bihermitian geometry is equivalent with SKT-geometry in two directions.

Bihermitian geometry plays an important role in the study of supersymmetric sigma models, as will be discussed in chapter 4. In the next section, the equivalence between bihermitian geometry and generalized Kähler geometry will be reviewed.

### 2.1.6 Hyperkähler geometry

A manifold with three integrable structures $(I,J,K)$ satisfying the quaternion algebra [Ham43]

$$I^2 = J^2 = K^2 = -1, \quad IJK = -1 \quad (2.37)$$

is called a *hypercomplex* manifold. All linear combinations of $(I, J, K)$, where the coefficients $a, b, c \in \mathbb{R}$ lie on a two-sphere,

$$\tilde{J} = aI + bJ + cK, \quad a^2 + b^2 + c^2 = 1, \tag{2.38}$$

is again a complex structure. Thus, a hypercomplex manifold has a two-sphere of complex structures and can be parametrized by a complex coordinate $\zeta$ using the stereographic projection $S^2 \to \mathbb{C}$ with the complex coordinate defined as $\zeta = (b + ic)/(1 + a)$ with $(a, b, c) \in S^2$ [HKLR87],

$$\tilde{J}(\zeta) = \tfrac{1 - \zeta\bar{\zeta}}{1 + \zeta\bar{\zeta}} I + \tfrac{\zeta + \bar{\zeta}}{1 + \zeta\bar{\zeta}} J + \tfrac{i(\bar{\zeta} - \zeta)}{1 + \zeta\bar{\zeta}} K. \tag{2.39}$$

If the metric is hermitian with respect to all three complex structures, the manifold is hyperhermitian. Further, if each Kähler form $\omega_i = (gI, gJ, gK)$ with $i = 1, 2, 3$, corresponding to the three complex structures is closed, or equivalently, if the complex structures are all parallel with respect to the Levi-Civita connection,

$$\nabla I = \nabla J = \nabla K = 0, \tag{2.40}$$

the manifold is *hyperkähler*. Choosing coordinates that are (anti-) holomorphic with respect to the complex structure $I$, the Kähler form corresponding to $I$ is $\omega_1 = -i\partial\bar{\partial}\tilde{K}$, where $\tilde{K}$ is the Kähler potential, and the combinations

$$\omega^{(\pm)} = \omega_2 \pm i\omega_3 \tag{2.41}$$

are holomorphic symplectic $(2,0)$ and anti-holomorphic $(0,2)$-forms, respectively. The three Kähler forms can be combined using the complex coordinate $\zeta$ into a holomorphic symplectic form with respect to the complex structure $\tilde{J}(\zeta)$ in (2.39),

$$\Omega(\zeta) = \omega^{(+)} + \zeta\omega_1 - \zeta^2\omega^{(-)}. \tag{2.42}$$

A Killing vector preserving all three symplectic forms, $\mathcal{L}_k(\omega_i) = 0$, is called *triholomorphic*. Triholomorphic Killing vectors will be relevant when gauging isometries of hyperkähler manifolds in later chapters.

The two-sphere of complex structures allows for an alternative definition of a hyperkähler manifold. A locally irreducible Riemannian manifold equipped with two complex structures $J^{(\pm)}$ is hyperkähler if the metric is Kähler with respect to both complex structures and the two complex structures are not proportional, $J^{(-)} \neq \pm J^{(+)}$ [Mor07]. It follows that the anti-commutator of the two complex structures is proportional to the identity,

$$\{J^{(+)}, J^{(-)}\} = 2c\mathbb{1}, \tag{2.43}$$

with the constant $c \in \mathbb{R}$ satisfying $|c| < 1$. This implies that a third complex structure can be defined as

$$K = \frac{1}{2\sqrt{1 - c^2}} [J^{(+)}, J^{(-)}]. \tag{2.44}$$

A four-dimensional Kähler manifold is hyperkähler if and only if there are holomorphic coordinates $(z, w)$ such that the metric $g_{ij} = \partial_i \partial_j \tilde{K}$ satisfies the Monge-Ampère equation [Yau78, Cal79]

$$\det(g) = g_{z\bar{z}} g_{w\bar{w}} - g_{z\bar{w}} g_{\bar{z}w} = 1. \tag{2.45}$$

In higher dimensions, this constraint generalizes to a system of partial differential equations for $\tilde{K}$. Corresponding to a non-vanishing constant $|c| < 1$ in (2.43), equation (2.45) can be generalized to $\det(g) = (1 - c^2)^2$.

An example of a hyperkähler geometry is the Eguchi-Hanson metric [EH79], which may be defined by the real function [Dyc11]

$$\tilde{K} = r + \frac{1}{2} \ln\left(\frac{r-1}{r+1}\right), \quad r^2 = 1 + 4z\bar{z}(1 + w\bar{w})^2. \tag{2.46}$$

The Kähler metric defined by $g_{ij} = \partial_i \partial_j \tilde{K}$ for $x^i = (z, \bar{z}, w, \bar{w})$ satisfies the Monge-Ampère equation $\det(g) = 1$.

The name hyperkähler originates from Calabi [Cal79], but the concept arose already in Berger's classification of the holonomy groups of Riemannian manifolds [Ber55, MS99]. Since the complex structures are covariantly constant (2.40), the holonomy group of a hyperkähler manifold is contained in both the orthogonal group $O(4n)$ and the group of quaternionic invertible matrices $GL(n, \mathbb{H})$. The maximal such intersection is the group of $n \times n$ quaternionic unitary matrices $Sp(n) \subset SU(n) \subset U(n)$. All hyperkähler manifolds are Ricci-flat Kähler and hence Calabi-Yau. Since the holonomy group $Sp(n)$ is also an intersection of $U(2n)$ and $Sp(2n, \mathbb{C})$, the linear transformations of $\mathbb{C}^{2n}$ that preserve a non-degenerate skew-symmetric form, a hyperkähler manifold is a complex manifold with a holomorphic symplectic form [Hit92].

If the connection includes torsion as in (2.35) and preserves both the metric and the three complex structures, the geometry is said to be *strong hyperkähler with torsion* (strong HKT) [HP96]. Of course, if the torsion vanishes, the connection reduces to the ordinary Levi-Civita connection and the geometry is hyperkähler.

### 2.1.7 Neutral hyperkähler geometry

A *pseudo-hypercomplex* manifold has three integrable structures $(I, S, T)$ satisfying the algebra of split quaternions,

$$-I^2 = S^2 = T^2 = 1, \quad IST = 1 \tag{2.47}$$

In other words, the manifold has two *local product structures* $S$ and $T$, squaring to one, and one complex structure $I$. The individual integrability of the structures is again equivalent to the vanishing of the Nijenhuis tensors (2.8).

Oriented four-dimensional manifolds with pseudo-hypercomplex structures always allow for a local skew-hermitian metric $g$ which must have signature $(2,2)$ [Dun02]. Such a metric is referred to as neutral; accordingly, pseudo-hypercomplex manifolds are sometimes referred to as *neutral* hypercomplex.

As for ordinary hypercomplex structures, the fundamental two-forms corresponding to the pseudo-hypercomplex structures are defined as in (2.29). Again, if the fundamental two-forms corresponding to $(I,S,T)$ are closed, or equivalently, if the three structures are covariantly constant with respect to the Levi-Civita connection, the manifold is *neutral hyperkähler* [Kam99], also known as *pseudo-hyperkähler* or *hypersymplectic* [Hit90]. Four-dimensional neutral hyperkähler manifolds may be either complex four-tori or Kodaira surfaces [Kam99].

## 2.2 Generalized complex geometry

Complex and symplectic geometry can be united in a larger framework called *generalized complex geometry*, introduced in [Hit03]. As will be seen in chapter 4, the most general sigma model with two manifest supersymmetries in each chirality has a target space which is bihermitian. This special case of generalized complex geometry is equivalent to *generalized Kähler geometry*, and the explicit map was given in [Gua03].

As was reviewed in the previous section, a complex structure is an integrable map $J : T_pM \to T_pM$ with $J^2 = -1$. This concept can be generalized by substituting the tangent bundle by the direct sum of the tangent bundle and the cotangent bundle $TM \oplus T^*M$ and the Lie bracket $[X,Y]$ by the Courant bracket

$$[X+\xi, Y+\eta]_C = [X,Y] + \mathcal{L}_X\eta - \mathcal{L}_Y\xi - \frac{1}{2}d(i_X\eta - i_Y\xi), \qquad (2.48)$$

where the vector fields $X, Y \in TM$ and the forms $\xi, \eta \in T^*M$ pair up as elements $X+\xi \in TM \oplus T^*M$. The Lie derivative $\mathcal{L}_X Y$ of a tensor $Y$ measures the change of the tensor along a flow generated by a vector field $X$. When $Y$ is a vector field, the Lie derivative is simply the Lie bracket $\mathcal{L}_X Y = [X,Y]$. The Lie derivative acting on a differential form is given by the Cartan formula [Car45],

$$\mathcal{L}_X \omega = \iota_X d\omega + d(\iota_X \omega), \qquad (2.49)$$

relating the exterior derivative $d$ with the interior derivative $\iota$.

The Courant bracket is skew-symmetric, but not a Lie bracket since it does not satisfy the Jacobi identity. The Jacobiator can be introduced to measure the Courant bracket's failure to satisfy the Jacobi identity. It does so by an exact form, namely the exterior derivative of the generalization of the Nijenhuis tensor in generalized complex geometry [Gua03]. When projected down onto $TM$, the Courant bracket reduces to the ordinary Lie bracket.

| Complex structure | Generalized complex structure |
|---|---|
| $J : TM \to TM$ | $\mathcal{J} : TM \oplus T^*M \to TM \oplus T^*M$ |
| $J^2 = -1$ | $\mathcal{J}^2 = -1, \quad \mathcal{J}^t \mathcal{I} \mathcal{J} = \mathcal{I}$ |
| $P^{\mp}[P^{\pm}X, P^{\pm}Y] = 0$ | $\Pi_{\mp}[\Pi_{\pm}\mathcal{X}, \Pi_{\pm}\mathcal{Y}]_C = 0$ |
| $X, Y \in TM$ | $\mathcal{X}, \mathcal{Y} \in TM \oplus T^*M$ |

*Figure 2.1:* Comparison between complex and generalized complex geometry.

An important property of the Courant bracket is that it allows an extra symmetry in addition to diffeomorphisms, namely *b*-field transformations involving a closed two-form *b* acting as

$$X + \xi \mapsto X + \xi + i_X b. \tag{2.50}$$

The Courant bracket may be twisted by a closed three-form *H*, defining the *H-twisted* Courant bracket as

$$[X + \xi, Y + \eta]_H = [X + \xi, Y + \eta]_C + i_X i_Y H. \tag{2.51}$$

If the three-form is exact, $H = db$, then the last term in (2.51) can be generated by a *b*-transform (2.50) with a non-closed two-form *b*. The metric on $TM$ can be extended to a *natural pairing* $\mathcal{I}$ on $TM \oplus T^*M$ defined by

$$\langle X + \xi, Y + \eta \rangle = \tfrac{1}{2}(i_X \eta + i_Y \xi). \tag{2.52}$$

The natural pairing is symmetric and non-degenerate and takes the form

$$\mathcal{I} = \begin{pmatrix} 0 & 1 \\ 1 & 0 \end{pmatrix} \tag{2.53}$$

in the local coordinates $(\partial_\mu, dx^\mu)$ [LMTZ05]. With these generalizations, summarized in the chart 2.1, a *generalized* almost complex structure $\mathcal{J}$ is defined as an automorphism of $TM \oplus T^*M$ which squares to minus one and preserves the natural pairing,

$$\mathcal{J}^2 = -1, \quad \mathcal{J}^t \mathcal{I} \mathcal{J} = \mathcal{I}. \tag{2.54}$$

The integrability condition is defined analogously as for complex structures. With projection operators defined as $\Pi_{\pm} = \tfrac{1}{2}(1 \mp i\mathcal{J})$, it can be written as

$$\Pi_{\mp}[\Pi_{\pm}(X + \xi), \Pi_{\pm}(Y + \eta)]_C = 0. \tag{2.55}$$

A map $\mathcal{J}$ fulfilling the conditions (2.54)-(2.55) is called a *generalized complex structure*, in analogy to the complex structures reviewed previously.

### 2.2.1 Generalized Kähler geometry

*Generalized Kähler geometry* is defined as a pair of two commuting generalized complex structures $\mathcal{J}_1, \mathcal{J}_2$ for which $\mathcal{G} = -\mathcal{J}_1 \mathcal{J}_2$ defines a positive definite metric on $TM \oplus T^*M$. Strictly speaking, $\mathcal{G}$ has the wrong index structure to be a metric, but it can be used to construct a metric $\mathcal{H} \propto \mathcal{G}$ [HHZ10]. Following conventional notation [Gua03], though, $\mathcal{G}$ will here be referred to as the generalized metric.

If $(J, g, \omega)$ defines a Kähler geometry with Kähler form $\omega$ and two generalized complex structures are defined by

$$\mathcal{J}_1 = \begin{pmatrix} J & 0 \\ 0 & -J^t \end{pmatrix}, \quad \mathcal{J}_2 = \begin{pmatrix} 0 & -\omega^{-1} \\ \omega & 0 \end{pmatrix}, \qquad (2.56)$$

then

$$\mathcal{G} = -\mathcal{J}_1 \mathcal{J}_2 = \begin{pmatrix} 0 & g^{-1} \\ g & 0 \end{pmatrix} \qquad (2.57)$$

defines a generalized Kähler geometry where $\mathcal{G}$ is constructed from the Kähler metric $g$. More generally, given a bihermitian structure $(J^{(\pm)}, g, b)$ with corresponding two-forms $\omega^{(\pm)} = g J^{(\pm)}$, a generalized Kähler structure can be defined by the two generalized complex structures [Gua03]

$$\mathcal{J}_{1,2} = \frac{1}{2} \begin{pmatrix} 1 & 0 \\ b & 1 \end{pmatrix} \begin{pmatrix} J^{(+)} \pm J^{(-)} & -[\omega^{(+)-1} \mp \omega^{(-)-1}] \\ \omega^{(+)} \mp \omega^{(-)} & -[J^{(+)t} \pm J^{(-)t}] \end{pmatrix} \begin{pmatrix} 1 & 0 \\ -b & 1 \end{pmatrix}. \qquad (2.58)$$

The generalized Kähler structures are integrable if and only if the Kähler forms satisfy [Gua03]

$$d^{c(+)} \omega^{(+)} + d^{c(-)} \omega^{(-)} = 0, \quad d d^{c(\pm)} \omega^{(\pm)} = 0, \qquad (2.59)$$

where $d^{c(\pm)} = i(\bar{\partial}^{(\pm)} - \partial^{(\pm)})$ and the $(\pm)$-index denotes holomorphicity with respect to the complex structure $J^{(\pm)}$ in respective canonical coordinates. The torsion is then given by $H = d^{c(+)} \omega^{(+)} = -d^{c(-)} \omega^{(-)}$. Note, that if the torsion vanishes, then $\partial^{(\pm)} \omega^{(\pm)}$ and $\bar{\partial}^{(\pm)} \omega^{(\pm)}$ vanish separately, implying that $d \omega^{(\pm)} = 0$ and the geometry is simply Kähler. This corresponds to the situation when $J^{(+)} = J^{(-)} = J$ in the case (2.56) above.

Equation (2.58) is is the explicit map between bihermitian geometry and generalized Kähler geometry. The inverse is true up to the symmetries of the Courant bracket; $b$-transforms and diffeomorphisms.

Real Poisson structures can be defined on a generalized Kähler manifold as [LZ02, Hit06]

$$\begin{aligned} \pi^{(\pm)} &= \left( J^{(+)} \pm J^{(-)} \right) g^{-1}, \\ \sigma &= [J^{(+)}, J^{(-)}] g^{-1} = \pi^{(-)} g \pi^{(+)}. \end{aligned} \qquad (2.60)$$

Generalized Kähler geometry can, like ordinary Kähler geometry, be described by one single generalized Kähler potential [LRvUZ07a]. The potential serves as the Lagrangian for two-dimensional sigma models with two manifest supersymmetries, as will be discussed in detail in section 4.4. The geometric structures are expressions of second order derivatives of the generalized Kähler potential and become linear when the Poisson structure $\sigma$ vanishes, but are non-linear in general. In fact, when $\sigma$ is invertible, $\Omega = \sigma^{-1}$ is a symplectic structure and the metric is given simply by

$$g = \Omega[J^{(+)}, J^{(-)}], \qquad (2.61)$$

which has important consequences for the geometry.

Recently, analogue relations between pseudo-hermitian geometries with indefinite metrics and corresponding structures in generalized (pseudo-) complex geometry have been developed [Gua07, HLMdS$^+$09, DGMY11].

For a generalized Kähler manifold with $\{J^{(+)}, J^{(-)}\} = 2c$ constant, the manifold is hyperkähler when $|c| < 1$ [LRvUZ07a]. This implies that the two complex structures are not proportional, $J^{(+)} \neq \pm J^{(-)}$. Actually, an equivalent statement for a generalized Kähler manifold to be hyperkähler is that the corresponding spaces $(M, g, J^{(\pm)})$ are Kähler and $J^{(+)} \neq \pm J^{(-)}$ [OP09].

The description of generalized complex geometry in terms of a generalized Kähler potential is valid locally away from irregular points. A regular point is defined as a point in the manifold for which a neighborhood exist where the ranks of the Poisson structures $\pi^{(\pm)}$ in (2.60) are constant, or equivalently, when the type of the generalized complex structure is constant. In this thesis, we restrict to descriptions away from irregular points.

# 3. Supersymmetry

> Scientific work must not be considered from the point of view of the direct usefulness of it. It must be done for itself, for the beauty of science, and then there is always the chance that a scientific discovery may become like the radium, a benefit.
>   *Marie Curie, physicist and chemist (1867-1934)*

Our understanding of physics relies on the framework of symmetries. Already Newton's laws of mechanics embodied symmetry principles, and later, Maxwell constructed the laws for classical electrodynamics, which were eventually understood to have both Lorentz invariance and gauge invariance. The special relativity developed by Einstein is basically Poincaré symmetry of space-time, and the underlying principle of general relativity is covariance under diffeomorphisms of space-time.

The implications of symmetries in nature became more well understood after Emmy Noether proved her famous theorem on conservation laws: that for every global continuos symmetry of a system, there is a conserved charge [Noe18]. For example, the total energy is conserved in a system that does not depend on time, and angular momentum is preserved if the system is rotationally invariant.

At the level of quantum mechanics, the significance of symmetries is even more profound. A quantum state is an irreducible representation of a symmetry group; in relativistic quantum mechanics the representations of the Poincaré group lead to a complete classification of elementary particles, labeled by their spin and mass, or, in the massless representations, helicity. This divides the elementary particles into two groups: the bosons with integer spin and wave functions that are invariant under the interchange of two identical particles, and fermions with half-integer spin and wave functions that receive a sign change.

The Standard Model of physics is a quantum field theory that unifies the electromagnetic force, the weak and the strong force into one framework and explains the dynamics of all known subatomic particles to an impressive precision. It is defined by the symmetry under the local (gauge) group $SU(3) \times SU(2) \times U(1)$. The latest achievement, and the missing piece in the puzzle, is the discovery of a Higgs-like boson at the Large Hadron Collider earlier this year (2012). But despite its many successes, the Standard Model fails to explain certain fundamental questions. It predicts the neutrinos to be massless, whereas new experiments have indicated that they do have mass. The Standard Model does not provide an explanation of gravity, and since it does not explain dark matter or dark energy, it only concerns 4% of the energy present in the universe. Further, the theory depends on 19 free parameters, but does not explain their value, and is unnatural, in the sense that it must be heavily fine tuned to neutralize quantum corrections. This and other deficits seem to hint that the theory is only the low energy limit of some deeper underlying theory.

Seeing that the search for symmetries have provided a fruitful route of understanding the physical laws of nature, a natural approach was to introduce a symmetry between bosons and fermions. This new extraordinary symmetry that unifies force and matter was given the intriguing name *supersymmetry* [GL71, VA73, WZ74].

But not only does supersymmetry intertwine fermions and bosons, the theory also suggests possible resolutions to problems in the Standard Model, such as explaining the hierarchy problem and converging the gauge couplings of the Standard Model at high energies. It moreover contains candidates for dark matter and is an essential feature of string theory.

However, supersymmetry predicts that every known elementary particle has a superpartner, and the existence of these particles have not yet been experimentally verified. If supersymmetry is a fundamental symmetry of the physical world we inhabit, it must be broken at a high energy scale; it remains to be seen if traces of supersymmetry can be observed in the future. Until then, supersymmetry continues to play an important role in physics and mathematics, providing us with tools and insights in quantum field theory, differential geometry, representation theory and string theory.

## 3.1 Supersymmetry and representations

Supersymmetry relates integer and half-integer spin particles by combining them in one multiplet. At the algebraic level, the supersymmetry algebra is a *Super-Poincaré* algebra, the extension of the Poincaré algebra to involve odd generators.

The elements in the Poincaré algebra generate translations $P_\mu$ and Lorentz transformations $M_{\mu\nu}$. In four-dimensional Minkowski space with metric $\eta_{\mu\nu}$, the commutators of the generators give rise to the Poincaré algebra

$$\begin{aligned}
[P_\mu, P_\nu] &= 0, \\
[P_\mu, M_{\nu\tau}] &= \eta_{\mu[\nu} P_{\tau]}, \\
[M_{\mu\nu}, M_{\tau\sigma}] &= M_{\mu[\sigma} \eta_{\tau]\nu} - M_{\nu[\sigma} \eta_{\tau]\mu},
\end{aligned} \quad (3.1)$$

see, e.g., [Wes86]. Coleman and Mandula showed on general grounds that, in order to be compatible with relativistic field theory, any larger group containing both the Poincaré group and an internal symmetry group with algebra $[B_I, B_J] = f_{IJ}{}^K B_K$, must be a direct product of the two groups, i.e., the internal symmetry generators commute trivially with the Poincaré generators [CM67].

In the superalgebra, this no-go theorem is circumvented by the introduction of fermionic (odd) generators that commute with the Poincaré generators and anti-commute with each other in a $\mathbb{Z}_2$-graded Lie algebra [HLS75]. The $\mathbb{Z}_2$-grading implies that the $N$ odd generators $Q$ commute with the Poincaré generators to odd elements

$$\begin{aligned}
[Q_\alpha^i, P_\nu] &= 0, \\
[Q_\alpha^i, M_{\mu\nu}] &= \tfrac{1}{2} (\sigma_{\mu\nu})_\alpha{}^\beta Q_\beta^i,
\end{aligned} \quad (3.2)$$

and anti-commute with each other to even elements. Note that the vanishing commutator between the supersymmetry and the momentum generators implies that $[P^2, Q] = 0$. Hence the mass operator $P^2$ is a Casimir operator, which means that all particles in an irreducible representation of the supersymmetry algebra have the same mass.

Requiring that the anti-commutator satisfies the generalized Jacobi identities and normalizing the momentum operator constrains the anti-commutation algebra to take the form

$$\{Q_\alpha^i, Q_\beta^j\} = 2(\gamma^\mu C)_{\alpha\beta} P_\mu \delta^{ij} + C_{\alpha\beta} Z^{ij} + (\gamma_5 C)_{\alpha\beta} Y^{ij}. \quad (3.3)$$

Here, the supercharges $Q_\alpha^i$, with spinor index $\alpha$ and $i = 1, \ldots, N$ in (3.2)-(3.3) are Majorana spinors in the $(0, \tfrac{1}{2}) \oplus (\tfrac{1}{2}, 0)$-representation of the Lorentz group, $C_{\alpha\beta}$ is the anti-symmetric charge conjugation matrix and $Z^{ij}$, $Y^{ij}$ are central charges. The central charges exist only in extended supersymmetry, $N > 1$, and commute with all generators $\mathcal{O}$,

$$[Z, \mathcal{O}] = [Y, \mathcal{O}] = 0. \quad (3.4)$$

In the Weyl representation, using two-component Weyl spinors, the algebra (3.3) takes the form

$$\begin{aligned}
\{Q_\alpha^i, \bar{Q}_{\dot{\alpha}}^j\} &= 2 P_{\alpha\dot{\alpha}} \delta^{ij}, \\
\{Q_\alpha^i, Q_\beta^j\} &= \varepsilon_{\alpha\beta} (Z^{ij} + Y^{ij}),
\end{aligned} \quad (3.5)$$

where vector and spinor indices are combined in a more compact notation as $P_{\alpha\dot\alpha} = (\sigma^\mu)_{\alpha\dot\alpha} P_\mu$ and $\varepsilon_{12} = \varepsilon_{\dot 1 \dot 2} = -1$ is anti-symmetric.

Introduce an operator $\mathcal{O} = (-1)^{N_f}$ where $N_f$ is the fermionic number operator. All bosonic states have eigenvalue $+1$ and all fermionic states $-1$ of this operator, which implies that the operator anti-commutes with the supersymmetry generators. The trace of the operator $\mathcal{O}$ will be the difference in the number of bosonic and fermionic states, $\text{Tr}\,\mathcal{O} = n_b - n_f$. Using the cyclic properties of the trace together with the supersymmetry algebra (3.5), one can see that the trace vanishes for all representations with nonzero momentum,

$$0 = \text{Tr}\big[\mathcal{O}\{Q_\alpha, \bar{Q}_{\dot\beta}\}\big] = \text{Tr}\big[\mathcal{O}\, 2(\sigma^\mu)_{\alpha\dot\beta} P_\mu\big] = 2(\sigma^\mu)_{\alpha\dot\beta} p_\mu \text{Tr}\big[\mathcal{O}\big]. \qquad (3.6)$$

This shows that every field theory representation of supersymmetry contains the same number of bosonic and fermionic states.

The papers [I-V] that are the basis for this thesis all focus on supersymmetry on two-dimensional world-sheets. In two dimensions, there is a Lorentz-invariant notion of whether a massless particle is moving to the left or to the right. For closed strings in two dimensions, supersymmetry thus induces two independent left- and right-moving supercurrents corresponding to the two chiralities [HW85], and the algebra is given simply by

$$\begin{aligned}
\{Q_+^i, Q_+^j\} &= 2\delta^{ij} P_{++}, \\
\{Q_-^{i'}, Q_-^{j'}\} &= 2\delta^{i'j'} P_{=}, \\
\{Q_+^i, Q_-^{j'}\} &= 0,
\end{aligned} \qquad (3.7)$$

where the momentum operator can be represented as a space-time translation $P_{\pm\pm} = i\partial_{\pm\pm}$ and the light-cone coordinates are defined in terms of the two coordinates of the world-sheet as $x^{\pm\pm} = \tau \pm \sigma$.

Just as the Poincaré algebra was here combined with odd generators according to a $\mathbb{Z}_2$-grading to form a Super-Poincaré algebra, other Lie algebras, such as the conformal algebra, can be combined with supersymmetry, in the latter case giving the superconformal algebra.

### 3.1.1 Twisted supersymmetry

The key feature of the supersymmetry algebra, that the supersymmetry generators anti-commute to a space-time translation, can be generalized to some of the generators anti-commuting to minus a translation, so called *pseudo-supersymmetry* [Hul98, AZH99]. This is made possible by the fact that the supercharges with opposite chirality anti-commute in two dimensions. An algebra with both supersymmetry and pseudo-supersymmetry is denoted *twisted* supersymmetry,

$$\{Q^i_+, Q^j_+\} = 2\eta^{ij} P_{++},$$
$$\{Q^{i'}_-, Q^{j'}_-\} = 2\eta^{i'j'} P_{=},$$
$$\{Q^i_+, Q^{j'}_-\} = 0, \qquad (3.8)$$

where $\eta^{ij} = \text{diag}(\delta^{uu}, -\delta^{vv})$ and $u+v$ equals the number of positive supercharges $Q^i_+$, and similar for $\eta^{i'j'}$.

The anti-commutator in the ordinary supersymmetry algebra (3.7) is a positive definite operator in a Hilbert space with positive definite metric, implying that supersymmetric systems have non-negative energy. The algebra of a twisted supersymmetry (3.8), on the other hand, generates a system with either constantly zero energy or with negative norm states [GH90]. Hence, twisted supersymmetry is essentially irrelevant from a physical context. However, twisted supersymmetry is interesting from a geometrical point of view, since imposed on sigma models it will require the target space to have indefinite metric, and such geometries have received attention both in the context of differential geometry [DJS05, AMT08, DGMY11, BB11] and in string theory [OV90, OV91, Hul98, AZH99].

## 3.2 Superspace and superfields

By construction, superspace is the space in which supersymmetry is inherently manifest, and superfields are functions defined on the superspace. The two concepts are needed to construct models with manifest supersymmetry and were introduced in [SS74] and [FWZ74].

Lorentz invariance is manifest in the Minkowski space, which can be defined as the coset of the Poincaré group and the Lorentz group. Superspace may be defined analogously, as the coset of the Super-Poincaré group and the Lorentz group. In addition to the even coordinates $x^\mu$ of the Minkowski space, superspace is equipped with odd *Grassmann* coordinates carrying spinor index, $\theta^\alpha$. The group element of the Super-Poincaré group are generated by the operators in the Super-Poincaré algebra as

$$g(x,\theta) = e^{i(x^\mu P_\mu + \theta^\alpha Q_\alpha)}. \qquad (3.9)$$

By definition, the Grassmann coordinates satisfy the anti-commutation relations $\{\theta^\alpha, \theta^\beta\} = 0$, implying nilpotency $(\theta^\alpha)^2 = 0$. The Grassmann differential operator is defined to act from the left and also satisfies anti-commutation relations,

$$\frac{\partial}{\partial \theta^\alpha}(\theta^\beta \theta^\gamma) = \frac{\partial \theta^\beta}{\partial \theta^\alpha} \theta^\gamma - \theta^\beta \frac{\partial \theta^\gamma}{\partial \theta^\alpha} = \delta^\beta_\alpha \theta^\gamma - \delta^\gamma_\alpha \theta^\beta. \qquad (3.10)$$

The integration over the odd Grassmann coordinates is denoted the *Berezin integral* [Ber66]. Requiring that the integral should be linear and invariant under translations implies that the Berezin integral must pick out the highest term in the component expansion of a superfield. In a one-dimensional anti-commuting space with only one Grassmann variable, a general superfield has the expansion $\phi = a + \theta b$, and the Berezin integral is thus

$$\int d\theta \, (a + \theta b) \sim b. \tag{3.11}$$

Normalization is chosen such that $\int d\theta \, \theta = 1$ and $\int d\theta \, 1 = 0$, which actually means that the Berezin integral is identical to differentiation:

$$\int d\theta = \frac{\partial}{\partial \theta}, \tag{3.12}$$

where the differentiation operator is defined to act from the left.

A *superfield* is a function both of the space-time coordinates and the Grassmann coordinates, $\phi = \phi(x, \theta)$. Due to the nilpotency of the Grassmann coordinates, superfields can be expanded in terminating Taylor series in $\theta$. As such, they can be viewed as a collection of component fields, a *supermultiplet*. For example, in two dimensions and for $N=(1,1)$ supersymmetry, there is only one Grassmann coordinate in each chirality, so the component expansion of a superfield has four terms,

$$\phi^\mu(x, \theta) = X^\mu(x) + \theta^+ \psi^\mu_+(x) + \theta^- \psi^\mu_-(x) + \theta^+ \theta^- F^\mu(x) \tag{3.13}$$

and can be viewed as the collection of the fields $(X^\mu(x), \psi^\mu_\pm(x), F^\mu(x))$. Obviously, linear combinations and products of superfields are again superfields.

Supersymmetry covariant derivatives are defined to anti-commute with the supersymmetry operators, $\{D, Q\} = 0$. Similarly to the momentum operator, the supersymmetry generator and the supersymmetry covariant derivatives may be represented as differential operators. In two dimensions, the $N=(1,1)$ operators can be represented as

$$Q_\pm = i\frac{\partial}{\partial \theta^\pm} + \theta^\pm \partial_{\pm\pm}, \quad D_\pm = \frac{\partial}{\partial \theta^\pm} + i\theta^\pm \partial_{\pm\pm}, \tag{3.14}$$

and satisfy the supersymmetry algebra $Q_\pm^2 = D_\pm^2 = i\partial_{\pm\pm}$. Projecting out the $\theta$-independent part of the covariant derivative in (3.14) implies the useful relation

$$D\big| = \frac{\partial}{\partial \theta}, \tag{3.15}$$

where $\big| = \big|_{\theta=0}$ is the notation for setting all Grassmann coordinates in an expression to zero. The components of the superfield (3.13) are obtained by projecting out the $\theta$-independent part. Then, the leading component field of (3.13) is a bosonic scalar field $\phi| = X$, whereas the component fields $D_\pm\phi| = \psi_\pm$ are fermionic (odd), since the covariant derivatives $D_\pm$ are odd. As will be clear later when discussing actions of superfields, $X$ and $\psi_\pm$ are physical fields, with propagating degrees of freedom, whereas $F$ is an auxiliary field in the sense that its equations of motion are purely algebraic and can be used to eliminate it from the action.

A supersymmetry transformation of a scalar superfield is generated by the supersymmetry operators $Q$ as

$$\delta\phi = \phi' - \phi = e^{i\epsilon Q}\phi(x,\theta)e^{-i\epsilon Q} - \phi(x,\theta), \qquad (3.16)$$

where $\epsilon$ is the supersymmetry parameter. Using the Baker-Hausdorff formula and discarding all infinitesimal terms smaller than or equal to $\epsilon^2$, the infinitesimal version of the supersymmetry transformations is

$$\delta\phi = i[\epsilon Q, \phi(x,\theta)]. \qquad (3.17)$$

The defining property of the supersymmetry covariant derivatives implies that $\delta(D\phi) = D(\delta\phi)$, hence the name *covariant* derivatives. The supersymmetry algebra (3.7) implies that two subsequent supersymmetry transformations on a field commute to a translation,

$$[\delta(\epsilon_1), \delta(\epsilon_2)]X = \epsilon_{[2}\epsilon_{1]}i\partial X. \qquad (3.18)$$

In the twisted supersymmetry algebra (3.8), some generators close to minus a translation, a pseudo-supersymmetry

$$[\delta(\epsilon_1), \delta(\epsilon_2)]X = -\epsilon_{[2}\epsilon_{1]}i\partial X. \qquad (3.19)$$

From (3.17) and the explicit form of the supersymmetry generators in (3.14), the supersymmetry transformations on the component fields can be read off as

$$\delta\phi = \delta X + \theta^+\delta\psi_+ + \theta^-\delta\psi_- + \theta^+\theta^-\delta F \qquad (3.20)$$

where the component fields transform as

$$\delta X = -\epsilon^+\psi_+ - \epsilon^-\psi_-, \qquad \delta\psi_+ = -i\epsilon^+\partial_{++}X - \epsilon^-F,$$
$$\delta F = i\epsilon^+\partial_{++}\psi_- - i\epsilon^-\partial_=\psi_+, \qquad \delta\psi_- = -i\epsilon^-\partial_=X + \epsilon^+F. \qquad (3.21)$$

Note that the bosonic fields are transformed into fermionic fields and vice versa, and that the $\theta^+\theta^-$-term is a total derivative. This will be true for the highest term in the $\theta$-expansion for all superfields. Schematically, expanding the transformation of a superfield as $\delta X + \cdots + \theta^q\delta F$, the transformation of the highest component is

$$\delta F = \delta(D^q\phi|) = D^q(\delta\phi)| = D^q(\epsilon Q\phi)| \propto D^q(\epsilon D\phi)|, \qquad (3.22)$$

and the supersymmetry algebra for more than $q$ supersymmetry generators is proportional to a total derivative. Since the Berezin integral picks out the highest superfield component (3.11), the fact that the last expression is a total derivative then implies that any action written in terms of superfields is invariant under supersymmetry transformations, up to boundary terms,

$$\delta S = \delta \int d^d x \int d^q \theta \mathcal{L}(\phi) = \int d^d x \, \delta F = 0. \tag{3.23}$$

The superfields form a linear representation of the supersymmetry algebra, but in general, the representations are reducible, that is, they contain a constrained superfield that serves as a non-trivial subrepresentation. The redundant component fields can be eliminated by imposing covariant constraints. In four dimensions, unconstrained superfields have 16 component fields, including a vector field $A_{\alpha\dot\alpha} = (\sigma^\mu)_{\alpha\dot\alpha} A_\mu$,

$$\begin{aligned}\Phi(x,\theta,\bar\theta) = {}& X + \theta^\alpha \psi_\alpha + \bar\theta^{\dot\alpha} \bar\lambda_{\dot\alpha} + \theta^2 M + \bar\theta^2 N \\ & + \theta^\alpha \bar\theta^{\dot\alpha} A_{\alpha\dot\alpha} + \theta^\alpha \bar\theta^2 \chi_\alpha + \bar\theta^{\dot\alpha} \theta^2 \bar\zeta_{\dot\alpha} + \theta^2 \bar\theta^2 F,\end{aligned} \tag{3.24}$$

where the notation $\theta^2 = \theta^\alpha \theta_\alpha$ is used. Two irreducible representations that can be constructed are given by the chiral superfield, constrained by the differential operator $\bar D_{\dot\alpha} \Phi = 0$, and by the vector superfield, constrained by the reality condition $\Phi^\dagger = \Phi$, see, e.g., [WB92]. The chiral multiplet is constrained to depend on only one scalar boson, two fermions and one scalar auxiliary field, and derivatives of the same fields. The vector multiplet, on the other hand, still contains a vector field $A_\mu$, hence the name. We will return to constrained superfields in the next chapter.

# 4. Supersymmetric sigma models

> It is impossible to be a mathematician without being a poet in soul.
> *Sonya Kovalevsky, mathematician (1850-1891)*

In the previous two chapters, the topics of complex geometry and supersymmetry have been reviewed. Now, the foundation is laid to merge the two subjects.

The non-linear sigma models relate the concepts of complex geometry and supersymmetry in a striking way. Imposing extended supersymmetry on sigma models results in geometric constraints on the target space. For example, a two-dimensional sigma model with extended $N=(2,2)$ supersymmetry has a target space geometry which is bihermitian, or generalized Kähler. This connection between sigma models and geometry provides a link between string theory and mathematics.

Sigma models were introduced to describe a phenomenological model involving a pion and a spin 0 meson denoted $\sigma$ [GML60]. The non-linear sigma models discussed here are something much more general. In the first section, the non-linear sigma model will be derived from the perspective of a propagating string. But, despite the fact that sigma models are fundamental objects in string theory, they can be discussed without the context of strings. A non-linear sigma model is simply a theory describing fields that span a manifold, that is usually curved. They can be used to describe effective field theories, quantum field theories or purely classical theories, the motion of higher-dimensional branes, or to model different geometries.

Focus in this thesis is on two-dimensional non-linear sigma models. The two-dimensional theories are special in several aspects, which will be discussed in more detail later.

Some details will be left out; the interested reader may consult any of the standard textbooks for further details, for example [Pol98a, Pol98b, GSW87] or [BBS07].

## 4.1 String theory and bosonic sigma models

A *sigma model* is a theory of maps from a $D$-dimensional parameter space $\Sigma$, denoted the world-volume, into a $d$-dimensional target manifold $M$,

$$X^\mu : \Sigma \to M$$
$$\xi^a \mapsto X^\mu(\xi^a). \qquad (4.1)$$

The set of maps $X^\mu$ with $\mu = 1, \ldots d$ are the coordinates of the the world-volume in the target space. Let the target manifold be equipped with a metric $g = g(X)$. The pullback of the target space metric by the embedding $X^\mu$,

$$(X^*g)_{ab} = \frac{\partial X^\mu}{\partial \xi^a}\frac{\partial X^\nu}{\partial \xi^b} g_{\mu\nu}(X), \qquad (4.2)$$

induces a metric $\gamma_{ab} = (X^*g)_{ab}$ on the world-volume. A measure that is invariant under diffeomorphisms on the world-volume can be constructed as

$$dV = d^D\xi\sqrt{-\det\gamma}. \qquad (4.3)$$

The invariant volume element can be used to construct an action. First, the one-dimensional case of a point particle will be considered, then we move over to the discussion of the string.

### 4.1.1 The relativistic point particle

A sigma model in one dimension describes the dynamics of a free particle. The volume element is simply the infinitesimal line element along which the particle is propagating,

$$ds^2 = -g_{\mu\nu}dX^\mu dX^\nu, \qquad (4.4)$$

which is real for a time-like curve. The dynamics of the particle is given by extremizing the action

$$S = -m\int ds, \qquad (4.5)$$

where the sign is chosen so that the action reproduces the action for a free particle in the non-relativistic limit $v \ll c$,

$$S \sim \int \frac{mv^2}{2} dt. \qquad (4.6)$$

Chosing time as the parameter of the world-line, the Euler-Lagrange equations resulting from (4.6) are the geodesic equations describing a free massive particle,

$$\nabla \dot{X}^\mu = \ddot{X}^\mu + \Gamma^\mu_{\nu\rho}\dot{X}^\nu\dot{X}^\rho = 0. \qquad (4.7)$$

The action (4.5) with the line element (4.4) has the disadvantages that it cannot describe a massless particle, and the square root in the integral obstructs quantization. Instead, a classically equivalent action can be introduced,

$$S = \frac{1}{2} \int dt \left( \frac{1}{e} g_{\mu\nu} \dot{X}^\mu \dot{X}^\nu - em^2 \right), \quad (4.8)$$

where $e = e(t)$ is an independent metric of the world-line. Varying the action with respect to $e$ implies that $(me)^2 = -g_{\mu\nu}\dot{X}^\mu \dot{X}^\nu$, which inserted into (4.8) recovers the initial action (4.5) and shows the equivalence of the two actions. The sigma model for a massless particle is obtained in the limit where $e = 1$ and $m = 0$.

### 4.1.2 The bosonic string

The two-dimensional analogue of the line element (4.4) and action of the particle in (4.5), is a one-dimensional string sweeping out a two-dimensional world-sheet, described by the action

$$S = -T \int d^2\xi \sqrt{-\det \gamma}, \quad \gamma_{ab} = g_{\mu\nu} \frac{\partial X^\mu}{\partial \xi^a} \frac{\partial X^\nu}{\partial \xi^b}. \quad (4.9)$$

The tension $T$ has dimension mass per unit length and is, for historical reasons, usually written in terms of the Regge slope parameter $\alpha'$ as $T = 1/(2\pi\alpha')$. Denote the two coordinates of the world-sheet as $\xi^a = (\tau, \sigma)$ where $\tau$ is time-like and $\sigma$ space-like, with derivatives with respect to the parameters written as $\partial_\tau X = \dot{X}$ and $\partial_\sigma X = X'$, respectively. Inserted, the *Nambu-Goto* action [Got71, Nam86] takes the form

$$S = -T \int d\tau d\sigma \sqrt{(\dot{X} \cdot X')^2 - (\dot{X})^2 (X')^2}. \quad (4.10)$$

As in the one-dimensional action discussed above, the square root obstructs quantization; instead, a classically equivalent action is introduced. Endowing the world-sheet with an independent metric $h_{ab}$, the *Polyakov* action takes the form

$$S = -\frac{T}{2} \int d\tau d\sigma \sqrt{-h} h^{ab} \gamma_{ab}. \quad (4.11)$$

This action was introduced independently in [DZ76] and [BDVH76] and later quantized in [Pol81a, Pol81b]. As in the one-dimensional case, varying the Polyakov action with respect to the world-sheet metric and reinserting the resulting expression $2\sqrt{-\gamma} = h^{cd}\gamma_{cd}\sqrt{-h}$, reproduces the Nambu-Goto action (4.10) and shows that the two actions are classically equivalent.

The two-dimensional sigma model differs from sigma models in other dimensions in several important ways. Under a local rescaling of the world-sheet metric of the form

$$h_{ab} \to e^{\phi(\tau,\sigma)} h_{ab}, \quad (4.12)$$

a so called *Weyl transformation*, the volume element $dV$ is invariant in two dimensions, which is not the case in other dimensions.

The Polyakov action (4.11) has two further symmetries: it is invariant under global Poincaré transformations and under local diffeomorphisms of the world-sheet, $\xi^a \to \xi^{b\prime}(\xi^a)$. These invariances can be used to choose conformal gauge in which the metric is conformally flat. First, a reparametrization of the world-sheet coordinates can be performed to put the metric on a diagonal form up to a scale change, $h_{ab} \to e^{\phi}\eta_{ab}$, where $\eta_{ab} = \text{diag}(-1,1)$. Then, the Weyl invariance (4.12) of the two-dimensional surface can be used to rescale the metric to the form $h_{ab} = \eta_{ab}$. In this *conformal gauge*, the Polyakov action takes the form

$$S = -\frac{T}{2} \int d\tau d\sigma \, \partial_a X^\mu \partial^a X^\nu g_{\mu\nu}(X), \tag{4.13}$$

where $g_{\mu\nu}(X)$ is again the metric of the target space.

Varying the action with respect to the embedding coordinates $X^\mu$ gives, after partial integration, a bulk term and a surface term,

$$\delta S = T \int d\tau d\sigma \left( h^{ab} \partial_a \partial_b X_\mu \right) \delta X^\mu - T \int d\tau X_\mu{}' \delta X^\mu \Big|_{\sigma=0}^{\sigma=\pi}. \tag{4.14}$$

Appropriate boundary constraints must be applied in order to eliminate the surface terms. For open strings, one of the two options is to constrain the end points of the string to be fixed, a so called Dirichlet boundary condition, $\dot{X}^\mu(\tau,0) = \dot{X}^\mu(\tau,\pi) = 0$. The other option is the Neumann condition, where the component of the momentum normal to the boundary of the world-sheet is required to vanish, $X^{\mu\prime}(\tau,0) = X^{\mu\prime}(\tau,\pi) = 0$. For closed strings, periodicity is required, i.e., $X^\mu(\tau,\sigma) = X^\mu(\tau,\sigma+\pi)$ and $X^{\mu\prime}(\tau,\sigma) = X^{\mu\prime}(\tau,\sigma+\pi)$. For the remainder of this thesis, focus is on closed, oriented strings with periodic boundary constraints.

The bulk equations of motion in (4.14) is the two-dimensional wave equation, which in light-cone coordinates $x^{\pm\pm} = \tau \pm \sigma$ takes the form $\partial_{++}\partial_{=}X^\mu = 0$. The wave equation is solved by a left- and right-going wave,

$$X^\mu(x^{++},x^{=}) = X_L^\mu(x^{++}) + X_R^\mu(x^{=}), \tag{4.15}$$

where the periodicity constraints of the closed string restricts the Taylor series expansion to take the form

$$X_L^\mu(x^{++}) = \frac{1}{2}X_0^\mu + \alpha' P_0^\mu x^{++} + i\sqrt{2\alpha'} \sum_{n \neq 0} \frac{\tilde{\alpha}_n^\mu}{n} e^{-2inx^{++}},$$

$$X_R^\mu(x^{=}) = \frac{1}{2}X_0^\mu + \alpha' P_0^\mu x^{=} + i\sqrt{2\alpha'} \sum_{n \neq 0} \frac{\alpha_n^\mu}{n} e^{-2inx^{=}}, \tag{4.16}$$

see, e.g., [BBS07]. The integration constants $X_0^\mu$ and $P_0^\mu$ represent the center of mass position and the momentum of the string, respectively, describing the free motion of the string center of mass. The $\tilde{\alpha}_n^\mu$ and $\alpha_n^\mu$ represent the oscillatory modes of the string.

The theory may now be quantized by defining raising and lowering operators in terms of the oscillatory modes. Although the details will be omitted here, some relevant implications of the quantization procedure should be stressed. First, the vacuum state of the closed bosonic string is a tachyon with negative mass squared. Secondly, the excited states must be massless and require the dimension of the target space to be $d = 26$. The excited states transform as a tensor product of two vectors under the group $SO(24)$. The representation is reducible and decomposes into three irreducible representations: a symmetric traceless tensor $g$, an anti-symmetric tensor $b$ and a scalar trace part $\Phi$,

$$g \oplus b \oplus \Phi. \tag{4.17}$$

These background fields correspond to the *graviton*, the *Kalb-Ramond form* and the *dilaton*. The dilaton is of one order higher in the perturbative $\alpha'$-expansion and can therefore be neglected to lowest order, but the $b$-field is of the same order as the metric and should be included for a general sigma model. In light-cone coordinates, the action (4.13) with background field $E = g + b$ takes the form

$$S = \int d^2x \, \partial_{++} X^\mu E_{\mu\nu} \partial_= X^\nu. \tag{4.18}$$

The field equations for a general bosonic sigma model with arbitrary metric and a $b$-field then generalize to

$$\partial_{++}\partial_= X^\mu + (\Gamma^{(0)\mu}_{\nu\rho} + T^\mu_{\nu\rho})\partial_{++} X^\nu \partial_= X^\rho = 0, \tag{4.19}$$

which in a more compact notation can be written as

$$\nabla^{(+)}_{++} \partial_= X^\mu = 0. \tag{4.20}$$

The metric gives rise to the ordinary Levi-Civita connection $\Gamma^{(0)}$ and the $b$-field gives rise to the torsion $T = \frac{1}{2}g^{-1}db$.

## 4.2 Imposing supersymmetry

When quantizing the bosonic sigma model, negative norm states appear, as was briefly discussed in the previous section; to remove this unwanted tachyon, fermions must be included into the theory.

### 4.2.1 N = 1 supersymmetric sigma model

By including the standard Dirac term for free massless fermions, $\bar{\psi}^\mu \rho^\alpha \partial_\alpha \psi_\mu$, spinorial fields can be added to the bosonic action, and the supersymmetric extension of the action (4.18) with flat metric and no b-field is

$$S = \int d^2x\, g_{\mu\nu} \left[ \partial_{++} X^\mu \partial_= X^\nu + \psi^\mu_+ i\partial_= \psi^\nu_+ + \psi^\mu_- i\partial_{++} \psi^\nu_- \right]. \tag{4.21}$$

The action is invariant up to a surface term (which is eliminated by periodicity constraints) under the supersymmetry transformations identified in (3.21), where the bosonic and fermionic field are transformed into each other,

$$\delta X = -\epsilon^+ \psi_+ - \epsilon^- \psi_-, \quad \delta \psi_\pm = -i\epsilon^\pm \partial_{\pm\pm} X. \tag{4.22}$$

The equations of motion for the sigma model (4.21) are

$$\partial_{++} \partial_= X = 0, \quad \partial_{++} \psi_- = \partial_= \psi_+ = 0. \tag{4.23}$$

The action in (4.21) contains no auxiliary field $F$, implying that the transformations on the fermionic fields close to a supersymmetry algebra (3.18) only on-shell,

$$[\delta(\epsilon_1), \delta(\epsilon_2)]X = \epsilon_{[2}\epsilon_{1]} i\partial X,$$
$$[\delta(\epsilon_1), \delta(\epsilon_2)]\psi = \epsilon_{[2}\epsilon_{1]} i\partial \psi + (\text{field eqns}). \tag{4.24}$$

The supersymmetry can be made manifest and close off-shell by going to superspace. This is achieved by replacing the fields in the bosonic sigma model (4.18) by superfields (3.13), the derivatives by covariant supersymmetry derivatives, and integrating over the full superspace,

$$S = \int d^2x\, d^2\theta\, D_+ \phi^\mu E_{\mu\nu} D_- \phi^\nu. \tag{4.25}$$

The sigma model has manifest $N = (1,1)$ supersymmetry and contains the action (4.21) when the superspace coordinates are integrated out. For a flat metric, the reduced action is

$$S = \int d^2x\, g_{\mu\nu} \left[ \partial_{++} X^\mu \partial_= X^\nu + \psi^\mu_+ i\partial_= \psi^\nu_+ + \psi^\mu_- i\partial_{++} \psi^\nu_- - F^\mu F^\nu \right], \tag{4.26}$$

which takes the form of the action (4.21) after the auxiliary fields are eliminated using their equations of motion, $F^\mu = 0$. For an arbitrary metric and b-field, terms involving derivatives of the metric and b-field will also be present. The supersymmetry transformations (3.21) of the component fields close to a supersymmetry algebra without the help of the field equations; the auxiliary fields are needed for off-shell closure of the supersymmetry algebra.

The fermions of the closed strings must be equipped with either periodic (Ramond) [Ram71] or anti-periodic (Neveu-Schwarz) [NS71] boundary conditions in the left- and/or right-moving sectors, which will give rise to four different closed-string sectors with differing background fields. The background fields in (4.17) correspond to the NS-NS sector of closed type IIA and IIB strings.

The field equations for the supersymmetric sigma model are of the same form as for the bosonic sigma model (4.20),

$$\nabla^{(+)}_+ D_- \phi^\mu = 0, \qquad (4.27)$$

and contain the geometric information that the target manifold is Riemannian with torsion.

### 4.2.2  N=2 supersymmetric sigma model

The non-linear sigma model in (4.25) has one manifest supersymmetry in each chirality. The field equations (4.27) reveal the target space geometry as Riemannian with non-trivial metric and torsion. Adding another supersymmetry to the model will constrain the possible geometries of the target space. This is where the interesting connection between supersymmetry and complex geometry begins.

Dimensional analysis shows that the unique ansatz for additional supersymmetry can be constructed of some arbitrary $(1,1)$-tensors $J^{(\pm)}$ and the covariant supersymmetry derivatives (3.14) to act on the superfields as

$$\delta \phi^\mu = \epsilon^+ J^{(+)\mu}_{\phantom{(+)}\nu} D_+ \phi^\nu + \epsilon^- J^{(-)\mu}_{\phantom{(-)}\nu} D_- \phi^\nu. \qquad (4.28)$$

The action (4.25) is invariant under the transformations if and only if the target space metric is hermitian with respect to the two structures $J^{(\pm)}$, and the structures are covariantly constant with respect to a connection with torsion,

$$J^{(\pm)t} g J^{(\pm)} = g, \quad \nabla^{(\pm)} J^{(\pm)} = 0. \qquad (4.29)$$

The transformations represent a supersymmetry if two subsequent transformations acting on a superfield close to the supersymmetry algebra (3.18). A sufficient requirement for this to happen is that the structures $J^{(\pm)}$ square to minus the identity, have vanishing Nijenhuis tensors (2.9), vanishing Magri-Morosi concomitant (2.12) and commute,

$$[\delta(\epsilon_1), \delta(\epsilon_2)]\phi^\mu = \epsilon^\pm_{[2} \epsilon^\pm_{1]} \left( -(J^{(\pm)})^{2\mu}_{\phantom{2}\nu} i \partial_{\pm\pm} \phi^\nu + \mathcal{N}(J^{(\pm)})^\mu_{\nu\rho} D_\pm \phi^\nu D_\pm \phi^\rho \right) \qquad (4.30)$$
$$+ \epsilon^+_{[2} \epsilon^-_{1]} \left( \mathcal{M}(J^{(+)}, J^{(-)})^\mu_{\nu\rho} D_+ \phi^\nu D_- \phi^\rho - [J^{(+)}, J^{(-)}]^\mu_\nu D_+ D_- \phi^\nu \right).$$

In other words, the supersymmetry algebra closes if $J^{(\pm)}$ are commuting, complex structures that are simultaneously integrable. Using the constraints (4.29) from invariance of the action, the simultaneous integrability reduces to a vanishing covariant term plus a connection, and the last terms in (4.30) take the form of a field equation,

$$[\delta(\epsilon_1), \delta(\epsilon_2)]\phi^\mu = \epsilon_{[2}^\pm \epsilon_{1]}^\pm \left(-(J^{(\pm)})^2 i\partial_\pm \phi + \mathcal{N}(J^{(\pm)})^\mu_{\nu\rho} D_\pm \phi^\nu D_\pm \phi^\rho\right)$$
$$- \epsilon_{[2}^+ \epsilon_{1]}^- [J^{(+)}, J^{(-)}]^\mu_\nu \nabla_+^{(+)} D_- \phi^\nu. \quad (4.31)$$

The transformations thus leave the action invariant and close to a supersymmetry algebra on-shell if and only if $J^{(\pm)}$ are two complex structures fulfilling (4.29). Off-shell closure further requires that the two complex structures commute.

Two covariantly constant complex structures that preserve the metric and the $b$-field define a bihermitian geometry, as discussed in chapter 2. Hence, the existing $N=(1,1)$ supersymmetry can be further extended to non-manifest $N=(2,2)$ supersymmetry if and only if the target space geometry is bihermitian [GHR84]. If the $b$-field is zero, the torsion vanishes and the connection reduces to the ordinary Levi-Civita connection, and the bihermitian geometry simplifies to Kähler geometry [Zum79].

Twisted $N=(2,2)$ supersymmetry will imply the same constraints as the supersymmetry, with the difference that the structures $J^{(\pm)}$ in (4.31) square to *plus* the identity. Accordingly, the target space will be equipped with a pair of product structures instead of complex structures. In the torsion-free case, the geometry reduces to pseudo-Kähler geometry [AZH99].

### 4.2.3 N=4 supersymmetric sigma model

The scheme of adding more supersymmetry to the sigma model and analyzing the arising geometrical constraints on the target manifold can be continued. An ansatz for $N=(4,4)$ supersymmetry can be constructed on the same form as the $N=(2,2)$ supersymmetry transformations,

$$\delta_{(i)}\phi^\mu = \epsilon_i^+ \left(J_i^{(+)}\right)^\mu_\nu D_+ \phi^\nu + \epsilon_i^- \left(J_i^{(-)}\right)^\mu_\nu D_- \phi^\nu, \quad (4.32)$$

for three independent structures in both chiralities $J_i^{(\pm)}$ with $i=1,2,3$. In the same way as before, the action is invariant under the transformations if the target space metric is hermitian with respect to all structures and all are covariantly constant with respect to a torsionful connection. The algebraic requirements from the supersymmetry algebra closure includes that all six $J_i^{(\pm)}$ must be complex structures, but also that they must satisfy the quaternionic algebra,

$$J_i^{(\pm)} J_j^{(\pm)} = -\delta_{ij} + \varepsilon_{ijk} J_k^{(\pm)}. \quad (4.33)$$

The conclusion is that $N=(4,4)$ supersymmetry requires the target space geometry of the sigma model to be bihyperhermitian, also known as strong hyperkähler with torsion (HKT) in both directions [GHR84]. In the case of zero torsion, this reduces to hyperkähler geometry investigated already in [Zum79, AGF80].

An ansatz for $N=(3,3)$ supersymmetry will imply the existence of two complex structures satisfying $J_i J_j + J_j J_i = -2\delta^{ij}$, and their product will automatically generate another supersymmetry [AGF81], so $N=3$ supersymmetry on the world-sheet implies $N=4$. Further, a supersymmetric sigma model on an irreducible manifold has at most four conserved spinor charges, so $N=4$ is the maximal extended supersymmetry of the two-dimensional sigma model [AGF81]. The only interesting options are the ones discussed here; $N=1$, $N=2$ and $N=4$, as well as combinations of these, such as sigma models with supersymmetry in only one direction $N=(2,0)$, $N=(4,0)$ or with different amount of supersymmetry in the directions, investigated, e.g., in [DS86, Hul98, AZH99, HLR$^+$12].

The geometrical constraints arising on the target space from extended world-sheet supersymmetry on non-linear sigma models with background fields consisting of a metric and possibly a *b*-field can be summarized as in the following chart.

|  | *(2,2) supersymmetry* | *(4,4) supersymmetry* |
|---|---|---|
| *no torsion* | Kähler | hyperkähler |
| *with torsion* | bihermitian | bihyperhermitian |

Analogously to twisted $N=(2,2)$ supersymmetry, twisted $N=(4,4)$ supersymmetry requires that the structures $J_i^{(\pm)}$ satisfy the algebra of split quaternions. In the case of vanishing torsion, the geometry reduces to pseudo-hyperkähler geometry, also denoted neutral hyperkähler.

|  | *(2,2) twisted supersymmetry* | *(4,4) twisted supersymmetry* |
|---|---|---|
| *no torsion* | pseudo-Kähler | pseudo-hyperkähler |

## 4.3 Constrained N = (2,2) superfields

As for the $N=(1,1)$ supersymmetry discussed in the previous section, the extended supersymmetry can be made manifest by going to $N=(2,2)$ superspace. The $N=(2,2)$ formalism enables the construction of new models with extended supersymmetry.

The extended superspace is parametrized by four fermionic coordinates together with the ordinary space-time coordinates, $(x^\mu, \theta^\alpha, \bar{\theta}^\alpha)$, where $\alpha = +, -$.

For later convenience, the real and the imaginary parts of the Grassmann coordinates are denoted $\theta^\alpha = \theta_1^\alpha + i\theta_2^\alpha$. A general superfield expanded in the Grassmann coordinates will contain 16 independent component fields,

$$\begin{aligned}\Phi(x,\theta,\bar{\theta}) = {} & X + \theta^+\psi_+ + \theta^-\psi_- + \bar{\theta}^+\chi_+ + \bar{\theta}^-\chi_- + \theta^+\bar{\theta}^+ A \\ & + \theta^+\bar{\theta}^- B + \theta^-\bar{\theta}^+ C + \theta^-\bar{\theta}^- D + \theta^2 M + \bar{\theta}^2 N \\ & + \bar{\theta}^2\theta^+\lambda_+ + \bar{\theta}^2\theta^-\lambda_- + \theta^2\bar{\theta}^+\varphi_+ + \theta^2\bar{\theta}^-\varphi_- + \theta^2\bar{\theta}^2 F,\end{aligned} \qquad (4.34)$$

where all the component fields are functions of the space-time coordinates of the world-sheet, $X = X(x)$, and $\theta^2 = \theta^+\theta^-$. The component fields can be projected out by covariant $N=(2,2)$ derivatives, defined analogously to (3.14) as

$$\mathbb{D}_\pm = \frac{\partial}{\partial\theta^\pm} + \frac{i}{2}\bar{\theta}^\pm\partial_{\pm\pm}, \quad \bar{\mathbb{D}}_\pm = \frac{\partial}{\partial\bar{\theta}^\pm} + \frac{i}{2}\theta^\pm\partial_{\pm\pm}. \qquad (4.35)$$

They satisfy the algebra $\mathbb{D}_\pm^2 = \bar{\mathbb{D}}_\pm^2 = 0$ and

$$\{\mathbb{D}_\pm, \bar{\mathbb{D}}_\pm\} = i\partial_{\pm\pm}, \qquad (4.36)$$

and their real and imaginary parts can be seen as independent real, commuting $N=(1,1)$ operators,

$$\begin{aligned}D_\pm &= \mathbb{D}_\pm + \bar{\mathbb{D}}_\pm = \frac{\partial}{\partial\theta_1^\pm} + i\theta_1^\pm\partial_{\pm\pm}, \\ Q_\pm &= i(\mathbb{D}_\pm - \bar{\mathbb{D}}_\pm) = \frac{\partial}{\partial\theta_2^\pm} + i\theta_2^\pm\partial_{\pm\pm}.\end{aligned} \qquad (4.37)$$

For an action in $N=(2,2)$ superspace that integrates over full superspace, from dimensional considerations, the Lagrangian can only be a scalar function of the superfields,

$$S = \int d^2x\, d^2\theta\, d^2\bar{\theta}\, K(\Phi, \bar{\Phi}). \qquad (4.38)$$

Like the superfields (3.24) discussed previously, the general $N=(2,2)$ superfields in (4.34) form reducible representations of the $N=(2,2)$ supersymmetry, and the redundant component fields can be eliminated by imposing covariant constraints. Since the action with unconstrained superfields (4.38) contains no derivatives, the dynamics of the sigma model with two manifest supersymmetries will arise from the differential constraints on the superfields, i.e., from the choice of representation.

The real operators in (4.37) can be used to impose differential constraints of the form

$$Q_\alpha \Phi = J^{(\alpha)} D_\alpha \Phi, \qquad (4.39)$$

where $J^{(\alpha)}$ is some tensor. Given the left-hand side, Lorentz invariance and dimensional analysis implies that (4.39) is the most general differential constraint [ST97]. One of the constraints reduces half of the degrees of freedom;

imposing constraints in both chiralities reduce the number of independent component fields to four, resulting in a superfield with the same degrees of freedom as a $N=(1,1)$ superfield. Chiral and twisted chiral superfields are constrained in both chiralities, whereas semichiral superfields are only constrained in one. In [ST97], it was conjectured that chiral, twisted chiral and semichiral superfields fully describe a sigma model with two manifest supersymmetries. This was later proven to be true in [LRvUZ07a] and will now be discussed in detail.

### 4.3.1 Chiral and twisted chiral superfields

The *chiral* $N=(2,2)$ superfields are defined by the covariant linear differential constraint

$$\bar{\mathbb{D}}_\pm \phi = 0, \qquad (4.40)$$

implying that half of the components in (4.34) are restricted to vanish and only four of the component fields are independent,

$$\phi(x,\theta,\bar{\theta}) = X + \theta^+\psi_+ + \theta^-\psi_- + \theta^+\bar{\theta}^+\tfrac{i}{2}\partial_{++}X + \theta^-\bar{\theta}^-\tfrac{i}{2}\partial_{=}X + \theta^2 M$$
$$- \theta^2\bar{\theta}^+\tfrac{i}{2}\partial_{++}\psi_- + \theta^2\bar{\theta}^-\tfrac{i}{2}\partial_{=}\psi_+ + \tfrac{1}{4}\theta^2\bar{\theta}^2 \partial_{++}\partial_{=}X. \qquad (4.41)$$

The component fields $(X(x), \psi_\pm(x), M(x))$ have helicity 0, 1/2 and 1, respectively. In terms of the real operators in (4.37), the chiral constraint reads

$$Q_\pm \phi = iD_\pm \phi. \qquad (4.42)$$

Splitting the Grassmann coordinates into their real and imaginary parts and collecting the component fields into $N=(1,1)$ superfields $\varphi = \varphi(x, \theta_1^\pm)$, it becomes clear that the chiral superfields depend on only *one* $N=(1,1)$ superfield,

$$\phi(x, \theta_1^\pm, \theta_2^\pm) = \varphi + \theta_2^+ iD_+\varphi + \theta_2^- iD_-\varphi + \theta_2^+\theta_2^- D_+D_-\varphi, \qquad (4.43)$$

a fact that could be seen already from the condition (4.42).

The *twisted chiral* superfields are defined analogously, in terms of two linear differential constraints,

$$\bar{\mathbb{D}}_+\chi = \mathbb{D}_-\chi = 0. \qquad (4.44)$$

As for the chiral superfields, this constraint reduces the number of independent component fields to four, or, equivalently, to one single $N=(1,1)$ superfield. The twisted chiral analogue of the chiral constraint in (4.42) reads

$$Q_\pm \chi = \pm iD_\pm \chi. \qquad (4.45)$$

The constraints on the chiral and twisted chiral fields (4.42) and (4.45) can be conveniently summarized in a unified notation $X = (\phi, \bar{\phi}, \chi, \bar{\chi})$ as

$$Q_\pm X = J^{(\pm)} D_\pm X, \quad J^{(\pm)} = \begin{pmatrix} J & 0 \\ 0 & \pm J \end{pmatrix} \quad (4.46)$$

where $J$ is the canonical complex structure (2.11).

### 4.3.2 Semichiral superfields

*Semichiral* superfields, on the other hand, are subject to only one differential constraint [BLR88],

$$\bar{\mathbb{D}}_+ \mathbb{X}^\ell = 0, \quad \bar{\mathbb{D}}_- \mathbb{X}^r = 0. \quad (4.47)$$

The resulting superfields are denoted left and right semichiral superfields, respectively. In terms of the real operators in (4.37), the constraints read

$$Q_+ \mathbb{X}^\ell = iD_+ \mathbb{X}^\ell, \quad Q_- \mathbb{X}^r = iD_- \mathbb{X}^r. \quad (4.48)$$

As compared to the analogue constraints for the chiral and twisted chiral fields in (4.42) and (4.45), it is clear that the semichiral fields depend on more independent component fields, since they are less constrained. Whereas the chiral and twisted chiral superfields depended on only one $N=(1,1)$ superfield each, the semichiral fields depend of two; one bosonic superfield $X(x, \theta_1^\pm)$ and one fermionic $\psi = \psi(x, \theta_1^\pm)$. Defining the fermionic superfields as

$$\psi_-^\ell = Q_- \mathbb{X}^\ell\big|, \quad \psi_+^r = Q_+ \mathbb{X}^r\big|, \quad (4.49)$$

where the notation | denotes projecting out the $\theta_2$-independent part, the expansion of the semichiral $N=(2,2)$ superfields in terms of $N=(1,1)$ superfields reads

$$\begin{aligned} \mathbb{X}^\ell &= X^\ell + \theta_2^+ iD_+ X^\ell + \theta_2^- \psi_-^\ell - \theta_2^+ \theta_2^- iD_+ \psi_-^\ell, \\ \mathbb{X}^r &= X^r + \theta_2^- iD_- X^r + \theta_2^+ \psi_+^r + \theta_2^+ \theta_2^- iD_- \psi_+^r. \end{aligned} \quad (4.50)$$

The fermionic superfields are auxiliary but their existence is necessary to make the supersymmetry algebra generated by $Q_\pm$ close off-shell, in analogue to the auxiliary component field in the $N=(1,1)$ superfield in (3.13).

Writing the semichiral fields in a collective notation as $\mathbb{X}^i = (\mathbb{X}^\ell, \bar{\mathbb{X}}^{\bar{\ell}}, \mathbb{X}^r, \bar{\mathbb{X}}^{\bar{r}})$, the constraints on the fields in the $N=(1,1)$ formalism are again of the form (4.46),

$$Q_\pm \mathbb{X} = J^{(\pm)} D_\pm \mathbb{X}, \quad (4.51)$$

but the chirality constraints on the semichiral fields are not sufficient to determine the full structure of the matrices $J^{(\pm)}$,

$$J^{(+)} = \begin{pmatrix} J & 0 \\ ? & ? \end{pmatrix}, \quad J^{(-)} = \begin{pmatrix} ? & ? \\ 0 & J \end{pmatrix}. \quad (4.52)$$

## 4.4 N = (2,2) sigma models and bihermitian geometry

The three kinds of superfields defined in the previous section parametrize a general two-dimensional sigma model with two manifest supersymmetries, a model whose target space geometry is bihermitian. As will be described in this section, the chiral and the twisted chiral superfields span the sector of the target space where the two complex structures of the bihermitian geometry commute, and the semichiral superfields span the complement.

### 4.4.1 Chiral and twisted chiral section

Consider a sigma model where the Lagrangian is a function of chiral $N=(2,2)$ fields,

$$S = \int d^2x d^2\theta d^2\bar{\theta} K(\phi, \bar{\phi}). \tag{4.53}$$

By reducing one of the supersymmetries, the action can be compared to the $N=(1,1)$ action (4.25) with additional non-manifest supersymmetry of the form (4.28). Using the properties of the Berezin integral, the $N=(2,2)$ superspace measure is

$$\mathbb{D}_+\mathbb{D}_-\bar{\mathbb{D}}_+\bar{\mathbb{D}}_-\big| = -\tfrac{1}{4}D_+D_-Q_+Q_-\big|, \tag{4.54}$$

where the vertical bar denotes reducing to $N=(1,1)$ superspace by setting $\theta_2^\pm = 0$. Using partial integration, the action (4.53) reduces to

$$S \propto \int d^2x d^2\theta_1 \left(g_{\mu\bar{\nu}} D_+\varphi^\mu D_-\bar{\varphi}^{\bar{\nu}} + g_{\bar{\mu}\nu} D_+\bar{\varphi}^{\bar{\mu}} D_-\varphi^\nu\right), \quad g_{\mu\bar{\nu}} = \frac{\partial^2 K}{\partial\varphi^\mu \partial\bar{\varphi}^{\bar{\nu}}}. \tag{4.55}$$

By collecting the holomorphic and anti-holomorphic indices as $\varphi^i = (\varphi^\mu, \bar{\varphi}^{\bar{\mu}})$, the action takes the well-known expression

$$S = \int d^2x d^2\theta_1 D_+\varphi^i g_{ij} D_-\varphi^j \tag{4.56}$$

and the $N=(2,2)$ supersymmetry transformations are given by

$$\delta\varphi^i = \epsilon^\alpha Q_\alpha \phi^i\big| = \epsilon^\alpha J^i_j D_\alpha \varphi^j, \tag{4.57}$$

where $\alpha = +, -$ and $J$ is the canonical complex structure (2.11). The hermitian metric is written in terms of second derivatives of the potential and there is no $b$-field. Hence, the target manifold of a manifest $N=(2,2)$ sigma model written in terms of chiral superfields is Kähler. The Kähler metric is invariant under *Kähler transformations*,

$$K(\phi, \bar{\phi}) \to K(\phi, \bar{\phi}) + f(\phi) + \bar{f}(\bar{\phi}). \tag{4.58}$$

This model can be generalized to include torsion and non-Kähler manifolds by including twisted chiral superfields. The sigma model

$$S = \int d^2x d^2\theta d^2\bar{\theta} K(\phi,\bar{\phi},\chi,\bar{\chi}) \tag{4.59}$$

is equivalent to the $N=(1,1)$ sigma model with torsion (4.25) when reduced to $N=(1,1)$ formalism [GHR84]. The mixed terms with chiral and twisted chiral fields give rise to the $b$-field; if only chiral *or* twisted chiral fields are present, this again reduces to ordinary Kähler geometry.

Now recall from section 4.2, that the non-manifest $N=(2,2)$ supersymmetry algebra closes only on-shell or if the two complex structures $J^{(\pm)}$ in the supersymmetry transformations commute. As just discussed, a manifest $N=(2,2)$ sigma model parametrized by chiral and twisted chiral fields reduces to a general $N=(1,1)$ sigma model with additional supersymmetry of the form

$$\delta X = \epsilon^{\pm} J^{(\pm)} D_{\pm} X, \tag{4.60}$$

with $J^{(\pm)}$ given in (4.46). Since $J^{(+)}$ and $J^{(-)}$ commute and are covariantly constant complex structures, the non-manifest $N=(2,2)$ supersymmetry closes off-shell. The vanishing commutator implies that an integrable almost product structure can be defined as $\Pi = J^{(+)} J^{(-)}$ [GHR84]. A bihermitian geometry with this property is referred to as an almost product space, or a bihermitian local product (BiLP) space. Since the two complex structures commute, coordinates $X = (\phi,\bar{\phi},\chi,\bar{\chi})$ can be found in which they are simultaneously diagonalizable and take the form (4.46). In a BiLP-space, the geometric structures can be expressed linearly in terms of second derivatives of the generalized Kähler potential [LRvUZ07b].

Any $N=(1,1)$ model with additional supersymmetry and commuting structures $J^{(\pm)}$ can be written in a manifest way as (4.59) using chiral and twisted chiral superfields [GHR84]. Locally, the sector of the tangent space where the two structures commute can be decomposed as

$$\ker[J^{(+)}, J^{(-)}] = \ker(J^{(+)} + J^{(-)}) \oplus \ker(J^{(+)} - J^{(-)}). \tag{4.61}$$

The subspace $\ker(J^{(+)} - J^{(-)})$ is always described by chiral fields and the subspace $\ker(J^{(+)} + J^{(-)})$ by twisted chiral fields [IKR95].

### 4.4.2 Semichiral section

Consider a sigma model parametrized by semichiral superfields, where the left semichiral superfields carry indices $\mathbb{X}^a = \{\mathbb{X}^{\ell_1}, \mathbb{X}^{\ell_2}, \ldots, \mathbb{X}^{\ell_n}\}$ and the right semichiral superfields $\mathbb{X}^{a'} = \{\mathbb{X}^{r_1}, \mathbb{X}^{r_2}, \ldots, \mathbb{X}^{r_n}\}$. A short-hand notation for left semichiral and left anti-semichiral fields $\mathbb{X}^L = \{\mathbb{X}^a, \bar{\mathbb{X}}^{\bar{a}}\}$ will be used, as well as the right semichiral analogue $\mathbb{X}^R = \{\mathbb{X}^{a'}, \bar{\mathbb{X}}^{\bar{a}'}\}$. The full set of left and right semichiral fields is labeled by

$$\mathbb{X}^i = \left(\mathbb{X}^a, \bar{\mathbb{X}}^{\bar{a}}, \mathbb{X}^{a'}, \bar{\mathbb{X}}^{\bar{a}'}\right), \quad i = 1, \ldots, 4n. \tag{4.62}$$

For an action of the semichiral fields to reproduce the sigma model (4.25) when reduced to $N = (1,1)$ superspace, the Lagrangian must be a function of both the left and the right semichiral fields, together with their complex conjugates, hence the target space of the semichiral sigma model is always $4d$-dimensional [BLR88]. The action is

$$S = \int d^2x d^2\theta d^2\bar{\theta} K(\mathbb{X}_L^a, \bar{\mathbb{X}}_L^{\bar{a}}, \mathbb{X}_R^{a'}, \bar{\mathbb{X}}_R^{\bar{a}'}). \tag{4.63}$$

Denote derivation with respect to the semichiral fields as $K_{ab} = \partial_{\mathbb{X}^a} \partial_{\mathbb{X}^b} K$ and define a matrix notation as

$$K_{LR} = \begin{pmatrix} K_{ab'} & K_{a\bar{b}'} \\ K_{\bar{a}b'} & K_{\bar{a}\bar{b}'} \end{pmatrix}, \quad K_{LL} = \begin{pmatrix} K_{ab} & K_{a\bar{b}} \\ K_{\bar{a}b} & K_{\bar{a}\bar{b}} \end{pmatrix}. \tag{4.64}$$

When the semichiral action is integrated over one of the superspace coordinates, the resulting $N = (1,1)$ action contains both the bosonic and the auxiliary $N = (1,1)$ superfield in (4.50). The fermionic superfields $\psi_-^\ell$ and $\psi_+^{r}$ are auxiliary, however, and can be eliminated by their equations of motion. From the definition in (4.49), the equations of motion for the auxiliary superfields determine the lower rows of $J^{(+)}$ and the upper rows of $J^{(-)}$ in (4.52). The remaining entries were already determined by the semichiral constraints. Assuming that the matrices in (4.64) are invertible, $J^{(\pm)}$ take the form [BLR88, ST97]

$$J^{(+)} = \begin{pmatrix} J & 0 \\ (K_{LR})^{-1} C_{LL} & (K_{LR})^{-1} J K_{LR} \end{pmatrix},$$

$$J^{(-)} = \begin{pmatrix} (K_{RL})^{-1} J K_{RL} & (K_{RL})^{-1} C_{RR} \\ 0 & J \end{pmatrix}, \tag{4.65}$$

where, using the notation from [LRvUZ07a], the definition $C_{LL} = [J, K_{LL}]$ is used, and similarly for $C_{RR}$.

Summarized, the semichiral sigma model (4.63) reduced to $N = (1,1)$ superspace yields an action (4.25) with one manifest supersymmetry of the form (3.17), and a second supersymmetry of the form (4.28). Note that the supersymmetry operator $Q$ in (3.17) represents the first supersymmetry generator, whereas the operator $Q$ in (4.37) generates the second supersymmetry.

$$\delta X = \epsilon^\alpha Q_\alpha \mathbb{X}^i \big| = \epsilon^\alpha J^{(\alpha)i}{}_j D_\alpha X^j \tag{4.66}$$

where the matrices $J^{(\pm)}$ take the expressions in (4.65).

The expressions for $J^{(\pm)}$ reveal important information on the target space geometry. First of all, $J^{(+)}$ and $J^{(-)}$ do not commute. For a non-degenerate metric $g$ and in the semichiral parametrization, a Poisson structure $\sigma$ can be defined as in (2.60), $\sigma = [J^{(+)}, J^{(-)}]g^{-1}$. Since the commutator is non-zero, $\sigma$ is invertible and $\Omega = \sigma^{-1}$ is a symplectic form with respect to both complex structures, satisfying $J^{(\pm)t}\Omega J^{(\pm)} = -\Omega$. Identifying the reduced action with the standard $N=(1,1)$ sigma model (4.25), the metric and the $b$-field can be read off. In the semichiral parametrization, they take the simple expressions [BSvdLVG99]

$$g = \Omega[J^{(+)}, J^{(-)}], \quad b = \Omega\{J^{(+)}, J^{(-)}\} \tag{4.67}$$

where the symplectic structure $\Omega$ is

$$\Omega = \frac{1}{2}\begin{pmatrix} 0 & K_{LR} \\ -K_{RL} & \end{pmatrix}. \tag{4.68}$$

The torsion can be defined only locally as $H = db$, but the $b$-field is globally defined as in (4.67), away from irregular points in the manifold [Gua03]. Non-degeneracy of the metric obviously requires that $[J^{(+)}, J^{(-)}] \neq 0$ everywhere and that the matrix $K_{LR}$ is invertible. As in Kähler geometry, the metric is a function of second derivatives of the potential $K$, but due to the non-linearity of the expressions for $J^{(\pm)}$, the metric is a non-linear function of $\partial\bar\partial K$.

The explicit form of the matrices $J^{(+)}$ and $J^{(-)}$ can also be found from another point of view [LRvUZ07a]. As already seen, in the section parametrized by the chiral and twisted chiral fields, the two complex structures are simultaneously diagonalizable, but this is not the case in the section parametrized by the semichiral fields. But it is possible to choose coordinates in which one of the matrices is diagonal. Denote by $(q,p) = (\mathbb{X}^L, Y_L)$ the coordinates in which $J^{(+)}$ is diagonal,

$$J^{(+)}_{\text{diag}} = \begin{pmatrix} J & 0 \\ 0 & J \end{pmatrix}, \tag{4.69}$$

where $J$ is the canonical complex structure (2.11). The symplectic structure can be decomposed into a holomorphic and anti-holomorphic symplectic structure, and the coordinates $(\mathbb{X}^L, Y_L)$ can be chosen as the Darboux coordinates,

$$\Omega = d\mathbb{X}^\ell \wedge dY_\ell + d\bar{\mathbb{X}}^{\bar\ell} \wedge d\bar{Y}_{\bar\ell}. \tag{4.70}$$

Similarly denote by $(P,Q) = (\mathbb{X}^R, Y_R)$ the Darboux coordinates for $\Omega$ in which $J^{(-)}$ take the same diagonal form. The coordinates $(q,p)$ and $(Q,P)$ are related by a canonical transformation, specified by a generating function $K(q,P)$ [LRvUZ10] satisfying

$$p = \frac{\partial K}{\partial q}, \quad Q = \frac{\partial K}{\partial P}. \tag{4.71}$$

In mixed coordinates $(q,P) = (\mathbb{X}^L, \mathbb{X}^R)$, the components of $J^{(+)}$ can be computed using the Jacobian for the coordinate transformation,

$$(\text{Jac})^{-1} = \frac{\partial(q,p)}{\partial(q,P)} = \begin{pmatrix} 1 & 0 \\ K_{LL} & K_{LR} \end{pmatrix}, \tag{4.72}$$

where the matrices $K_{LL}$ and $K_{LR}$ are defined as in (4.64). In coordinates $(\mathbb{X}^L, \mathbb{X}^R)$, the matrix $J^{(+)}$ now takes the form

$$J^{(+)} = (\text{Jac})\, J^{(+)}_{\text{diag}}\, (\text{Jac})^{-1}, \tag{4.73}$$

which reproduces the matrix in (4.65). Starting from coordinates $(\mathbb{X}^R, Y_R)$ in which $J^{(-)}$ is diagonal, the matrix in mixed coordinates is found after a similar transformation. In the mixed coordinates $(\mathbb{X}^L, \mathbb{X}^R)$, the symplectic structure takes the form given in (4.68). From this discussion, it is clear that the generalized Kähler potential $K(\mathbb{X}^L, \mathbb{X}^R)$ is simply the generating function between the Darboux coordinates $(\mathbb{X}^L, Y_L)$, holomorphic with respect to the complex structure $J^{(+)}$, and $(\mathbb{X}^R, Y_R)$, holomorphic with respect to the other complex structure $J^{(-)}$.

The non-commutative property of the complex structures in the target space spanned by semichiral coordinates allows for the construction of additional structures. Since the kernel of the commutator $[J^{(+)}, J^{(-)}]$ is empty, the two complex structures are not proportional, $J^{(-)} \neq \pm J^{(+)}$. Recall the discussion on hyperkähler manifolds in chapter 2 and in particular the definition in (2.43). The manifold is hyperkähler if the anti-commutator is proportional to the identity,

$$\{J^{(+)}, J^{(-)}\} = 2c\mathbb{1}, \tag{4.74}$$

with $|c| < 1$ constant. The manifold is then equipped with a two-sphere of complex structures (2.38). Choosing the first structure as $I = J^{(+)}$, the remaining two structures can be defined as [LRvUZ07a]

$$J = \frac{1}{\sqrt{1-c^2}}(J^{(-)} + cJ^{(+)}), \quad K = \frac{1}{2\sqrt{1-c^2}}[J^{(+)}, J^{(-)}], \tag{4.75}$$

such that $(I, J, K)$ is a hypercomplex structure. It is easy to see that the torsion vanishes since the $b$-field in (4.67) is constant, and that the covariant constancy of $J^{(\pm)}$ implies that all structures are covariantly constant with respect to the Levi-Civita connection, hence the target space geometry is hyperkähler.

Simultaneously, if $c$ is a number with absolute value greater than one, two product structures can be defined as

$$S = \frac{1}{\sqrt{c^2-1}}(J^{(-)} + cJ^{(+)}), \quad T = \frac{1}{2\sqrt{c^2-1}}[J^{(+)}, J^{(-)}], \tag{4.76}$$

and the structures $(I, S, T)$ satisfy the algebra of split quaternions (2.47), endowing the target manifold with a pseudo-hyperkähler structure.

In terms of the generalized Kähler potential, the anti-commutator in (4.74) is equivalent to the system of partial differential equations

$$\{(K_{LR})^{-1}C_{LL}(K_{RL})^{-1}, J\} = 0,$$
$$J(K_{RL})^{-1}JK_{RL} + (K_{RL})^{-1}JK_{RL}J + (K_{RL})^{-1}C_{RR}(K_{LR})^{-1}C_{LL} = c, \quad (4.77)$$

together with the corresponding equations where all $L$ and $R$-indices are interchanged. In a four-dimensional manifold, the first of these equations is identically satisfied, and the second reduces to the partial differential equation

$$(1+c)|K_{\ell r}|^2 + (1-c)|K_{\ell\bar{r}}|^2 = 2K_{\ell\bar{\ell}}K_{r\bar{r}}. \quad (4.78)$$

In the limit where $c = 0$, this is simply the equation $\det(K_{LR}) = 1$. By performing a suitable coordinate transformation, this equation can be shown to be equivalent to the Monge-Ampère equation $\det(g) = 1$ in (2.45), where $g$ is the metric [BSvdLVG99].

### 4.4.3 General sigma model with manifest N=2 supersymmetry

The full tangent space of the target manifold is a direct sum of the kernel of the two complex structures and the complement,

$$TM = \ker[J^{(+)}, J^{(-)}] \oplus (\ker[J^{(+)}, J^{(-)}])^\perp. \quad (4.79)$$

The kernel decomposes as (4.61), and is parametrized by chiral and twisted chiral fields [IKR95]. The dimension of the complement is always a multiple of four, and can always be spanned by semichiral superfields [LRvUZ07a].

Conjectured in [ST97] and later proven in [LRvUZ07a], the most general manifest $N=(2,2)$ sigma model can thus be written in terms of chiral, twisted chiral and semichiral superfields,

$$S = \int d^2x d^2\theta d^2\bar{\theta} \, K(\phi, \bar{\phi}, \chi, \bar{\chi}, \mathbb{X}^\ell, \bar{\mathbb{X}}^{\bar{\ell}}, \mathbb{X}^r, \bar{\mathbb{X}}^{\bar{r}}). \quad (4.80)$$

The generalized Kähler potential $K$ is defined modulo generalized Kähler gauge transformations, a generalization of the Kähler transformations in (4.58),

$$K \sim K + f(\phi, \chi, \mathbb{X}^\ell) + \bar{f}(\bar{\phi}, \bar{\chi}, \bar{\mathbb{X}}^{\bar{\ell}}) + g(\phi, \bar{\chi}, \mathbb{X}^r) + \bar{g}(\bar{\phi}, \chi, \bar{\mathbb{X}}^{\bar{r}}). \quad (4.81)$$

Reducing the general $N=(2,2)$ sigma model to $N=(1,1)$ superspace with additional non-manifest supersymmetry of the form (4.28) again gives the structure of the matrices $J^{(\pm)}$ as well as the metric and $b$-field. With the kernel of $[J^{(+)}, J^{(-)}]$ non-empty, the metric and the $b$-field are obviously not given by the simple expressions in (4.67), since commuting structures would give a degenerate metric. The full structure of the metric and *b*-field including the chiral and twisted chiral parametrizations was calculated in [LRvUZ07a].

As reviewed in section 2.2, bihermitian geometry is in one-to-one correspondence with generalized Kähler geometry. Since the target space of an $N=(2,2)$ supersymmetric sigma model with torsion must be bihermitian, and (4.80) is the most general manifest $N=(2,2)$ sigma model, the single function $K(\phi,\bar{\phi},\chi,\bar{\chi},\mathbb{X}^\ell,\bar{\mathbb{X}}^{\bar{\ell}},\mathbb{X}^r,\bar{\mathbb{X}}^{\bar{r}})$ describes generalized Kähler geometry. Conversely, generalized Kähler geometry can be described locally, away from irregular points, by the sigma model (4.80). The generalized Kähler potential encodes the full geometry of the model, including the metric, the *b*-field and the complex structures [ST97]. The generalized complex structures $\mathcal{J}_{1,2}$ are given as in (2.58) and the corresponding generalized Kähler metric is $\mathcal{G} = -\mathcal{J}_1\mathcal{J}_2$.

# 5. T-duality of sigma models

> Reserve your right to think, for even to think wrongly is better than not to think at all.
> *Hypatia of Alexandria, philosopher, mathematician, astronomer (370-415)*

Dualities arise everywhere in physics. Usually, their existence reveal crucial insights about the underlying properties of the physical system. In electromagnetism, for example, the fact that the source-free Maxwell's equations are symmetric under the interchange of the electric and the magnetic field relies on the symmetry between the field strength and its dual form, that satisfies the Bianchi identity. This duality between field equations and Bianchi identities will be introduced in the setting of bosonic sigma models in section 5.2 and will be one of the key ingredients for discussing the dualities between sigma models with extended supersymmetry in chapter 8.

In string theory, dualities have continuously played a profound role. The discovery that type IIA and type IIB string theory are actually equivalent descriptions of the same theory related by T-duality preceded the so called *second string revolution*, when the seemingly unrelated string theories were unified into one single theory. Another duality that has had a major impact in string theory is the AdS/CFT correspondence [Mal98]. This duality between physics at strong and weak coupling has influenced our understanding of how gauge theories and gravity relate to each other in general. As in the AdS/CFT correspondence, a duality typically exchanges coupling regimes. In the simplest setting of T-duality, the physical system of a closed string compactified on a circle of radius $R$ is equivalent to one compactified on a circle of inverse radius $1/R$, as will be discussed briefly soon.

Recently, T-duality has received renewed interest due to new developments in flux compactifications and generalized geometry. It is known that mirror symmetry for Calabi-Yau manifolds can be interpreted as T-duality on toroidal fibers [SYZ96], and non-geometric backgrounds arising in flux compactifications can be understood and generated by T-duality. T-duality is also related to several other constructions in mathematics, such as Takai duality and Fourier-Mukai transform, see, e.g., [Bou10].

In the context of supersymmetric sigma models, T-duality is a key tool for understanding geometries and generating new ones. A related method is the quotient reduction which will be introduced in section 5.4.

Locally, T-duality mixes the metric and the *b*-field and relates different sigma model backgrounds. Globally, it has been shown that T-duality also relates different topologies [AAGBL94, GLMW03, KSTT03, BEM04].

The $N = (2,2)$ supersymmetric sigma models developed in the previous chapter possess a rich variety of dualities that relate not only their target space geometries, but also the corresponding different supersymmetry representations. This is the main aspect of T-duality in this thesis and will be discussed in detail in this chapter and further developed in the chapters 7 and 8.

## 5.1 T-duality and double field theory

T-duality arises naturally in string theory due to the extended nature of one-dimensional strings, as opposed to zero-dimensional point particles. The duality was first described for closed strings compactified on a torus $X \sim X + 2\pi R$, where $R$ is the radius of the circle [GSB82, KY84, SS86]. Denoting by $\omega$ the number of times the string winds the circle, the boundary condition for the closed string will be given by

$$X(\tau, \sigma + 2\pi) = X(\tau, \sigma) + 2\pi R \omega. \tag{5.1}$$

The classical solutions to the wave equation subject to closed string boundary constraints are given in (4.15)-(4.16) and split into a left- and a right-going wave. The momentum in the two directions quantizes for a compactified string and is given by

$$p_{L,R} = \alpha' \frac{K}{R} \pm \omega R, \tag{5.2}$$

where $K \in \mathbb{Z}$ is the Kaluza-Klein excitation number. The mass spectrum of the string,

$$M^2 = \left(\frac{K}{R}\right)^2 + \left(\frac{\omega R}{\alpha'}\right)^2 + \dots \tag{5.3}$$

is invariant under the interchange of the Kaluza-Klein excitation number with the winding number simultaneously with the interchange $R \leftrightarrow \alpha'/R$ [KY84, SS86]. The physical interpretation is that a closed string compactified on a circle of radius $R$ is equivalent to one compactified on a circle of radius $\alpha'/R$. From (5.2), one can see that the momentum of the right-going mode switches sign under the duality, whereas the left-mode remains invariant,

$$X = X_L + X_R \quad \underset{\text{dual}}{\longleftrightarrow} \quad \tilde{X} = X_L - X_R. \tag{5.4}$$

The duality can be generalized to non-flat arbitrary backgrounds [Bus87] and arbitrary toroidal compactifications [Nar86, NSW87, RV92], and is the starting point for the formulation of closed string theory as a double field theory, described by the two dual coordinates $(X, \tilde{X})$ [Tse90, Sie93, Hul05, HZ09]. An action can be defined in the coordinates $X^M = (X^i, \tilde{X}_i)$ that is invariant under $O(d, d, \mathbb{Z})$ transformations. On a circle of radius $R$, the transformation reduces to $O(1, 1, \mathbb{Z}) = \mathbb{Z}_2$, or simply $R \leftrightarrow 1/R$. Reformulating string theory in terms of double geometry is thus a general approach to finding a formalism for string theory in which the T-duality symmetry is manifest.

## 5.2 Gauging isometries and T-duality

### 5.2.1 T-duality of bosonic sigma models

As a first illustrative example of T-duality of a sigma model, consider a bosonic sigma model with constant metric and vanishing $b$-field,

$$S = \int d^2x\, \partial_a X \partial^a X, \tag{5.5}$$

where $x^a$ are the coordinates on the two-dimensional world-sheet. The action is invariant under a constant shift, $X \to X + s$. If the isometry is taken to be local, $s \to s(x)$, to keep the action invariant, a gauge potential $V = V_a dx^a$ must be introduced transforming as $V_a \to V_a - \partial_a s$, and the derivative must be replaced by the covariant derivative $\nabla_a X = \partial_a X + V_a$. The gauge invariant action is

$$S = \int d^2x\, \nabla_a X \nabla^a X. \tag{5.6}$$

As in ordinary Yang-Mills theory, the gauge invariant field strengths can be interpreted as a curvature,

$$F_{ab} = [\nabla_a, \nabla_b] = \partial_{[a} V_{b]}. \tag{5.7}$$

For simplicity, the considered isometry is abelian; non-abelian isometries will be discussed later in this chapter. Introducing a Lagrange multiplier $\tilde{X}$ gives the first order action, which after gauge fix $\partial_a X = 0$ takes the form

$$S_{1\text{st}} = \int d^2x \left[ V^a V_a + \varepsilon^{ab} \tilde{X} F_{ab} \right] = \int d^2x \left[ V^a V_a - 2\varepsilon^{ab} \partial_b \tilde{X} V_a \right], \tag{5.8}$$

where $\varepsilon^{ab}$ is the totally anti-symmetric tensor. Extremizing the action with respect to the Lagrange multipliers implies that the curvature vanishes, $F = 0$. The vanishing gauge field strength implies, for simply connected world-sheets, that $V$ is pure gauge and no extra degrees of freedom have been introduced.

$$
\begin{array}{|lcl|}
\hline
S = \int (\partial \phi)^2 & \longleftrightarrow & \tilde{S} = \int (\partial \tilde{X})^2 \\
\text{field equations} & \partial_a V^a = 0 & \text{Bianchi identities} \\
\text{Bianchi identities} & \partial_{[a} V_{b]} = 0 & \text{field equations.} \\
\hline
\end{array}
$$

*Figure 5.1:* The field equations and the Bianchi identities are dual.

This is solved by $V_a = \partial_a X$, which inserted into the first order action recovers the original action (5.5). The field equations for the original sigma model are simply $\partial_a \partial^a X = \partial_a V^a = 0$. Defining the Hodge star operation in terms of $\varepsilon^{ab}$, the Bianchi identities are

$$ 0 = \partial_a (*V^a) = \partial_a \varepsilon^{ab} V_b = \partial_{[a} V_{b]}. \tag{5.9} $$

It is clear from the form of the potential that the Bianchi identities are identically satisfied, since partial differentials commute.

On the other hand, integrating with respect to the gauge potential $V$ gives $V_a = \varepsilon_{ab} \partial^b \tilde{X}$. Substituting this expression for $V^a$ into the first-order action gives the dual action,

$$ \tilde{S} = \int d^2 x \, \partial_a \tilde{X} \partial^a \tilde{X}. \tag{5.10} $$

The dual fields are related through the gauge potential as $\partial_a X = \varepsilon_{ab} \partial^b \tilde{X}$. This is in analogue to the duality condition (5.4) for the closed string winded around a circle. The field equations for the dual model are given by varying (5.10) with respect to $\tilde{X}$ and are $\partial_{[a} V_{b]} = 0$. The Bianchi identities then take the form

$$ 0 = \partial_a (*V^a) = \partial_a V^a. \tag{5.11} $$

Again, from the expression of the potential, the Bianchi identities are automatically satisfied since partial derivatives commute.

Hence, the field equations for the original model take the same form as the Bianchi identities for the dual model, (5.11), and the Bianchi identities for the original model (5.9) are replaced by the the field equations in the dual model. To summarize, the dual models are related as in chart 5.1. These relations will be generalized and studied for a sigma model written in terms of manifestly $N = (2,2)$ superfields in chapter 8. Several important subtleties arise when the target space contains more geometrical data, such as torsion and complex structures, as will now be discussed.

### 5.2.2 T-duality of supersymmetric sigma models

As seen above, if the target space of a sigma model possesses an isometry, T-duality can be used to construct a map to another sigma model describing the same physical system. In general, the dual model may have drastically different geometry and topology. This will now be studied for arbitrary isometries and backgrounds.

Consider a manifold $M$ with an isometry group $G$. The elements in the corresponding Lie algebra can be written in a basis of Killing vectors $k$ that specify the isometry direction in the tangent space. The isometry group acts infinitesimally on the coordinates of the manifold as

$$\delta \phi^\mu = [\lambda k, \phi]^\mu = \lambda^A k_A^\mu, \tag{5.12}$$

where $\lambda^A$ are constant parameters and the basis of the Killing vectors $k_A = k_A^\mu \partial_\mu$ generate a Lie algebra $[k_A, k_B] = f_{AB}{}^C k_C$. Exponentiation yields the finite transformation $\phi^\mu \to e^{\mathcal{L}_{\lambda \cdot k}} \phi^\mu$; if the Lie group is compact, all elements connected to the identity can be written in this way.

Per definition, an isometry of a manifold is an endomorphism that leaves the metric invariant,

$$g_{f(p)}(f_* X, f_* Y) = g_p(X, Y) \tag{5.13}$$

for $p \in M$ and $X, Y \in T_p M$. In terms of the Lie derivative acting along the Killing vector field, this can be rewritten as the defining property of a Killing vector,

$$\mathcal{L}_k g = 0, \tag{5.14}$$

which in turn is equivalent to the Killing equation, $\nabla_{(\mu} k_{\nu)} = 0$.

Consider a supersymmetric sigma model defined as in (4.25),

$$S = \int d^2 x d^2 \theta D_+ \phi^\mu E_{\mu\nu} D_- \phi^\nu, \tag{5.15}$$

with a target space geometry $E_{\mu\nu} = g_{\mu\nu} + b_{\mu\nu}$, subject to an isometry generated by a Killing vector $k$. Locally, coordinates can be chosen such that the geometric structures are independent of the direction $\phi^0$ and the Killing vector can be written as

$$k = \frac{\partial}{\partial \phi^0}, \quad \phi^\mu = (\phi^0, \phi^i). \tag{5.16}$$

If the rigid isometry transformation in (5.12) is taken to be local, the gauged action is obtained by introducing a gauge potential $V_\pm$ and replacing the derivative by a covariant derivative, $\nabla_\pm \phi^0 = D_\pm \phi^0 + V_\pm$. Choosing a gauge such that $\phi^0$ vanishes and adding a Lagrange multiplier $\tilde{\phi}$ to impose pure gauge, the first order action is obtained [IKR95],

$$S_{1st} = \int d^2 x d^2 \theta \big[ E_{00} V_+ V_- + E_{i0} D_+ \phi^i V_- + E_{0i} V_+ D_- \phi^i + E_{ij} D_+ \phi^i D_- \phi^j$$
$$+ \tilde{\phi}(D_+ V_- + D_- V_+) \big]. \tag{5.17}$$

As in the bosonic case, integrating over the Lagrange multipliers implies pure gauge, $D_\pm V_\mp = 0$. By inserting the solution $V_\pm = D_\pm \tilde{\phi}$, the original action (5.15) is recovered. On the other hand, eliminating $V_\pm$ by their equations of motion gives the dual model

$$\tilde{S} = \int d^2x d^2\theta D_+ \tilde{\phi}^\mu \tilde{E}_{\mu\nu} D_- \tilde{\phi}^\nu, \tag{5.18}$$

in terms of the dual coordinates $\tilde{\phi}^\mu = (\tilde{\phi}, \phi^i)$ and the dual geometric structures $\tilde{E} = \tilde{g} + \tilde{b}$, given by the Buscher rules [Bus87]

$$\tilde{g}_{0i} = \frac{b_{0i}}{g_{00}}, \quad \tilde{g}_{ij} = g_{ij} - \frac{1}{g_{00}}(g_{i0}g_{0j} + b_{i0}b_{0j}), \quad \tilde{g}_{00} = \frac{1}{g_{00}},$$
$$\tilde{b}_{0i} = \frac{g_{0i}}{g_{00}}, \quad \tilde{b}_{ij} = b_{ij} + \frac{1}{g_{00}}(g_{i0}b_{0j} - b_{i0}g_{0j}). \tag{5.19}$$

If, as here, the NS-NS flux is present in the background, the constraint that the isometry leaves the metric invariant is not sufficient; the torsion must also be preserved,

$$\mathcal{L}_k(db) = 0. \tag{5.20}$$

Since the torsion is an exact form, this implies that locally, $\iota_k(db) = du$, where $u$ is a one-form determined up to an exact, Lie-algebra valued one-form.

If the original sigma model has an additional non-manifest supersymmetry defined in terms of two complex structures $J^{(\pm)}$ of the form (4.28), and the Killing vector is holomorphic with respect to these, the dual complex structures can be derived and take the form [IKR95]

$$\tilde{J}^{(+)} = \frac{1}{E_{00}} \begin{pmatrix} E_{\mu 0}(J^{(+)})^\mu_0 & -E_{\mu 0}(J^{(+)})^\mu_0 E_{j0} + E_{00} E_{\mu 0}(J^{(+)})^\mu_j \\ (J^{(+)})^i_0 & E_{00}(J^{(+)})^i_j - (J^{(+)})^i_0 E_{j0} \end{pmatrix},$$
$$\tilde{J}^{(-)} = \frac{1}{E_{00}} \begin{pmatrix} E_{0\mu}(J^{(-)})^\mu_0 & E_{0\mu}(J^{(-)})^\mu E_{0j} - E_{00} E_{0\mu}(J^{(-)})^\mu_j \\ -(J^{(-)})^i_0 & E_{00}(J^{(-)})^i_j - (J^{(-)})^i_0 E_{0j} \end{pmatrix}. \tag{5.21}$$

Hence, the dual complex structure $\tilde{J}^{(\pm)}$ mixes the original complex structures $J^{(\pm)}$, the metric and the $b$-field. An important consequence is that even if the original complex structures commute, the dual complex structures do, in general, not [IKR95].

In a complex manifold, the isometry must, in addition to (5.14), and possibly (5.20) if torsion is present, also respect the complex structures and their corresponding two-forms,

$$\mathcal{L}_k J^{(\pm)} = 0, \quad \mathcal{L}_k \omega^{(\pm)} = 0. \tag{5.22}$$

Actually, one of these constraints together with (5.14) implies the other one [HKLR87]. In the Kähler geometry case, the two-form $\omega$ is closed, implying the existence of a moment map $\mu$ such that locally,

$$\iota_k \omega = d\mu. \tag{5.23}$$

By going to holomorphic coordinates that diagonalize the complex structure, this expression can be integrated to give the moment map up to a constant. In the bihermitian setting, however, the two-forms corresponding to the complex structures $J^{(\pm)}$ are not closed, but instead satisfy the condition

$$d\omega^{(\pm)}(J^{(\pm)}X, J^{(\pm)}Y, J^{(\pm)}Z) = \pm db(X,Y,Z), \tag{5.24}$$

from which it follows that the combination $\omega^{(\pm)}k \mp (J^{(\pm)})^t u$ is closed. Locally, two Killing potentials can then be found such that [MPZV07]

$$\omega^{(\pm)}k \mp (J^{(\pm)})^t u = d\mu^{(\pm)}. \tag{5.25}$$

When $\mu^{(\pm)}$ can be defined globally, they are called moment maps. Since sigma models with extended supersymmetry have constrained target spaces with complex structures, moment maps are needed to gauge sigma models with extended supersymmetry [HKLR87, HPS91].

## 5.3   N = 2 vector multiplets

In the previous section, sigma model isometries were gauged by the method of minimal coupling, replacing partial derivatives with covariant derivatives. In $N=(2,2)$ superspace, the gauging of a sigma model with isometries can be achieved either by minimal coupling or by introducing potentials of the gauge multiplet directly into the generalized Kähler potential. The latter method keeps all supersymmetries manifest and is useful for discussing dualities between manifest $N=(2,2)$ sigma models, and will here be discussed for a duality between chiral and twisted chiral superfields.

Consider a sigma model with two manifest supersymmetries where the generalized Kähler potential is a function only of the real part of a chiral superfield,

$$S = \int d^2x d^2\theta d^2\bar\theta K(\phi + \bar\phi). \tag{5.26}$$

The sigma model has a translational isometry defined by the Killing vector $k$, leaving the Lagrangian invariant,

$$k = i(\partial_\phi - \partial_{\bar\phi}), \quad kK = 0. \tag{5.27}$$

Actually, as seen in previous chapter, the action (5.26) is invariant under Kähler transformations (4.58); hence, the Lagrangian $K$ may transform under the isometry up to a generalized Kähler transformation.

The target space of a sigma model parametrized by (anti-) chiral superfields is Kähler, and the constraint that the isometry should preserve the complex structure implies the existence of a moment map as in (5.23). Choosing holomorphic coordinates, the condition can be integrated and gives the moment map as $\mu(\phi, \bar{\phi}) = K_\phi = K'$. The chiral and anti-chiral fields transform under the isometry according to (5.12) as $\delta \phi = i\lambda$ and $\delta \bar{\phi} = -i\lambda$, where $\lambda$ is a real constant parameter. These isometry transformations are gauged by promoting the parameters to be local (anti-) chiral functions,

$$\delta \phi = i\Lambda(\phi), \quad \delta \bar{\phi} = -i\bar{\Lambda}(\bar{\phi}). \tag{5.28}$$

To ensure that the action remains invariant under the gauged isometry transformations, an $N=(2,2)$ gauge potential is introduced into the Lagrangian,

$$S = \int d^2x \, d^2\theta \, d^2\bar{\theta} \, K(\phi + \bar{\phi} + V). \tag{5.29}$$

The gauge potential $V$ transforms under the local isometry as $\delta V = i(\bar{\Lambda} - \Lambda)$ and enables the construction of a gauge invariant field strength $F = i\bar{\mathbb{D}}_+ \mathbb{D}_- V$.

Introducing unconstrained Lagrange multipliers $X$ and choosing gauge such that $\phi + \bar{\phi} = 0$, a first order action can be written as

$$S_{1st} = \int \left[ K(V, z) - (XF + \bar{X}\bar{F}) \right] = \int \left[ K(V, z) - (\chi + \bar{\chi})V \right], \tag{5.30}$$

where the integration measure $d^2x \, d^2\theta \, d^2\bar{\theta}$ is implicit and $\chi = i\bar{\mathbb{D}}_+ \mathbb{D}_- X$ is a twisted chiral field. Varying the first order action with respect to the Lagrange multipliers constrains the gauge field strength $F$ to vanish, and the original chiral model (5.26) is recovered. Varying instead with respect to the gauge potential implies that $K_V = \chi + \bar{\chi}$, hence the gauge potential is a function of the sum of a twisted chiral and twisted anti-chiral field, and the dual twisted chiral model $\tilde{K}$ is obtained as a Legendre transform,

$$S_{dual} = \int \left[ K(V(\chi + \bar{\chi})) - (\chi + \bar{\chi}) \cdot V(\chi + \bar{\chi}) \right] = \int \tilde{K}(\chi + \bar{\chi}). \tag{5.31}$$

Since $K_V = K' = \mu$, the T-duality embeds the moment map as the real part of the dual twisted chiral coordinates [MV08].

The transformations in (5.28) are infinitesimal. The finite version of the transformations are

$$\phi \to e^{i\Lambda}\phi, \quad \bar{\phi} \to \bar{\phi}e^{-i\bar{\Lambda}}. \tag{5.32}$$

In the chiral representation, the covariant derivatives transform with the chiral parameter $\Lambda$ as

$$\nabla \to e^{i\Lambda} \nabla e^{-i\Lambda}. \tag{5.33}$$

Gauge invariant objects can be obtained by inserting the converter $e^V$ between the chiral field and its complex conjugate,

$$\bar{\phi}e^V\phi \to \bar{\phi}e^{-i\bar{\Lambda}}e^{V+i(\bar{\Lambda}-\Lambda)}e^{i\Lambda}\phi = \bar{\phi}e^V\phi. \quad (5.34)$$

This holds only in the abelian case. In the non-abelian, the fields transform as $\phi \to e^{i\Lambda^A T_A}\phi$, where $T_A$ are the elements in a Lie algebra. The non-abelian version of the transformation of $e^V$ in (5.34) is

$$e^V \to e^{i\bar{\Lambda}}e^V e^{-i\Lambda}, \quad (5.35)$$

where the parameters are Lie-algebra valued, $\Lambda = \Lambda^A T_A$ and $V = V^A T_A$.

The vector potential $V$ described here may be used to gauge an isometry of chiral superfields in manifest $N=(2,2)$ superspace. Likewise, a vector potential transforming into (anti-) twisted chiral parameters can be used to gauge the isometry of twisted chiral fields, $K = K(\chi + \bar{\chi})$.

In chapter 7, two multiplets gauging isometries mixing chiral and twisted chiral directions on the one hand, and semichiral directions on the other hand will be introduced, the so called large vector multiplet and the semichiral multiplet, respectively. The two multiplets have different properties; e.g., one can incorporate additional supersymmetry, whereas the other cannot. The large vector multiplet will further be used in chapter 8 to understand the discrepancy between one model with linear off-shell $N=(4,4)$ supersymmetry, and the dual model with non-linear on-shell $N=(4,4)$ supersymmetry.

## 5.4 Quotient reduction

In the dualization process of the sigma models discussed earlier in this chapter, the gauging of the isometry with gauge potentials was followed by the introduction of gauge field strengths and Lagrange multipliers to ensure pure gauge. The resulting dual action depends on the dual coordinates and may have very different geometry as compared to the original model.

A related method that gauges isometries using gauge potentials to construct geometries is the quotient reduction [HKLR86, HKLR87]. In the quotient construction, the gauge degrees of freedom are reduced by restricting to the space of orbits $M/G$, where $G$ is the isometry group acting on the manifold $M$.

### 5.4.1 Reduction of sigma models

Consider the sigma model with one manifest supersymmetry in (4.25), but restrict to the case with no $b$-field,

$$S = \int d^2x d^2\theta D_+\phi^\mu g_{\mu\nu} D_-\phi^\nu, \quad (5.36)$$

and let the target manifold $M$ be subject to an isometry group $G$ acting on the field as in (5.12), $\delta\phi^\mu = \lambda^A k_A^\mu$. The gauging $\lambda \to \lambda(x)$ is performed as before by the minimal coupling procedure $D_\pm\phi^\mu \to \nabla_\pm\phi^\mu = D_\pm\phi^\mu + V_\pm^\mu$, where $V_\pm^\mu$ are Lie-algebra valued gauge potentials, $V_\pm^\mu = V_\pm^A k_A^\mu$ with appropriate gauge transformations. The gauged action, analogous to (5.6) in the bosonic case, is

$$S_g = \int d^2x d^2\theta \nabla_+\phi^\mu g_{\mu\nu} \nabla_-\phi^\nu. \tag{5.37}$$

Now, instead of introducing a field strength and Lagrange multipliers, the connection that extremizes the gauged action is chosen, $V_\pm^A = -g_{\mu\nu}k_B^\nu H^{BA} D_\pm\phi^\mu$, where $H^{AB} = (g_{\mu\nu}k_A^\mu k_B^\nu)^{-1}$. The reduced quotient action is obtained by inserting this connection [HKLR87],

$$S_{\text{red}} = \int d^2x d^2\theta D_+\phi^\mu \hat{g}_{\mu\nu} D_-\phi^\nu, \quad \hat{g}_{\mu\nu} = g_{\mu\nu} - g_{\mu\rho}g_{\nu\sigma}k_A^\rho k_B^\sigma H^{AB}. \tag{5.38}$$

The quotient metric $\hat{g}$ is defined on the reduced space $M_{\text{red}} = M/G$.

However, if the manifold is equipped with additional geometrical structures, the quotient space does not, in general, inherit the same structures. If, for example, $M$ is a symplectic manifold, then in general the quotient space is not. As seen previously in this chapter, the gauging of a sigma model with a target space equipped with complex structures implies the existence of a moment map (5.23). The existence of the moment map, also known as the Killing potential, enables the quotient reduction to a subspace with preserved additional structures, as will now be discussed from a geometrical viewpoint.

### 5.4.2 *Geometric interpretation*

Recall the discussion on fiber bundles in chapter 2. A manifold $M$ with isometries is a principal $G$-bundle over the quotient space $M/G$. The points in the manifold can be projected to points on the quotient space by a projection map $\pi : M \to M/G$, and the quotient space can be given the structure of a manifold [HKLR87]. As seen in section 2.1.3, given a connection, for each point $p \in M$, the tangent space $T_pM$ can be split uniquely into (2.23), a vertical subspace $V_pM$, tangent bundle to the orbit of $G$ through $p$, and a horizontal subspace $H_pM$. If the Lie group acts by isometries, the Killing vectors of the Lie algebra form a basis of the vertical subspace.

Given the metric $g$ of the manifold, the quotient space $M/G$ can be endowed with an induced metric $\hat{g}$. At a point $\pi(p) \in M/G$ in the quotient space, two tangent vectors $X, Y \in T_{\pi(p)}M$ can be lifted uniquely to the horizontal subspace $\tilde{X}, \tilde{Y} \in H_pM$, and a metric can be defined on the quotient space as $\hat{g}(X,Y) = g(\tilde{X}, \tilde{Y})$ [HKLR87]. The splitting of the tangent space (2.23) defines a connection one-form as a projection of the tangent space onto the vertical subspace. The quotient space inherits the connection by pullback of a local section, i.e. local gauge, $A_i = s_i^* \omega$, as seen in (2.26). This is the geometrical understanding of the gauge connection that minimizes the gauged action above.

Assume that $M$ is a symplectic manifold. Given the moment map $\mu : M \to \mathbf{g}^*$ and choosing a subspace in $M$ that consists of elements that are mapped onto the same element in the dual Lie algebra (without loss of generality, usually the element $0 \in \mathbf{g}^*$ is considered), the quotient space

$$M_{\text{red}} = \mu^{-1}(0)/G \tag{5.39}$$

is a symplectic manifold [HKLR87]. This is the symplectic reduction [MW74].

Similarly, if the manifold $M$ is Kähler, the quotient space constructed in this way is also Kähler. For example, the target manifold of the chiral sigma model with two supersymmetries discussed in section 5.3 is Kähler; a Kähler quotient can be defined using the existence of a moment map, which ensures that the reduced sigma model $\hat{K}$ defined on the orbit space is also Kähler [HKLR87].

The same construction can be used to construct hyperkähler metrics by the hyperkähler quotient [HKLR87]. Consider a hyperkähler manifold $M$. There exists three complex structures and their corresponding closed Kähler forms $\omega_i$, $i = 1, 2, 3$. If a triholomorphic Killing vector exists, i.e., $\mathcal{L}_k \omega_i = 0$, then three Killing potentials exist as in (5.23) such that locally, $\iota_k \omega_i = d\mu_i$. By defining the subset $\mu^{-1}(0)$ as the intersection of all three subsets $\mu_i^{-1}(0) \subset M$, the quotient space defined as in (5.39) is hyperkähler [HKLR87].

Isometries mixing other superfields require the use of more involved gauge multiplets, e.g., the semichiral vector multiplet that gauges isometries mixing semichiral directions, and enables the semichiral quotient reduction. We will return to $N = 2$ vector multiplets in chapter 7.

# Part II:
# Developments

# 6. Semichiral sigma models with N = (4,4) supersymmetry

> Sometimes I find [mathematical problems] difficult, but my old obstinacy remains, for if I do not succeed today, I attack them again on the morrow.
> *Mary Somerville, physicist, astronomer, mathematician (1780-1872)*

Adding supersymmetry to non-linear sigma models has proven to be an effective route for investigating new geometries, as has been discussed in the previous chapters. Sigma models with manifest $N = (2,2)$ supersymmetry were reviewed in chapter 4. Their target space geometries are bihermitian, or generalized Kähler, and can be parametrized by three kinds of superfields; chiral, twisted chiral and left and right semichiral.

It has long been known [GHR84] that a two-dimensional sigma model parametrized by (anti-) chiral and twisted (anti-) chiral fields admits $N = (4,4)$ supersymmetry if and only if the generalized Kähler potential $K(\phi, \bar{\phi}, \chi, \bar{\chi})$ satisfies the Laplace equation. But the analogous situation for a semichiral sigma model $K(\mathbb{X}^\ell, \bar{\mathbb{X}}^\ell, \mathbb{X}^r, \bar{\mathbb{X}}^r)$ was previously not known, although similar sigma models had been studied previously in harmonic or projective superspace, e.g., in [LIR94, Iva96, GK98].

To understand the details and the geometry of this model was the main goal of the papers [I], [II] and [IV], that this chapter is based on. Initially, the question was believed to be simply resolved, but the scope of the problem grew and can be summarized as follows.

RESEARCH QUESTIONS:
⋆ Does a sigma model parametrized by one set of semichiral fields admit off-shell $N = (4,4)$ supersymmetry, in an analogous way to the chiral and twisted chiral model? If not, what are the obstructions? Can the model incorporate additional twisted supersymmetry instead, and what are the resulting constraints on the target space geometry?

⋆ Does the situation change if the model has more than one set of left and right semichiral fields, i.e. if the target space dimension is enlarged? How can the supersymmetry transformations be described geometrically?
⋆ Can the field equations be used to make the supersymmetry algebra close on-shell? What additional constraints are imposed on the generalized Kähler potential, and what is the geometrical interpretation?
⋆ Can interesting examples of the found sigma models and their corresponding geometries be constructed?

The material in the chapter is not presented chronologically, instead a systematical overview over the full area is given, as displayed in chart 6.1 at the end of the chapter. Section 6.2 is a review of paper [I] and summarizes the first approach to solve the first question posed above, which resulted in a description of a new sigma model with twisted supersymmetry and neutral hyperkähler target space. In sections 6.1 and 6.3, the generalization to an enlarged target space is discussed, based on paper [II]. The on-shell supersymmetric sigma model and the corresponding target space geometry are summarized in section 6.5 and 6.6, respectively, based on the results of paper [II] and [IV]. The precise relation to the chiral and twisted chiral sigma model with off-shell $N=(4,4)$ supersymmetry was saved for paper [V] and will be treated separately in chapter 8.

## 6.1 General ansatz

Consider a sigma model parametrized by left and right semichiral superfields labeled as in section 4.4 by $\mathbb{X}^L = \{\mathbb{X}^a, \bar{\mathbb{X}}^{\bar{a}}\}$ and $\mathbb{X}^R = \{\mathbb{X}^{a'}, \bar{\mathbb{X}}^{\bar{a}'}\}$, respectively, and $\mathbb{X}^i = \{\mathbb{X}^L, \mathbb{X}^R\}$ denoting the full set of semichiral fields,

$$S = \int d^2x d^2\theta d^2\bar{\theta} K(\mathbb{X}^a, \bar{\mathbb{X}}^{\bar{a}}, \mathbb{X}^{a'}, \bar{\mathbb{X}}^{\bar{a}'}). \tag{6.1}$$

When the action is reduced to $N=(1,1)$ superspace formalism, the underlying geometry is revealed as bihermitian, as described in section 4.4.2. Hence, two complex structures $J^{(\pm)}$ exist, that in the basis of left and right semichiral fields $\{\mathbb{X}^L, \mathbb{X}^R\}$ take the form (4.65). They are both covariantly constant with respect to a torsionful connection and the target space metric is hermitian with respect to both of them. An important feature of the geometry is also that the two complex structures do not commute.

The $N=(4,4)$ supersymmetry transformations should act covariantly, that is, the ansatz should respect the chirality properties of the semichiral fields,

$$\bar{\mathbb{D}}_+(\delta \mathbb{X}^a) = 0, \quad \bar{\mathbb{D}}_-(\delta \mathbb{X}^{a'}) = 0. \tag{6.2}$$

Based on this simple requirement, a general ansatz for additional supersymmetry transformations can be constructed as [II]

$$\delta \mathbb{X}^a = \bar{\epsilon}^+ \bar{\mathbb{D}}_+ f^a(\mathbb{X}^c, \bar{\mathbb{X}}^{\bar{c}}, \mathbb{X}^{c'}, \bar{\mathbb{X}}^{\bar{c}'}) + g_b^a(\mathbb{X}^c)\bar{\epsilon}^- \bar{\mathbb{D}}_- \mathbb{X}^b + h_b^a(\mathbb{X}^c)\epsilon^- \mathbb{D}_- \mathbb{X}^b,$$
$$\delta \bar{\mathbb{X}}^{\bar{a}} = \epsilon^+ \mathbb{D}_+ \bar{f}^{\bar{a}}(\mathbb{X}^c, \bar{\mathbb{X}}^{\bar{c}}, \mathbb{X}^{c'}, \bar{\mathbb{X}}^{\bar{c}'}) + \bar{g}_{\bar{b}}^{\bar{a}}(\bar{\mathbb{X}}^{\bar{c}})\epsilon^- \mathbb{D}_- \bar{\mathbb{X}}^{\bar{b}} + \bar{h}_{\bar{b}}^{\bar{a}}(\bar{\mathbb{X}}^{\bar{c}})\bar{\epsilon}^- \bar{\mathbb{D}}_- \bar{\mathbb{X}}^{\bar{b}},$$
$$\delta \mathbb{X}^{a'} = \bar{\epsilon}^- \bar{\mathbb{D}}_- \tilde{f}^{a'}(\mathbb{X}^c, \bar{\mathbb{X}}^{\bar{c}}, \mathbb{X}^{c'}, \bar{\mathbb{X}}^{\bar{c}'}) + \tilde{g}_{b'}^{a'}(\mathbb{X}^{c'})\bar{\epsilon}^+ \bar{\mathbb{D}}_+ \mathbb{X}^{b'} + \tilde{h}_{b'}^{a'}(\mathbb{X}^{c'})\epsilon^+ \mathbb{D}_+ \mathbb{X}^{b'},$$
$$\delta \bar{\mathbb{X}}^{\bar{a}'} = \epsilon^- \mathbb{D}_- \bar{\tilde{f}}^{\bar{a}'}(\mathbb{X}^c, \bar{\mathbb{X}}^{\bar{c}}, \mathbb{X}^{c'}, \bar{\mathbb{X}}^{\bar{c}'}) + \bar{\tilde{g}}_{\bar{b}'}^{\bar{a}'}(\bar{\mathbb{X}}^{\bar{c}'})\epsilon^+ \mathbb{D}_+ \bar{\mathbb{X}}^{\bar{b}'} + \bar{\tilde{h}}_{\bar{b}'}^{\bar{a}'}(\bar{\mathbb{X}}^{\bar{c}'})\bar{\epsilon}^+ \bar{\mathbb{D}}_+ \bar{\mathbb{X}}^{\bar{b}'}.$$
(6.3)

The question of whether this ansatz can be further generalized to include central charges will be discussed in section 6.7 at the end of this chapter. Note that, in order to satisfy the chirality properties, the functions $f$ and $\tilde{f}$ may depend on all the semichiral fields, whereas $g, h$ and $\tilde{g}, \tilde{h}$ are functions of only the left semichiral and right semichiral fields, respectively. The ansatz is covariant under holomorphic coordinate transformations of a form that preserves the separation into left and right semichiral coordinates,

$$\mathbb{X}^a \mapsto \mathbb{X}^a(\mathbb{X}^b), \quad \mathbb{X}^{a'} \mapsto \mathbb{X}^{a'}(\mathbb{X}^{b'}). \tag{6.4}$$

It will prove very useful to write the ansatz (6.3) in a more compact notation. Writing all the semichiral fields in the unified notation $\mathbb{X}^i$ and defining transformation matrices as

$$U^{(+)} = \begin{pmatrix} * & f_{\bar{b}}^a & f_{b'}^a & f_{\bar{b}'}^a \\ * & 0 & 0 & 0 \\ * & 0 & \tilde{g}_{b'}^{a'} & 0 \\ * & 0 & 0 & \bar{\tilde{h}}_{\bar{b}'}^{\bar{a}'} \end{pmatrix}, \quad U^{(-)} = \begin{pmatrix} g_b^a & 0 & * & 0 \\ 0 & \bar{h}_{\bar{b}}^{\bar{a}} & * & 0 \\ \tilde{f}_b^{a'} & \tilde{f}_{\bar{b}}^{a'} & * & \tilde{f}_{\bar{b}'}^{a'} \\ 0 & 0 & * & 0 \end{pmatrix}, \tag{6.5}$$

$$V^{(+)} = \begin{pmatrix} 0 & * & 0 & 0 \\ \bar{f}_b^{\bar{a}} & * & \bar{f}_{b'}^{\bar{a}} & \bar{f}_{\bar{b}'}^{\bar{a}} \\ 0 & * & \tilde{h}_{b'}^{a'} & 0 \\ 0 & * & 0 & \bar{\tilde{g}}_{\bar{b}'}^{\bar{a}'} \end{pmatrix}, \quad V^{(-)} = \begin{pmatrix} h_b^a & 0 & 0 & * \\ 0 & \bar{g}_{\bar{b}}^{\bar{a}} & 0 & * \\ 0 & 0 & 0 & * \\ \bar{\tilde{f}}_b^{\bar{a}'} & \bar{\tilde{f}}_{\bar{b}}^{\bar{a}'} & \bar{\tilde{f}}_{b'}^{\bar{a}'} & * \end{pmatrix}, \tag{6.6}$$

the $N=(4,4)$ supersymmetry ansatz (6.3) for all semichiral fields simply reads

$$\delta \mathbb{X}^i = \bar{\epsilon}^\alpha U^{(\alpha)i}{}_j \bar{\mathbb{D}}_\alpha \mathbb{X}^j + \epsilon^\alpha V^{(\alpha)i}{}_j \mathbb{D}_\alpha \mathbb{X}^j, \tag{6.7}$$

where the spinor indices $\alpha = +, -$ are summed over. There are a few important properties to note about the transformation matrices. The first is that the lower indices of the functions $f$ and $\tilde{f}$ are derivatives, whereas this is not necessarily the case for the functions $g, h, \tilde{g}$ and $\tilde{h}$. Secondly, one column in each of the matrices is arbitrary, due to the semichiral constraints. Third, the matrices $V^{(\pm)}$ are the complex conjugates of $U^{(\pm)}$ in the sense that the coordinates $\mathbb{X}^\ell$ and $\bar{\mathbb{X}}^{\bar\ell}$ as well as $\mathbb{X}^r$ and $\bar{\mathbb{X}}^{\bar r}$ are interchanged under the complex conjugation,

$$V^{(\pm)} = \begin{pmatrix} 0 & 1 & 0 & 0 \\ 1 & 0 & 0 & 0 \\ 0 & 0 & 0 & 1 \\ 0 & 0 & 1 & 0 \end{pmatrix} \bar U^\pm \begin{pmatrix} 0 & 1 & 0 & 0 \\ 1 & 0 & 0 & 0 \\ 0 & 0 & 0 & 1 \\ 0 & 0 & 1 & 0 \end{pmatrix}. \tag{6.8}$$

The invariance of the action (6.1) under the transformations (6.3) and the closing of the transformations to a supersymmetry algebra will now be discussed. First, a special case with constant transformation parameters and four-dimensional target space will be considered, then the general transformations will be discussed in section 6.3 and finally the on-shell analysis will follow in section 6.5.

## 6.2 Off-shell pseudo-supersymmetry

### 6.2.1 First approach

To attack the problem, two simplifications that can be made are to restrict to a four-dimensional target space, and/or to linear transformations. In this section, corresponding to paper [I], both these simplifications are considered.

A four-dimensional target space is parametrized by only one set of left and right semichiral fields, so the indices labeling the fields can be dropped,

$$K = K(\mathbb{X}^\ell, \bar{\mathbb{X}}^{\bar\ell}, \mathbb{X}^r, \bar{\mathbb{X}}^{\bar r}). \tag{6.9}$$

Linear transformations mean that all the functions $f_i^a$, $g_b^a$, $h_b^a$ and the tilde-versions are constants. In paper [I], the following notation was used for the constant parameters:

$$\varepsilon = -if_b^a, \quad b = -if_{b'}^a, \quad c = -if_{\bar b'}^a, \quad \kappa = -ig_b^a, \quad \lambda = ih_b^a \tag{6.10}$$

and analogously for the parameters $\tilde\varepsilon$, $\tilde b$, $\tilde c$, $\tilde\kappa$ and $\tilde\lambda$. The ansatz

$$\delta\mathbb{X}^\ell = i\bar\epsilon^+ \bar{\mathbb{D}}_+(\varepsilon\bar{\mathbb{X}}^{\bar\ell} + b\mathbb{X}^r + c\bar{\mathbb{X}}^{\bar r}) + i\kappa\bar\epsilon^- \bar{\mathbb{D}}_- \mathbb{X}^\ell - i\lambda\epsilon^- \mathbb{D}_- \mathbb{X}^\ell,$$
$$\delta\mathbb{X}^r = i\bar\epsilon^- \bar{\mathbb{D}}_-(\tilde\varepsilon\bar{\mathbb{X}}^{\bar r} + \tilde b\mathbb{X}^\ell + \tilde c\bar{\mathbb{X}}^{\bar\ell}) + i\tilde\kappa\bar\epsilon^+ \bar{\mathbb{D}}_+ \mathbb{X}^r - i\tilde\lambda\epsilon^+ \mathbb{D}_+ \mathbb{X}^r, \tag{6.11}$$

can close to a supersymmetry only if the algebra decouples, meaning that the left semichiral fields possess only a left-going symmetry, and the right semichiral fields a right-going. However, an action is invariant under the de-coupled supersymmetry only if the Lagrangian (6.9) is linear, implying that the action vanishes. Hence, the ansatz *cannot* close to a supersymmetry off-shell.

The transformations can, however, close to a *pseudo*-supersymmetry (3.19),

$$[\delta(\epsilon_1), \delta(\epsilon_2)]\mathbb{X} = -i\bar{\epsilon}_{[2}^{\pm}\epsilon_{1]}^{\pm}\partial_{\pm}\mathbb{X} \qquad (6.12)$$

for all the semichiral fields, provided that some of the transformation parameters are solved in terms of the others. All remaining parameters but one can be absorbed in a rescaling of the semichiral fields, leaving only one constant $\kappa$ arbitrary.

The action is invariant under the transformations if and only if the Lagrangian satisfies a system of partial differential equations,

$$\begin{aligned} K_{\ell\bar{\ell}} - K_{\ell r} - \bar{\kappa}K_{\bar{\ell}r} &= 0, \\ (\kappa\bar{\kappa} - 1)K_{r\bar{r}} + K_{\ell r} - \kappa K_{\ell\bar{r}} &= 0, \end{aligned} \qquad (6.13)$$

where the indices denote derivatives with respect to the semichiral fields. A family of solutions to this system is given by the generalized Kähler potential

$$K = F(y) + \bar{F}(\bar{y}), \quad y = \alpha\mathbb{X}^\ell + \beta\bar{\mathbb{X}}^{\bar{\ell}} + \gamma\mathbb{X}^r + \delta\bar{\mathbb{X}}^{\bar{r}}, \qquad (6.14)$$

where two of the coefficients in the variable $y$ are determined in terms of the others as

$$\gamma = \frac{\alpha\beta}{\alpha + \bar{\kappa}\beta}, \quad \delta = \frac{\alpha\beta}{\kappa\alpha + \beta}. \qquad (6.15)$$

The metric of this solution is non-degenerate if the parameters further satisfy $|\alpha|^2 \neq |\beta|^2$ and $|\kappa|^2 \neq 1$. Due to the linearity of the system, a general solution is a superposition of potentials of the form (6.14),

$$S = \int d^2x d^2\theta d^2\bar{\theta} \int d\alpha d\beta K(\alpha, \beta; y, \bar{y}), \qquad (6.16)$$

where the free parameters are integrated over.

### 6.2.2 Twisted supersymmetry and neutral hyperkähler geometry

As just revealed, the semichiral sigma model with four-dimensional target space cannot incorporate linear $N = (4,4)$ supersymmetry off-shell. Although not presented here, the same result holds for non-linear transformations [IV]. The off-shell algebra can only close to twisted supersymmetry, where the additional $N = (4,4)$ transformation is a pseudo-supersymmetry. This has important consequences on the geometry of the target space.

A neutral hypercomplex structure is defined by a set of integrable endomorphisms on the tangent bundle, satisfying the algebra of split quaternions (2.47), as reviewed in section 2.1.7. In chapter 2, it was shown that every complex manifold admits a hermitian metric. But unlike the positive definite case, in a pseudo-Riemannian manifold, given a complex structure, a global pseudo-hermitian metric does not always exist, despite the fact that a local one can always be defined [DGMY11]. However, by going to a global cover, the metric can be defined globally [DGMY09].

An oriented four-dimensional smooth manifold admits a neutral hypercomplex structure if and only if there exist two complex structures $J^{(\pm)}$ with the same orientation, such that their anti-commutator is proportional to the identity,

$$\{J^{(+)}, J^{(-)}\} = 2c \tag{6.17}$$

with $|c| > 1$ [LRvUZ07a]. If this condition holds, the complex structure can be chosen as $I = J^{(+)}$, and the two product structures can be constructed explicitly as in (4.76) [LRvUZ07a]

$$S = \frac{1}{\sqrt{c^2-1}}(J^{(-)} + cJ^{(+)}), \quad T = \frac{1}{2\sqrt{c^2-1}}[J^{(+)}, J^{(-)}], \tag{6.18}$$

as was discussed in section 4.4.2. In four dimensions, the anti-commutator (6.17) of the complex structures $J^{(\pm)}$, that in the semichiral parametrization take the form (4.65), gives the generalization of the Monge-Ampère equation (4.78),

$$(1+c)|K_{\ell r}|^2 + (1-c)|K_{\ell \bar{r}}|^2 = 2K_{\ell \bar{\ell}} K_{r\bar{r}},$$

and reduces to an equation that is equivalent to the Monge-Ampère equation for $c = 0$. Inserting the solution (6.14) into this equation gives

$$c = -\frac{\kappa\bar{\kappa}+1}{\kappa\bar{\kappa}-1}. \tag{6.19}$$

The function $c$ is a constant in the sense that it is independent of the semichiral fields, thus, this results show that the torsion vanishes, $db = 2c \cdot d\Omega = 0$. The transformations collapse for $|\kappa| \to 1$ as well as $|\kappa| \to \infty$, so $c$ is a well-defined number with $|c| > 1$.

Hence, the target manifold is neutral hyperkähler. The neutral hyperkähler structures can be constructed using (6.18) and (6.19) and take the form

$$I = J^{(+)}, \quad S = \frac{1}{|\kappa|}\left((\kappa\bar{\kappa}-1)J^{(-)} - (\kappa\bar{\kappa}+1)J^{(+)}\right), \quad T = \frac{\kappa\bar{\kappa}-1}{|\kappa|}[J^{(+)}, J^{(-)}]. \tag{6.20}$$

The corresponding fundamental two-forms can then be constructed as in (2.29).

The metric has neutral signature $(--++)$ and can be derived for the models with Lagrangian (6.14). The quadratic solution $F(y) = y^2$ with the parameters chosen as $\kappa = \sqrt{2}$, $\alpha = 1$ and $\beta = -(1+\sqrt{2})$, for example, gives the

metric

$$g = 8(1+\sqrt{2}) \begin{pmatrix} 0 & -1 & 1 & 0 \\ -1 & 0 & 0 & 1 \\ 1 & 0 & 0 & 1 \\ 0 & 1 & 1 & 0 \end{pmatrix}. \qquad (6.21)$$

To summarize, a sigma model parametrized by one set of left and right semichiral fields cannot incorporate linear $N=(4,4)$ supersymmetry off-shell, but linear pseudo-supersymmetry may be imposed if and only if the target space is neutral hyperkähler [I].

## 6.3 Enlarging the target space

Approaching the problem by simplifying it to a four-dimensional target space and linear transformations showed that off-shell linear supersymmetry was not possible for a single set of left and right semichiral fields [I]. Constraining the target space to be four-dimensional but leaving the transformations arbitrary gives the same result; a semichiral sigma model with four-dimensional target space cannot incorporate off-shell $N=(4,4)$ supersymmetry [IV]. To find off-shell supersymmetry, a larger target space with at least two sets of semichiral fields must be considered. This was investigated in paper [II] and will be reviewed in this section.

### 6.3.1 Off-shell algebra closure

Two subsequent transformations given by the general ansatz in (6.3) acting on a semichiral field commute to[†]

$$[\delta_1, \delta_2]\mathbb{X}^i = \bar{\epsilon}^{\pm}_{[2}\bar{\epsilon}^{\pm}_{1]} \left[ \mathcal{N}(U^{(\pm)})^i_{jk} \bar{\mathbb{D}}_{\pm}\mathbb{X}^j \bar{\mathbb{D}}_{\pm}\mathbb{X}^k \right] + \epsilon^{\pm}_{[2}\epsilon^{\pm}_{1]} \left[ \mathcal{N}(V^{(\pm)})^i_{jk} \mathbb{D}_{\pm}\mathbb{X}^j \mathbb{D}_{\pm}\mathbb{X}^k \right] \qquad (6.22)$$

$$+ \bar{\epsilon}^{+}_{[2}\bar{\epsilon}^{-}_{1]} \left[ \mathcal{M}(U^{(+)}, U^{(-)})^i_{jk} \bar{\mathbb{D}}_{+}\mathbb{X}^j \bar{\mathbb{D}}_{-}\mathbb{X}^k - [U^{(+)}, U^{(-)}]^i_{j} \bar{\mathbb{D}}_{+}\bar{\mathbb{D}}_{-}\mathbb{X}^j \right]$$

$$+ \bar{\epsilon}^{+}_{[2}\epsilon^{-}_{1]} \left[ \mathcal{M}(U^{(+)}, V^{(-)})^i_{jk} \bar{\mathbb{D}}_{+}\mathbb{X}^j \mathbb{D}_{-}\mathbb{X}^k - [U^{(+)}, V^{(-)}]^i_{j} \bar{\mathbb{D}}_{+}\mathbb{D}_{-}\mathbb{X}^j \right] \qquad (6.23)$$

$$+ \bar{\epsilon}^{\pm}_{[2}\epsilon^{\pm}_{1]} \Big[ \mathcal{M}(U^{(\pm)}, V^{(\pm)})^i_{jk} \bar{\mathbb{D}}_{\pm}\mathbb{X}^j \mathbb{D}_{\pm}\mathbb{X}^k$$

$$- (U^{(\pm)}V^{(\pm)})^i_{j} \bar{\mathbb{D}}_{\pm}\mathbb{D}_{\pm}\mathbb{X}^j - (V^{(\pm)}U^{(\pm)})^i_{j} \mathbb{D}_{\pm}\bar{\mathbb{D}}_{\pm}\mathbb{X}^j \Big]. \qquad (6.24)$$

Closing this expression to a supersymmetry (3.18) off-shell implies four kinds of constraints. First, the vanishing of the first line (6.22) implies the vanishing of an expression involving the Nijenhuis tensor for all transformation matrices,
$$\mathcal{N}(U^{(\pm)})^i_{jk}\bar{\mathbb{D}}_\pm\mathbb{X}^j\bar{\mathbb{D}}_\pm\mathbb{X}^k = 0, \qquad (6.25)$$
and similarly for $V^{(\pm)}$. Note, that the Nijenhuis tensor multiplies derivatives acting on semichiral fields, so the vanishing of $\mathcal{N}$ is only required in all sections that are not projected out by the semichiral constraints. As discussed in chapter 2, the vanishing of the Nijenhuis tensor for the transformation matrices corresponds to them being integrable.

---

[†] Complex conjugate expressions of (6.23) corresponding to $\epsilon^{+}_{[2}\epsilon^{-}_{1]}$ and $\epsilon^{+}_{[2}\bar{\epsilon}^{-}_{1]}$ also appear but have been omitted here for clarity.

Secondly, requiring that the second two lines (6.23) vanish and the third line (6.24) closes to a translation implies a constraint of the generic form
$$\mathcal{M}(I_1, I_2)D_1XD_2X = (I_1I_2 + \delta)D_1D_2X + (I_2I_1 + \delta)D_2D_1X, \qquad (6.26)$$
where $I_1, I_2$ are any of the transformation matrices $U^{(\pm)}, V^{(\pm)}$ and $D_1, D_2$ the corresponding derivatives. Since, off-shell, second order differentials $\mathbb{DDX}$ and products of first order differentials $\mathbb{DXDX}$ are unrelated, the relation in (6.26) implies that both sides must vanish independently.

The vanishing of the right-hand side in (6.26) when the transformation matrices are $I_1 = U^{(\pm)}$ and $I_2 = V^{(\pm)}$ implies that the products of the matrices equal minus one, since their corresponding derivatives $\bar{\mathbb{D}}_\pm$ and $\mathbb{D}_\pm$ do not commute,
$$\begin{aligned}(U^{(\pm)}V^{(\pm)})^i_j\bar{\mathbb{D}}_\pm\mathbb{D}_\pm\mathbb{X}^j &= -\bar{\mathbb{D}}_\pm\mathbb{D}_\pm\mathbb{X}^i,\\ (V^{(\pm)}U^{(\pm)})^i_j\mathbb{D}_\pm\bar{\mathbb{D}}_\pm\mathbb{X}^j &= -\mathbb{D}_\pm\bar{\mathbb{D}}_\pm\mathbb{X}^i.\end{aligned} \qquad (6.27)$$
Evaluating the constraint $U^{(\pm)}V^{(\pm)} = -1$ at the level of the transformation functions in (6.5)-(6.6) shows that it requires, e.g., that $f^a_b \bar{f}^{\bar{b}}_c = -\delta^a_b$. In a four-dimensional target space, $f \cdot \bar{f}$ can never equal minus one, but it is possible in larger target spaces with $a > 1$. This is the reason why the supersymmetry algebra can close off-shell in a larger target space with at least two sets of semichiral fields, but not in a four-dimensional space parametrized by only one set of semichiral fields.

The remaining constraints from (6.26) state that the Magri-Morosi concomitant must vanish for all pairs of transformation matrices, and that they all commute. The Magri-Morosi concomitant relating matrices of the same chirality can be rewritten in terms of derivatives of products plus curl-terms,
$$\mathcal{M}(U,V)^i_{jk} = -(UV)^i_{j,k} + (VU)^i_{k,j} - U^i_{[j,l]}V^l_k + V^i_{[k,l]}U^l_j, \qquad (6.28)$$
where matrices of the same chirality are combined, $UV = U^{(\pm)}V^{(\pm)}$. The first two terms in this expression vanish due to (6.27). Using also the integrability (6.25), the remaining two terms can be shown to vanish. Hence, the Magri-Morosi concomitant $\mathcal{M}(U^{(\pm)}, V^{(\pm)})$ vanishes without further constraints. Moving on to the constraints (6.23) relating matrices of different chirality, requiring that the transformation matrices of opposite chirality commute,
$$[U^{(+)}, U^{(-)}]^i_j\bar{\mathbb{D}}_+\bar{\mathbb{D}}_-\mathbb{X}^j = 0, \quad [U^{(+)}, V^{(-)}]^i_j\bar{\mathbb{D}}_+\mathbb{D}_-\mathbb{X}^j = 0, \qquad (6.29)$$

implies that the corresponding Magri-Morosi concomitants contain only curl-terms,

$$\mathcal{M}(U^{(+)}, U^{(-)})^i_{jk} = U^{(+)l}{}_j U^{(-)i}{}_{[k,l]} - U^{(-)l}{}_k U^{(+)i}{}_{[j,l]} - (U^{(+)} U^{(-)})^i_{[k,l]}. \qquad (6.30)$$

In summary, the off-shell closure of the $N = (4,4)$ supersymmetry algebra in $4d$-dimensional target space with $d > 1$ implies that, in the sections that are not multiplying zeros, the transformation matrices are commuting (6.29), integrable (6.25) and simultaneously integrable (6.30) structures, and further that the products $UV$ and $VU$ equal minus the identity (6.27) [II].

### 6.3.2 Yano f-structures

As just discussed, the algebraic constraints hold only in the sections that are not projected out by the semichiral constraints. By inserting zeros in the arbitrary columns of the transformation matrices (6.5)-(6.6), the constraints hold for the full structures. The drawback of full integrability is degeneracy of the matrices,

$$\begin{aligned} V^{(+)} U^{(+)} &= -\text{diag}\,(0,1,1,1), & U^{(+)} V^{(+)} &= -\text{diag}\,(1,0,1,1), \\ V^{(-)} U^{(-)} &= -\text{diag}\,(1,1,0,1), & U^{(-)} V^{(-)} &= -\text{diag}\,(1,1,1,0). \end{aligned} \qquad (6.31)$$

Instead of complex structures, one is lead to define a generalization of complex structures allowing for degeneracy,

$$f(f^2 + 1) = 0, \qquad (6.32)$$

so called Yano $f$-structures [Yan61]. The $f$-structures are endomorphisms of $TM \oplus TM$ and can be defined as the $8d \times 8d$ matrices

$$f^{(\pm)} = \begin{pmatrix} 0 & U^{(\pm)} \\ V^{(\pm)} & 0 \end{pmatrix}. \qquad (6.33)$$

Two complementary projection operators can be defined from the $f$-structures as $m = 1 + f^2$ and $l = -f^2$ for both directions $f^{(\pm)}$, satisfying $l + m = 1$ and $lm = 0$ in addition to $fl = lf = f$ and $fm = mf = 0$. The operator $l$ defines the so called first fundamental distribution with dimension $6d$ and $m$ the second fundamental distribution, with dimension $2d$. The integrability of the distributions is given by the vanishing of the Nijenhuis tensor contracted with the projection operators [IY64]; the first and the second fundamental distributions are integrable if and only if

$$m^i_l \mathcal{N}(f)^l_{jk} = 0, \quad \mathcal{N}(f)^i_{jk} m^j_l m^k_n = 0 \qquad (6.34)$$

hold, respectively. Given the off-shell algebra of the transformation matrices, these integrability constraints are satisfied, such that the $f$-structures are integrable [I].

## 6.3.3 Twisted supersymmetry

As in the case with four-dimensional target space treated in the previous section, the algebra could just as well be closed to a pseudo-supersymmetry. The difference would be that the products of the transformation matrices (6.27) would have to equal one instead of minus one. The corresponding structures (6.33) would then be generalizations of product structures instead of complex structures [I],

$$f(f^2 - 1) = 0, \tag{6.35}$$

so called $f$-structures of hyperbolic type [Kir88].

## 6.4 Action invariance

### 6.4.1 Action invariance in arbitrary dimensional target space

The action (6.1) is invariant under the supersymmetry transformations (6.3) if and only if the generalized Kähler potential satisfies the partial differential equations

$$\left(K_{i[j}U^{(\pm)i}{}_{k]} + K_i U^{(\pm)i}{}_{[j,k]}\right)\bar{\mathbb{D}}_\pm \mathbb{X}^j \bar{\mathbb{D}}_\pm \mathbb{X}^k = 0 \tag{6.36}$$

and analogously for $V^{(\pm)}$. Expressed at the level of the component functions, the equations are

$$(K_a f^a_{[j} + K_{a'} \tilde{g}^{a'}_{[j} + K_{\bar{a}'} \bar{\tilde{h}}^{\bar{a}'}_{[j})_{k]} = 0 \tag{6.37}$$

and similar for $U^{(-)}$ and $V^{(\pm)}$ [II]. In the case of linear transformations closing to a pseudo-supersymmetry and four-dimensional target space, this reduces to the system of equations (6.13) which were solved in section 6.2.

The system of partial differential equations (6.37) simplify and are more transparent in a four-dimensional target space, where one can explicitly see how they are required for the underlying bihermitian structures to be covariantly constant and satisfy hermiticity conditions, and solutions can be found.

### 6.4.2 Action invariance in four-dimensional target space

The system of partial differential equations simplify in a four-dimensional target space. In a target space parametrized by only one left and one right semichiral field (and their complex counterparts), the indices $i = \{a, \bar{a}, a', \bar{a}'\}$ can be omitted. As a consequence, all lower indices will be derivatives and thus commute, no curl-terms of the kind $g_{[a,b]}$ will arise. The invariance of the action under the $U^{(+)}$ supersymmetry transformation will thus result in the partial differential equations

$$f_r K_{\ell\bar\ell} - f_{\bar\ell} K_{\ell r} + \tilde g K_{\bar\ell r} = 0,$$
$$f_{\bar r} K_{\ell\bar\ell} - f_{\bar\ell} K_{\ell\bar r} + \bar{\tilde h} K_{\bar\ell\bar r} = 0,$$
$$(\tilde g - \bar{\tilde h}) K_{r\bar r} - f_{\bar r} K_{r\ell} + f_r K_{\bar r \ell} = 0, \qquad (6.38)$$

and the $U^{(-)}$ transformations will imply the equations

$$\tilde f_\ell K_{r\bar r} - \tilde f_{\bar r} K_{r\ell} + g K_{\bar r \ell} = 0,$$
$$\tilde f_{\bar\ell} K_{r\bar r} - \tilde f_{\bar r} K_{r\bar\ell} + \bar h K_{\bar r \bar\ell} = 0,$$
$$(g - \bar h) K_{\ell\bar\ell} - \tilde f_{\bar\ell} K_{\ell r} + \tilde f_\ell K_{\bar\ell r} = 0. \qquad (6.39)$$

The invariance of the transformations defined by the matrices $V^{(\pm)}$ imply the complex conjugates of these equations. Assuming that the system of equations (6.38)-(6.39) have a solution and rewriting them for some of the parameters in terms of the others gives

$$f_r = \frac{1}{K_{\ell\bar\ell}}\left(K_{\ell r} f_{\bar\ell} - K_{\bar\ell r} \tilde g\right), \qquad \tilde f_\ell = \frac{1}{K_{r\bar r}}(K_{\ell r} \tilde f_{\bar r} - K_{\ell \bar r} g),$$
$$f_{\bar r} = \frac{1}{K_{\ell\bar\ell}}\left(K_{\ell\bar r} f_{\bar\ell} - K_{\bar\ell\bar r} \bar{\tilde h}\right), \qquad \tilde f_{\bar\ell} = \frac{1}{K_{r\bar r}}(K_{\bar\ell r} \tilde f_{\bar r} - K_{\bar\ell\bar r} \bar h),$$
$$\bar{\tilde h} = \frac{K_{\ell\bar\ell} K_{r\bar r} - |K_{\ell\bar r}|^2}{K_{\ell\bar\ell} K_{r\bar r} - |K_{\ell r}|^2} \tilde g, \qquad \bar h = \frac{K_{\ell\bar\ell} K_{r\bar r} - |K_{\ell\bar r}|^2}{K_{\ell\bar\ell} K_{r\bar r} - |K_{\ell r}|^2} g. \qquad (6.40)$$

From the definition of the transformations in (6.3), the lower indices are derivatives and must satisfy, e.g., $f_{[r,\bar r]} = 0$. Leaving $f_{\bar\ell}$ and $\tilde g$ to be free parameters as in (6.40) and assuming that partial derivatives commute, $f_{[\bar\ell,i]} = 0$, one can check that the solutions in (6.40) satisfy $f_{[i,j]} = 0$ for all indices. The constraints in (6.40) are necessary to close the algebra on-shell, as will be discussed in section 6.5.2.

Action invariance is usually related to hermiticity and covariant constancy of the geometrical structures. The same is true here, as will be clear in section 6.6; the connection to underlying structures that are covariantly constant and bihermitian can be made if the relations in (6.40) are used [IV]. The same is valid for the integrability; integrability of the transformations matrices $U^{(\pm)}$, and the identification with the integrable complex structures $J_i^{(\pm)}$, will follow if the relations (6.40) hold, as will be discussed in sections 6.5.2 and 6.6.

## 6.5 On-shell supersymmetry

Sections 6.2-6.3 showed that off-shell closure of the $N=(4,4)$ supersymmetry is only possible in a target space parametrized by at least two sets of left and right semichiral fields. In this section, the results from paper [II] and paper [IV] will be reviewed, showing that the supersymmetry algebra closes on-shell independently of the target space dimension.

The field equations for the semichiral sigma model (6.1) is

$$\bar{\mathbb{D}}_+ K_a = 0, \quad \bar{\mathbb{D}}_- K_{a'} = 0, \tag{6.41}$$

plus the complex conjugate expressions. As before, the indices are derivatives with respect to the left and right semichiral fields $\mathbb{X}^a$ and $\mathbb{X}^{a'}$, respectively.

### 6.5.1 Using the underlying bihermitian structures

As reviewed in section 4.2, a $N=(1,1)$ supersymmetric sigma model admits $N=(4,4)$ supersymmetry if and only if the target space has a set of six complex structures $J_i^{(\pm)}$, $i = 1,2,3$, satisfying the quaternion algebra (4.33). Invariance of the action under the supersymmetry further requires the complex structures to be covariantly constant with respect to a torsionful connection, and the metric to be hermitian with respect to all of them. In other words, $N=(4,4)$ supersymmetry requires the target space to be bihyperhermitian.

In paper [II], field equations were used to relate the transformation matrices of the $N=(4,4)$ supersymmetry to the underlying bihyperhermitian structures. This identification then provides a proof for the on-shell closure of the $N=(4,4)$ supersymmetry. When analyzing the closure of the algebra, the existence of the underlying complex structures can be used without restriction. The hermiticity of the target space metric and the covariant constancy of the complex structures can only be used, however, if the action is assumed to be invariant.

The semichiral constraints can be written using the real operators defined in (4.48) as

$$Q_+ \mathbb{X}^L = J D_+ \mathbb{X}^L, \quad Q_- \mathbb{X}^R = J D_- \mathbb{X}^R, \tag{6.42}$$

as discussed in section 4.3.2. Writing the field equations (6.41) in terms of the real operators defined in (4.37) implies the relations

$$J K_{LR} Q_+ \mathbb{X}^R = \frac{1}{2} J C_{LL} D_+ \mathbb{X}^L - K_{LR} D_+ \mathbb{X}^L - \frac{1}{2} C_{LL} Q_+ \mathbb{X}^L,$$

$$J K_{RL} Q_- \mathbb{X}^L = \frac{1}{2} J C_{RR} D_- \mathbb{X}^R - K_{RL} D_- \mathbb{X}^L - \frac{1}{2} C_{RR} Q_- \mathbb{X}^R, \tag{6.43}$$

where the matrices $K_{LR}$ and $K_{RL}$ are defined as in (4.64), and the remaining matrices are defined as $C_{LL} = [J, K_{LL}]$ and $C_{RR} = [J, K_{RR}]$. Using the semichiral constraints (6.42) and solving for $Q_+ \mathbb{X}^R$ and $Q_- \mathbb{X}^L$ gives

$$Q_\pm \mathbb{X}^i = J^{(\pm)i}{}_j D_\pm \mathbb{X}^j, \tag{6.44}$$

where the complex structures $J^{(\pm)}$ take the well-known form as in (4.65). The obtained expression can further be used to relate the complex derivatives $\mathbb{D}$ and the real derivatives $D$ using a projection operator $\pi$ as

$$\bar{\mathbb{D}}\mathbb{X} = \pi D\mathbb{X}, \quad \pi^{(\pm)} = \frac{1}{2}(1 + iJ^{(\pm)}). \tag{6.45}$$

Note that both relations (6.44) and (6.45) hold only on-shell.

The supersymmetry transformations written in terms of the transformation matrices $U^{(\pm)}$ in (6.7) can now be related to the $N=(4,4)$ transformations written in terms of the bihypercomplex structures in (4.32). In $N=(1,1)$ superspace formalism, the non-manifest $N=(4,4)$ supersymmetry transformations take the form $\delta X = \epsilon_i^\alpha J_i^{(\alpha)} D_\alpha X$, where $\alpha = +, -$ and $i = 1,2,3$. The three complex structures form the basis of an $SU(2)$ worth of complex structures. Choosing the complex structures corresponding to the $N=(2,2)$ transformations as $J_3^{(\pm)} = J^{(\pm)}$, the additional transformations for the $N=(4,4)$ supersymmetry are

$$\delta X = \epsilon_1^\alpha J_1^{(\alpha)} D_\alpha X + \epsilon_2^\alpha J_2^{(\alpha)} D_\alpha X. \tag{6.46}$$

Using the on-shell conditions (6.44)-(6.45), the $N=(4,4)$ supersymmetry transformations (6.7) on the semichiral fields are

$$\begin{aligned}\delta\mathbb{X} &= \bar{\epsilon}^\alpha U^{(\alpha)} \bar{\mathbb{D}}_\alpha \mathbb{X} + \epsilon^\alpha V^{(\alpha)} \mathbb{D}_\alpha \mathbb{X} \\ &= \bar{\epsilon}^\alpha U^{(\alpha)} \pi^{(\alpha)} D_\alpha \mathbb{X} + \epsilon^\alpha V^{(\alpha)} \bar{\pi}^{(\alpha)} D_\alpha \mathbb{X} \\ &= \epsilon_1^\alpha \left(U^{(\alpha)} \pi^{(\alpha)} + V^{(\alpha)} \bar{\pi}^{(\alpha)}\right) D_\alpha \mathbb{X} + \epsilon_2^\alpha i\left(U^{(\alpha)} \pi^{(\alpha)} - V^{(\alpha)} \bar{\pi}^{(\alpha)}\right) D_\alpha \mathbb{X}. \end{aligned} \tag{6.47}$$

Denoting the $N=(1,1)$ superspace fields of the semichiral fields as in (4.50) and comparing the lowest $N=(1,1)$ superspace component $D\mathbb{X}| = DX$ with the transformations in (6.46) enables the identification of the transformation matrices in terms of the underlying bihermitian structures as

$$\begin{aligned}J_1^{(\pm)} &= U^{(\pm)} \pi^{(\pm)} + V^{(\pm)} \bar{\pi}^{(\pm)}, \\ J_2^{(\pm)} &= i\left(U^{(\pm)} \pi^{(\pm)} - V^{(\pm)} \bar{\pi}^{(\pm)}\right). \end{aligned} \tag{6.48}$$

The bihyperhermitian structures $J_i^{(\pm)}$ satisfy the quaternion algebra (4.33) and are all integrable. Further, assuming that the action is invariant under the supersymmetry transformations, they are all covariantly constant with respect to a torsionful connection. These facts together with the on-shell constraints can now be used to show that the algebra in (6.22)-(6.24) closes to a supersymmetry on-shell.

The identification in (6.48) together with the fact that the $J_i^{(\pm)}$ are hypercomplex structures imply a number of useful identities for the projection operators in (6.45) and the transformation matrices,

$$\begin{array}{llll} \pi\bar{\pi} = 0, & J_1\pi = U\pi, & \bar{\pi}U\pi = U\pi, & VU\pi = -\pi, \\ \bar{\pi}\pi = 0 & J_1\bar{\pi} = V\bar{\pi}, & \pi V\bar{\pi} = V\bar{\pi}, & UV\bar{\pi} = -\bar{\pi}, \end{array} \tag{6.49}$$

where all identities are valid for structures of the same chirality, $\pi\bar{\pi} = \pi^{(\pm)}\bar{\pi}^{(\pm)}$. Using the on-shell identification in (6.48) and the related identities in (6.49) together with the fact that $J_i^{(\pm)}$ are hypercomplex structures, implies that the expression in (6.24) closes to a translation,

$$\mathcal{M}(U,V)^i_{jk}\bar{\mathbb{D}}\mathbb{X}^j\mathbb{X}^k - (UV)^i_j\bar{\mathbb{D}}\mathbb{D}\mathbb{X}^j - (VU)^i_j\mathbb{D}\bar{\mathbb{D}}\mathbb{X}^j$$
$$= -\bar{\mathbb{D}}\left[(UV)^i_j\mathbb{D}\mathbb{X}^j\right] - \mathbb{D}\left[(VU)^i_j\bar{\mathbb{D}}\mathbb{X}^j\right] + \left(U^l_j V^i_{[k,l]} - V^l_k U^i_{[j,l]}\right)\bar{\mathbb{D}}\mathbb{X}^j\mathbb{D}\mathbb{X}^k$$
$$= i\partial\mathbb{X}^i, \qquad (6.50)$$

where the last terms vanish due to integrability of the complex structures.

For the vanishing of the terms (6.23), the invariance of the action must also be evoked. A covariant expression for the Magri-Morosi concomitant can be introduced as

$$\widehat{\mathcal{M}}(I_1,I_2)^i_{jk} = (I_1)^l_j \nabla^{(2)}_l (I_2)^i_k - (I_2)^l_k \nabla^{(1)}_l (I_1)^i_j - (I_1)^i_l \nabla^{(2)}_j (I_2)^l_k + (I_2)^i_l \nabla^{(1)}_k (I_1)^l_j$$
$$= \mathcal{M}(I_1,I_2)^i_{jk} + \tfrac{1}{2}[I_1,I_2]^i_l \left(\Gamma^{(2)l}_{jk} + \Gamma^{(1)l}_{kj}\right), \qquad (6.51)$$

where $I_{1,2}$ are any of the bihyperhermitian structures $J_i^{(\pm)}$ and the covariant derivatives $\nabla^{(\pm)}$ differ on the sign of the torsion and act on the corresponding complex structure, $\nabla^{(\pm)}J_i^{(\pm)}$, i.e., if $I_1 = U^{(+)}$, then $\nabla^{(1)} = \nabla^{(+)}$. The covariance of the hermitian structures implies that both $\widehat{\mathcal{M}}(J^{(+)},J^{(-)}) = 0$ and that $\widehat{\mathcal{M}}(U^{(+)},U^{(-)}) = \widehat{\mathcal{M}}(U^{(+)},V^{(-)}) = 0$. Using this object, the vanishing of the algebraic expression (6.23) can be rewritten as

$$\widehat{\mathcal{M}}(U^{(\pm)},U^{(\mp)})^i_{jk}\bar{\mathbb{D}}_\pm\mathbb{X}^j\bar{\mathbb{D}}_\mp\mathbb{X}^k = [U^{(\pm)},U^{(\mp)}]^i_j \bar{\nabla}^{(\mp)}_\pm \bar{\mathbb{D}}_\mp\mathbb{X}^j, \qquad (6.52)$$

where the left-hand side vanishes due to the covariance with respect to a torsionful connection. The right-hand side is a commutator contracted with a field equation, defined in terms of the supersymmetry derivatives as

$$\bar{\nabla}^{(\mp)}_\pm \bar{\mathbb{D}}_\mp\mathbb{X} = \bar{\mathbb{D}}_\pm\bar{\mathbb{D}}_\mp\mathbb{X} + \Gamma^{(\mp)}\bar{\mathbb{D}}_\pm\mathbb{X}\bar{\mathbb{D}}_\mp\mathbb{X}. \qquad (6.53)$$

That the vanishing of this expression is a field equation can be seen by first rewriting it in terms of the real derivatives as

$$\bar{\nabla}^{(\mp)}_\pm \bar{\mathbb{D}}_\mp\mathbb{X}^i = \tfrac{1}{2}\{\pi^{(\pm)},\pi^{(\mp)}\}^i_j \nabla^{(\mp)}_\pm D_\mp\mathbb{X}^j. \qquad (6.54)$$

From the anti-commutator of supersymmetry generators of opposite chirality (3.7) one can deduce that the expression vanishes on-shell,

$$0 = \{Q_+, Q_-\}\mathbb{X} = [J^{(-)},J^{(+)}]\nabla^{(-)}_+ D_-\mathbb{X}. \qquad (6.55)$$

Since the $J^{(+)}$ and $J^{(-)}$ do not commute in a manifold parametrized by semichiral coordinates, $\nabla D\mathbb{X} = 0$ is satisfied on-shell, and the right-hand side of (6.52) vanishes. Thus, the algebraic constraints (6.23) are satisfied on-shell, using the identification with the underlying bihyperhermitian structures and requiring that the action is invariant.

Finally, the Nijenhuis expressions (6.22) have to vanish. It is expected that the closure should follow due to the on-shell identification of the transformation matrices with the underlying bihyperhermitian structures. However, the attempts to show this failed in paper [II], and the vanishing of the Nijenhuis tensor for the transformation matrices had to be imposed as an extra condition.

All in all, the on-shell identification with the underlying bihyperhermitian structure implies that the algebra closes on-shell, provided that the transformation matrices are integrable in terms of a vanishing Nijenhuis tensor. The additional constraint of integrability is unsatisfactory, however, and it would be desirable to be able to derive the results explicitly in the $N=(2,2)$ formalism, without making use of real $N=(1,1)$ operators. This is precisely what was done in paper [IV], where it was shown that the Nijenhuis tensor indeed vanishes on-shell, and will be reviewed next.

### 6.5.2 On-shell closure explicitly

The main quest of paper [IV] was to see explicitly how the $N=(4,4)$ algebra closes on-shell for a semichiral sigma model. Since paper [II] showed that off-shell closure cannot occur in a four-dimensional target space, paper [IV] focused on a semichiral sigma model with four-dimensional target space.

The ansatz for the $N=(4,4)$ supersymmetry takes the general form (6.3), but since there is only one set of left and right semichiral field, the indices can be omitted,

$$\delta \mathbb{X}^\ell = \bar{\epsilon}^+ \bar{\mathbb{D}}_+ f(\mathbb{X}^\ell, \bar{\mathbb{X}}^{\bar{\ell}}, \mathbb{X}^r, \bar{\mathbb{X}}^{\bar{r}}) + g(\mathbb{X}^\ell)\bar{\epsilon}^- \bar{\mathbb{D}}_- \mathbb{X}^\ell + h(\mathbb{X}^\ell)\epsilon^- \mathbb{D}_- \mathbb{X}^\ell,$$
$$\delta \mathbb{X}^r = \bar{\epsilon}^- \bar{\mathbb{D}}_- \tilde{f}(\mathbb{X}^\ell, \bar{\mathbb{X}}^{\bar{\ell}}, \mathbb{X}^r, \bar{\mathbb{X}}^{\bar{r}}) + \tilde{g}(\mathbb{X}^r)\bar{\epsilon}^+ \bar{\mathbb{D}}_+ \mathbb{X}^r + \tilde{h}(\mathbb{X}^r)\epsilon^+ \mathbb{D}_+ \mathbb{X}^r. \quad (6.56)$$

As seen in section 6.4, the invariance of the action under these transformations constrains the Lagrangian $K(\mathbb{X}^i)$ to satisfy a system of partial differential equations (6.38)-(6.39). In (6.40), some of the transformation parameters were solved in terms of the others and second derivatives of $K$. This is the key feature that together with the field equations provides on-shell closure of the algebra.

First, two subsequent transformations in the minus direction commute on the left semichiral field to

$$[\delta_1^{(-)}, \bar{\delta}_2^{(-)}]\mathbb{X}^\ell = \bar{\epsilon}_{[2}^- \epsilon_{1]}^- (-gh) i\partial_= \mathbb{X}^\ell. \quad (6.57)$$

Closing this to a translation, and similarly closing the plus-part of the algebra for the right semichiral field, implies

$$gh = -1, \quad \tilde{g}\tilde{h} = -1. \quad (6.58)$$

These algebraic constraints together with the relation between $\bar{h}$ and $g$ in (6.40) then imply an important result regarding the geometry of the model. The function $c$ in (4.78) takes the expression

$$c = \frac{1-|g|^2}{1+|g|^2} = \frac{1-|\tilde{g}|^2}{1+|\tilde{g}|^2} \qquad (6.59)$$

and further, the geometry is non-degenerate, $\det(K_{LR}) \neq 0$. Since $g$ and $\tilde{g}$ are functions of only the left and the right semichiral field, respectively, equation (6.59) implies that $c$ must be constant, hence the geometry is torsion-free.

The rest of the algebra closes on-shell without any further constraints. The field equations (6.41) provide a relation between the derivatives of the semichiral fields,

$$K_{\ell\bar{\ell}}\bar{\mathbb{D}}_+\bar{\mathbb{X}}^{\bar{\ell}} + K_{\ell r}\bar{\mathbb{D}}_+\mathbb{X}^r + K_{\ell\bar{r}}\bar{\mathbb{D}}_+\bar{\mathbb{X}}^{\bar{r}} = 0,$$
$$K_{r\bar{r}}\bar{\mathbb{D}}_-\bar{\mathbb{X}}^{\bar{r}} + K_{r\ell}\bar{\mathbb{D}}_-\mathbb{X}^\ell + K_{r\bar{\ell}}\bar{\mathbb{D}}_-\bar{\mathbb{X}}^{\bar{\ell}} = 0. \qquad (6.60)$$

The terms (6.22) involve the Nijenhuis tensors for the transformation matrices and close on-shell using the identifications in (6.40) and the fact that $c$ is constant,

$$[\bar{\delta}_1^{(+)}, \bar{\delta}_2^{(+)}]\mathbb{X}^\ell = \bar{\epsilon}_{[2}^+\bar{\epsilon}_{1]}^+ \left[\mathcal{N}(U^{(+)})^\ell_{jk}\bar{\mathbb{D}}_+\mathbb{X}^j\bar{\mathbb{D}}_+\mathbb{X}^k\right]$$
$$= \bar{\epsilon}_{[2}^+\bar{\epsilon}_{1]}^+ f_{\bar{\ell}} \frac{K_{r\bar{r}\ell}}{K_{\ell\bar{\ell}}}\left(\frac{2\tilde{g}}{c-1}\right)_{,\ell} \bar{\mathbb{D}}_+\mathbb{X}^r\bar{\mathbb{D}}_+\mathbb{X}^{\bar{r}}$$
$$= 0. \qquad (6.61)$$

To investigate the remaining terms (6.23)-(6.24), useful combinations of the transformation parameters can be defined as

$$\mu = f_{\bar{\ell}}\bar{f}_r + f_r\tilde{h}, \qquad \nu = f_{\bar{\ell}}\bar{f}_{\bar{r}} + f_{\bar{r}}\tilde{\bar{g}},$$
$$\tau = f_{\bar{\ell}}(g-\bar{h}) - f_r\tilde{f}_{\bar{\ell}}, \qquad \omega = f_{\bar{r}}g - \tilde{f}_{\bar{r}}f_r. \qquad (6.62)$$

The mixed terms in the algebra close as

$$[\bar{\delta}_1^{(+)}, \bar{\delta}_2^{(-)}]\mathbb{X}^\ell = \bar{\epsilon}_{[2}^+\bar{\epsilon}_{1]}^- \left[\mathcal{M}(U^{(+)}, U^{(-)})^\ell_{jk}\bar{\mathbb{D}}_+\mathbb{X}^j\bar{\mathbb{D}}_-\mathbb{X}^k - [U^{(+)}, U^{(-)}]^\ell_j\bar{\mathbb{D}}_+\bar{\mathbb{D}}_-\mathbb{X}^j\right]$$
$$= \bar{\epsilon}_{[2}^+\bar{\epsilon}_{1]}^- \bar{\mathbb{D}}_+ \left[\tau\bar{\mathbb{D}}_-\mathbb{X}^\ell + \omega\bar{\mathbb{D}}_-\mathbb{X}^{\bar{r}} + (-f_r\tilde{f}_{\bar{\ell}})\bar{\mathbb{D}}_-\mathbb{X}^\ell\right]$$
$$= \bar{\epsilon}_{[2}^+\bar{\epsilon}_{1]}^- \bar{\mathbb{D}}_+(g\tilde{g} - \tilde{g}\bar{h})\bar{\mathbb{D}}_-\mathbb{X}^\ell$$
$$= 0 \qquad (6.63)$$

and the plus-part of the transformations for $\mathbb{X}^\ell$ closes to a translation,

$$[\bar{\delta}_1^{(+)}, \delta_2^{(+)}]\mathbb{X}^\ell = \bar{\epsilon}_{[2}^+\epsilon_{1]}^+ \left[\mathcal{M}(U^{(+)}, V^{(+)})^\ell_{jk}\bar{\mathbb{D}}_+\mathbb{X}^j\mathbb{D}_+\mathbb{X}^k + (U^{(+)}V^{(+)})^\ell_j\bar{\mathbb{D}}_+\mathbb{D}_+\mathbb{X}^j\right]$$
$$= \bar{\epsilon}_{[2}^+\epsilon_{1]}^+ \bar{\mathbb{D}}_+ \left(-|f_{\bar{\ell}}|^2\mathbb{D}_+\mathbb{X}^\ell - \mu\mathbb{D}_+\mathbb{X}^r - \nu\mathbb{D}_+\mathbb{X}^{\bar{r}}\right)$$
$$= \bar{\epsilon}_{[2}^+\epsilon_{1]}^+ i\partial_{++}\mathbb{X}^\ell. \qquad (6.64)$$

The on-shell closure of the right semichiral fields follows completely analogously. To summarize, the action is invariant under the non-linear transformations (6.56) and the transformations closes to a $N=(4,4)$ supersymmetry algebra on-shell if and only if the partial differential equations in (6.38)-(6.39) are satisfied, together with the constraint (6.58). A result of this system of equations is that the metric is non-degenerate and that the target space geometry is torsion-free.

### 6.5.3 Linear on-shell supersymmetry

As presented in section 6.2, additional linear off-shell pseudo-supersymmetry leads to a semichiral sigma model with four-dimensional neutral hyperkähler target space, and non-trivial examples of this model could be constructed. However, supersymmetry imposed on the same model closes only on-shell, and the supersymmetry transformations (6.56) are non-linear. In contrast to the off-shell pseudo-supersymmetry, linear on-shell supersymmetry transformations necessarily lead to trivial solutions.

Consider a linear transformation that closes to a supersymmetry, schematically depicted as

$$\delta X = QX, \quad Q^2 X = [\delta, \delta]X = \partial X + F \tag{6.65}$$

where $F$ is a field equation that vanishes on-shell,

$$F = (Q^2 - \partial)X = 0. \tag{6.66}$$

This implies that the Lagrangian of the model must be quadratic. The same fact can be derived explicitly in the notation from the previous section. The partial differential equations (6.38)-(6.39) from the invariance of the action implies that the Lagrangian must satisfy, e.g.,

$$\mu K_{\ell r} = \bar{\nu} K_{\bar{\ell} r}. \tag{6.67}$$

If the transformations are linear, the coefficients $\mu$ and $\nu$ defined in (6.62) are constant, so (6.67) implies, when taking derivative with respect to $\bar{\mathbb{X}}^{\bar{r}}$ and again using the partial differential equations, that either $|\mu|^2 = |\nu|^2$, which leads to degenerate metric, or that $K_{\ell r \bar{r}} = 0$. The same can be shown for remaining third derivatives on $K$, hence a non-degenerate Lagrangian is quadratic.

Hence, linear transformations that close only on-shell will imply that the model is trivial in the sense that the metric is flat, and so any non-trivial semichiral sigma model with $N=(4,4)$ supersymmetry and four-dimensional target space has non-linear supersymmetry transformations. This will be an important fact when discussing the duality between a chiral and twisted chiral model with linear off-shell $N=(4,4)$ supersymmetry and a semichiral model with on-shell supersymmetry in chapter 8.

## 6.6 Hyperkähler geometry

As reviewed in section 4.4, a necessary condition for a four-dimensional manifold parametrized by semichiral fields to be hyperkähler is that $\{J^{(+)}, J^{(-)}\} = 2c$ with $c$ constant and $|c| < 1$.

In the first section of this chapter, we saw that linear pseudo-supersymmetry constrains a four-dimensional target space to have $c$ constant with $|c| > 1$, implying a pseudo-hyperkähler geometry.

In the four-dimensional target space analyzed in section 6.5.2, the algebraic constraints from the ordinary on-shell supersymmetry algebra closure implied a constant $c$ with $|c| < 1$. Hence, the geometry is hyperkähler and one should be able to identify the hyperkähler structures. This is straight-forward using the connection with the underlying bihyperhermitian geometry developed in section 6.5.1. There, the identification of the transformation matrices in terms of complex structures was given as

$$U^{(\pm)}\pi^{(\pm)} = \tfrac{1}{2}(J_1^{(\pm)} - iJ_2^{(\pm)}),$$
$$V^{(\pm)}\bar{\pi}^{(\pm)} = \tfrac{1}{2}(J_1^{(\pm)} + iJ_2^{(\pm)}). \qquad (6.68)$$

Choosing $J_3 = J^{(+)}$ and $J_1, J_2$ according to the recipe given in (4.75), the first of the structures in (6.68) takes the expression

$$\tfrac{1}{2}(J_1^{(+)} - iJ_2^{(+)}) =$$

$$\frac{1}{\Delta\sqrt{1-c^2}}\begin{pmatrix} -2K_{\ell\bar{\ell}}K_{r\bar{r}} & -2K_{\bar{\ell}r}K_{\ell\bar{r}} & -2K_{r\bar{r}}K_{\ell r} & -2K_{r\bar{r}}K_{\bar{l}\bar{r}} \\ 0 & 0 & 0 & 0 \\ (1-c)K_{\ell\bar{\ell}}K_{\ell\bar{r}} & (1-c)K_{\ell\bar{\ell}}K_{\bar{\ell}\bar{r}} & (1-c)K_{\ell\bar{r}}K_{\bar{\ell}r} & (1-c)K_{\bar{\ell}\bar{r}}K_{\ell\bar{r}} \\ (1+c)K_{\ell\bar{\ell}}K_{\ell r} & (1+c)K_{\ell\bar{\ell}}K_{\bar{\ell}r} & (1+c)K_{\ell r}K_{\bar{\ell}r} & (1+c)K_{\ell r}K_{\bar{\ell}\bar{r}} \end{pmatrix}$$
$$(6.69)$$

where $\Delta = \det(K_{LR})$. Using the equations from the invariance of the action (6.38)-(6.39), one can show that the matrix $U^{(+)}\pi^{(+)}$ takes the same expression as in (6.69) if the parameters $\tilde{g}$ and $\tilde{h}$ are phases, $\tilde{g} = -\tilde{h} = e^{i\alpha}$.

But there is an ambiguity in the choice of direction; if instead one chooses $J_3 = J^{(-)}$ and $J_1, J_2$ according to (4.75) but with $J^{(+)}$ and $J^{(-)}$ interchanged, then one can analogously identify $\tfrac{1}{2}(J_1^{(-)} - iJ_2^{(-)})$ with $U^{(-)}\pi^{(-)}$, again if the parameters $g$ and $h$ are phases, $g = -h = e^{i\beta}$. The ambiguity in the choice of the complex structure $J_3$ does not result in six independent complex structures; the latter structures $J_i^{(-)}$ are simply the former ones $J_i^{(+)}$ rotated among each other,

$$J_1^{(-)} = \sqrt{1-c^2}J_3^{(+)} + cJ_1^{(+)},$$
$$J_2^{(-)} = -J_2^{(+)},$$
$$J_3^{(-)} = \sqrt{1-c^2}J_1^{(+)} - cJ_3^{(+)}. \qquad (6.70)$$

Hence, the relation (6.68) has been verified. The fact that $g$ and $\tilde{g}$ are phases implies that $c$ vanishes, which in turn has the consequence that the generalized Kähler potential satisfies the Monge-Ampère equation.

### 6.6.1 Example

It is not evident that a non-trivial generalized Kähler potential and transformation parameters can be found such that the partial differential equations in (6.38)-(6.39) are satisfied. In paper [IV] an example fulfilling these constraints was found by the Legendre transform related to the construction developed in [LR83, HKLR87]. More examples of semichiral sigma models with $N=(4,4)$ supersymmetry and hyperkähler target space were constructed in paper [V]. These examples and more throughout discussions of dualities between supersymmetric sigma models will be given in chapter 8.

Starting from a function $F(x,v,\bar{v})$ of one real coordinate $x$ and one complex coordinate $v$, satisfying the Laplace equation, $F_{xx}+F_{v\bar{v}}=0$, a Kähler potential that is a function of combinations of semichiral fields can be constructed by a Legendre transform [BSvdLVG99],

$$K(z,y,\bar{y}) = F(x,v,\bar{v}) - xz - vy - \bar{v}\bar{y}. \tag{6.71}$$

Choosing $F$ to be the function $F = r - x\ln(x+r)$ where $r^2 = x^2 + 4v\bar{v}$, and solving the identities originating from the Legendre transform, the dual potential is

$$K = \tfrac{1}{2}e^{-\tfrac{i}{2}z}(1 - \tfrac{1}{4}y\bar{y}), \tag{6.72}$$

where $y$ and $z$ are combinations of the semichiral fields and define the isometry of the model,

$$y = \mathbb{X}^\ell + \bar{\mathbb{X}}^{\bar{\ell}} + 2\mathbb{X}^r, \quad z = \mathbb{X}^\ell - \bar{\mathbb{X}}^{\bar{\ell}}. \tag{6.73}$$

Transformation parameters that together with the potential (6.72) satisfy the system of partial differential equations (6.38)-(6.39) arising from the invariance of the action, can be chosen as

$$f = 2i\ln\left(\frac{2-i\bar{y}}{2-iy}\right), \quad \tilde{f} = -\tfrac{1}{2}z + \tfrac{i}{8}y^2, \quad g = \tilde{g} = -h = -\tilde{h} = 1. \tag{6.74}$$

The sigma model is invariant under the transformations (6.56) with the parameters defined as above, and the transformations close to a $N=(4,4)$ supersymmetry on-shell. The hyperkähler metric of this model is non-trivial and given by

$$g = \begin{pmatrix} g_{LL} & g_{LR} \\ g_{RL} & g_{RR} \end{pmatrix} \tag{6.75}$$

where the entries are the matrices

$$g_{LL} = \frac{ie^{-iz/2}(4+y\bar{y})}{16(y-\bar{y})} \begin{pmatrix} 4 - y\bar{y} - 2i(y+\bar{y}) & 4+y\bar{y} \\ 4+y\bar{y} & 4 - y\bar{y} + 2i(y+\bar{y}) \end{pmatrix},$$

$$g_{LR} = \frac{ie^{-iz/2}}{4(y-\bar{y})} \begin{pmatrix} (2-i\bar{y})(4+y\bar{y}-i(y-\bar{y})) & (2-iy)(4+y\bar{y}+i(y-\bar{y})) \\ (2+i\bar{y})(4+y\bar{y}+i(y-\bar{y})) & (2+iy)(4+y\bar{y}-i(y-\bar{y})) \end{pmatrix},$$

$$g_{RL} = (g_{LR})^t,$$

$$g_{RR} = \frac{ie^{-iz/2}}{(y-\bar{y})} \begin{pmatrix} 4+\bar{y}^2 & 4+y\bar{y} \\ 4+y\bar{y} & 4+y^2 \end{pmatrix}. \tag{6.76}$$

## 6.7 Results

In three separate papers, the research questions posed at the beginning of this chapter were analyzed and solved. The general ansatz for supersymmetry transformations on a semichiral sigma model is given in (6.3), or in a more compact notation in (6.5)-(6.7). It proved fruitful to analyze certain special cases of this ansatz and model.

In paper [I], it was found that a sigma model parametrized by one set of semichiral fields does not admit linear off-shell $N = (4,4)$ supersymmetry. (The same result for non-linear transformations was shown in [IV].) Only twisted supersymmetry can be imposed, resulting in a target space geometry that is neutral hyperkähler. The situation changes if the target space is enlarged. A sigma model parametrized by at least two sets of left and right semichiral fields admits off-shell $N = (4,4)$ supersymmetry [II]. The transformations can be analyzed geometrically in terms of Yano $f$-structures.

For a semichiral sigma model with four-dimensional target space, the additional supersymmetry can close only on-shell. For non-quadratic generalized Kähler potentials, the transformations are necessarily non-linear. The on-shell $N = (4,4)$ algebra is realized if and only if the potential satisfies a system of non-linear partial differential equations together with an additional algebraic constraint (6.58), implying that the target space geometry is hyperkähler [II, IV].

The scope of the papers analyzing semichiral sigma models with $N = (4,4)$ (twisted) supersymmetry can be summarized as in chart 6.1. Non-trivial examples of models with neutral hyperkähler geometry and also with on-shell $N = (4,4)$ supersymmetry and hyperkähler geometry were constructed.

Attempts were made to generalize the transformations in (6.56) to involve central charges and arrive at a four-dimensional target geometry with semichiral coordinates and non-vanishing torsion. In [HKLR86], an ansatz involving

|          | *Off-shell*                | *On-shell*                  |
|----------|----------------------------|-----------------------------|
| *4-dim*  | **Paper I**                | **Paper IV**                |
|          | (Section 6.2)              | (Section 6.4.2, 6.5.2, 6.6) |
|          | Only pseudo-susy           | On-shell susy               |
|          | $\to$ neutral hyperkähler  | $\to$ hyperkähler           |
| $\geq 8\ dim$ | **Paper II**          | **Paper II**                |
|          | (Section 6.1, 6.3, 6.4.1)  | (Section 6.5.1)             |
|          | Off-shell susy             | On-shell susy               |
|          | $\to$ $f$-structures       | $\to$ bihyperhermitian      |

*Figure 6.1:* Overview of the papers [I, II, IV], investigating semichiral sigma models and $N=(4,4)$ (twisted) supersymmetry.

central charges and acting on chiral superfield coordinates was constructed as

$$\delta \Phi^i = \bar{D}^2(\epsilon \Omega^i). \tag{6.77}$$

These transformations on the chiral fields close only on-shell and, as for the transformations of the semichiral fields considered here, the invariance of action has to be used to make the transformations (6.77) close to a supersymmetry. However, no generalization of the transformations in (6.56) similar to the one in [HKLR86] was found.

In paper [V], reviewed and discussed in chapter 8, a semichiral sigma model with $N=(4,4)$ supersymmetry is obtained by a duality transformation along a translational isometry from a chiral and twisted chiral sigma model. The resulting semichiral sigma model also has a hyperkähler target space geometry. A way to find a semichiral sigma model with $N=(4,4)$ supersymmetry and a torsionful four-dimensional target space geometry could be to dualize a chiral and twisted chiral sigma model along a rescaling isometry [Cri12].

# 7. Vector multiplets and N=(4,4) supersymmetry

> What has been done is little–scarcely a beginning; yet it is much in comparison with the total blank of a century past. And our knowledge will, we are easily persuaded, appear in turn the merest ignorance to those who come after us. Yet it is not to be despised, since by it we reach up groping to touch the hem of the garment of the Most High.
> *Agnes Mary Clerke, astronomer (1842-1907)*

In the previous chapter, results from the three papers [I, II, IV] analyzing semichiral models with $N=(4,4)$ (twisted) supersymmetry were presented. The question of how a semichiral model with non-linear on-shell $N=(4,4)$ supersymmetry can be related by T-duality to a chiral and twisted chiral model with linear off-shell supersymmetry was briefly touched upon, a question that will be studied in detail in the next chapter, based on paper [V].

T-duality between sigma models is obtained by gauging isometries of the Lagrangian, as was reviewed in chapter 5. The corresponding gauge potentials must be introduced in the Lagrangian to ensure invariance under the gauged isometry. The vector multiplets needed to gauge a translational isometry mixing chiral and twisted chiral fields, and one mixing semichiral fields, is the large vector multiplet, and the semichiral vector multiplet, respectively. The dualities corresponding to these multiplets were introduced in [GMST99] and [BSvdLVG99]; the vector multiplets were introduced in [GM07] and [LRR[+]07] and further analyzed in [Ryb07, LRR[+]09].

The conditions for additional $N=(4,4)$ supersymmetry of these two multiplets was investigated in paper [III]. The aim of the paper can be summarized in the following research questions.

RESEARCH QUESTIONS:
⋆ Does the semichiral vector multiplet allow for $N=(4,4)$ supersymmetry? How can the transformations be constructed, both in the case of abelian and non-abelian field strengths?
⋆ Can the large vector multiplet have $N=(4,4)$ supersymmetry? How can the similarities/differences from the semichiral vector multiplet be explained?

## 7.1 The semichiral vector multiplet

Recall the duality between an (anti-) chiral and a twisted (anti-) chiral sigma model, reviewed in section 5.3. Consider now a sigma model with an isometry mixing the semichiral fields, described by the Killing vector

$$k_{LR} = i(\partial_\ell - \partial_{\bar{\ell}} + \partial_r - \partial_{\bar{r}}), \tag{7.1}$$

specifying a direction in the tangent bundle. A generalized Kähler potential that is invariant along this direction can only depend on certain combinations of the semichiral fields, e.g.,

$$K = K\big(\phi, \bar{\phi}, \chi, \bar{\chi}, \mathbb{X}^\ell + \bar{\mathbb{X}}^{\bar{\ell}}, \mathbb{X}^r + \bar{\mathbb{X}}^{\bar{r}}, -i(\mathbb{X}^\ell - \bar{\mathbb{X}}^{\bar{\ell}} - \mathbb{X}^r + \bar{\mathbb{X}}^{\bar{r}})\big). \tag{7.2}$$

Under the gauged isometry, the semichiral fields transform into local semichiral gauge parameters, analogously to the gauge transformations in (5.28),

$$\delta \mathbb{X}^\ell = i\Lambda^\ell(\mathbb{X}^\ell), \quad \delta \mathbb{X}^r = i\Lambda^r(\mathbb{X}^r). \tag{7.3}$$

Invariance of the action under the gauged isometries is obtained by introducing the *semichiral vector multiplet* $(\mathbb{V}^\ell, \mathbb{V}^r, \mathbb{V}')$ [LRR+07, GM07]

$$K\big(\phi, \bar{\phi}, \chi, \bar{\chi}, \mathbb{X}^\ell + \bar{\mathbb{X}}^{\bar{\ell}} + \mathbb{V}^\ell, \mathbb{X}^r + \bar{\mathbb{X}}^{\bar{r}} + \mathbb{V}^r, -i(\mathbb{X}^\ell - \bar{\mathbb{X}}^{\bar{\ell}} - \mathbb{X}^r + \bar{\mathbb{X}}^{\bar{r}}) + \mathbb{V}'\big) \tag{7.4}$$

with transformation properties defined to cancel the gauge transformations of the semichiral fields,

$$\begin{aligned} \delta \mathbb{V}^\ell &= i(\bar{\Lambda}^{\bar{\ell}} - \Lambda^\ell), \\ \delta \mathbb{V}^r &= i(\bar{\Lambda}^{\bar{r}} - \Lambda^r), \\ \delta \mathbb{V}' &= -\Lambda^\ell - \bar{\Lambda}^{\bar{\ell}} + \Lambda^r + \bar{\Lambda}^{\bar{r}}. \end{aligned} \tag{7.5}$$

To construct gauge invariant field strengths, gauge potentials transforming into parameters with definite chirality properties must be defined,

$$\begin{aligned} \mathbb{V}_\phi &= \tfrac{1}{2}\big(i\mathbb{V}' + \mathbb{V}^\ell - \mathbb{V}^r\big), & \delta \mathbb{V}_\phi &= i(\Lambda^r - \Lambda^\ell), \\ \mathbb{V}_\chi &= \tfrac{1}{2}\big(i\mathbb{V}' + \mathbb{V}^\ell + \mathbb{V}^r\big), & \delta \mathbb{V}_\chi &= i(\bar{\Lambda}^{\bar{r}} - \Lambda^\ell). \end{aligned} \tag{7.6}$$

The potentials satisfy a reality constraint saying that their imaginary parts are the same,

$$\bar{\mathbb{V}}_\phi - \mathbb{V}_\phi = \bar{\mathbb{V}}_\chi - \mathbb{V}_\chi. \tag{7.7}$$

Field strengths can now be defined as

$$\mathbb{F} = i\bar{\mathbb{D}}_+ \mathbb{D}_- \mathbb{V}_\phi, \quad \tilde{\mathbb{F}} = i\bar{\mathbb{D}}_+ \mathbb{D}_- \mathbb{V}_\chi \tag{7.8}$$

together with their complex conjugates. The field strengths are chiral and twisted chiral, respectively. Reducing to $N = (1,1)$ superspace, the semichiral gauge multiplet is described by one $N = (1,1)$ gauge multiplet and three real unconstrained $N = (1,1)$ scalar superfields [LRR+07].

### 7.1.1 Non-abelian semichiral multiplet

The finite version of the infinitesimal gauge transformations of the fields in (7.3) is $\mathbb{X}^\ell \to e^{i\Lambda^\ell} \mathbb{X}^\ell$ and similar for the right semichiral field. If the gauge parameters take value in a non-commutative Lie group, analogous to the transformations in (5.35), the non-abelian extension of the gauge transformations in (7.5) are [LRR+07]

$$e^{\mathbb{V}^\ell} \to e^{i\bar{\Lambda}^{\bar{\ell}}} e^{\mathbb{V}^\ell} e^{-i\Lambda^\ell}, \quad e^{\mathbb{V}^r} \to e^{i\bar{\Lambda}^{\bar{r}}} e^{\mathbb{V}^r} e^{-i\Lambda^r}. \tag{7.9}$$

The non-abelian generalization of the gauge transformations for $e^{\mathbb{V}'}$ is not as straight-forward. Instead, the equivalent description

$$e^{\mathbb{V}_\phi} \to e^{i\Lambda^r} e^{\mathbb{V}_\phi} e^{-i\Lambda^\ell}, \quad e^{\mathbb{V}_\chi} \to e^{i\bar{\Lambda}^{\bar{r}}} e^{\mathbb{V}_\chi} e^{-i\Lambda^\ell}. \tag{7.10}$$

is used, subject to a reality constraint as in the abelian case. The non-abelian version of the reality constraint (7.7) on the gauge potentials reads

$$e^{\mathbb{V}^\ell} = e^{\bar{\mathbb{V}}_\chi} e^{\mathbb{V}_\phi} = e^{\bar{\mathbb{V}}_\phi} e^{\mathbb{V}_\chi},$$
$$e^{\mathbb{V}^r} = e^{-\bar{\mathbb{V}}_\phi} e^{\bar{\mathbb{V}}_\chi} = e^{\mathbb{V}_\chi} e^{-\mathbb{V}_\phi}. \tag{7.11}$$

The two expressions are equivalent and can be obtained from each other.

Covariant derivatives can generally be written in terms of the gauge potentials $V$ and the ordinary (supersymmetry) derivatives as

$$\nabla = e^{-V} D e^V = D + \Gamma, \tag{7.12}$$

where $\Gamma$ is the connection. In the left semichiral representation, covariant derivatives are defined to transform as $\nabla \to e^{i\Lambda^\ell} \nabla e^{-i\Lambda^\ell}$, and can be constructed as

$$\bar{\nabla}_+ = \bar{\mathbb{D}}_+, \qquad \bar{\nabla}_- = e^{-\mathbb{V}_\phi} \bar{\mathbb{D}}_- e^{\mathbb{V}_\phi},$$
$$\nabla_+ = e^{-\mathbb{V}^\ell} \mathbb{D}_+ e^{\mathbb{V}^\ell}, \qquad \nabla_- = e^{-\mathbb{V}_\chi} \mathbb{D}_- e^{\mathbb{V}_\chi}. \tag{7.13}$$

A symmetric and real representation can be obtained by introducing two gauge potentials satisfying

$$e^{\mathbb{V}_\phi} = e^{-\mathbb{U}_R} e^{\mathbb{U}_L}, \quad e^{\mathbb{V}_\chi} = e^{\bar{\mathbb{U}}_R} e^{\mathbb{U}_L} \tag{7.14}$$

with corresponding transformation properties

$$e^{\mathbb{U}_L} \to e^{iK} e^{\mathbb{U}_L} e^{-i\Lambda^\ell}, \quad e^{\mathbb{U}_R} \to e^{iK} e^{\mathbb{U}_L} e^{-i\Lambda^r}, \tag{7.15}$$

where $K$ is an arbitrary real Lie-algebra valued parameter. Covariant derivatives transforming as $\nabla \to e^{iK}\nabla e^{-iK}$ can be defined in a symmetric setting using these gauge potentials as

$$\bar{\nabla}_+ = e^{\mathbb{U}_L}\bar{\mathbb{D}}_+ e^{-\mathbb{U}_L}, \qquad \bar{\nabla}_- = e^{\mathbb{U}_R}\bar{\mathbb{D}}_- e^{-\mathbb{U}_R},$$
$$\nabla_+ = e^{-\bar{\mathbb{U}}_L}\mathbb{D}_+ e^{\bar{\mathbb{U}}_L}, \qquad \nabla_- = e^{-\bar{\mathbb{U}}_R}\mathbb{D}_- e^{\bar{\mathbb{U}}_R}. \qquad (7.16)$$

The non-abelian form of the field strengths in (7.8) can be expressed in terms of the covariant derivatives as

$$\mathbb{F} = i\{\bar{\nabla}_+, \bar{\nabla}_-\}, \quad \tilde{\mathbb{F}} = i\{\bar{\nabla}_+, \nabla_-\}. \qquad (7.17)$$

The expressions are independent of the representation and gives the usual interpretation of the field strengths as a curvature.

### 7.1.2 Additional supersymmetry

In the abelian case, linear transformations acting on the field strengths and closing to a $N=(4,4)$ supersymmetry can be constructed as

$$\delta\mathbb{F} = \epsilon^+\bar{\mathbb{D}}_+\tilde{\bar{\mathbb{F}}} + \epsilon^-\bar{\mathbb{D}}_-\tilde{\mathbb{F}},$$
$$\delta\tilde{\mathbb{F}} = -\epsilon^+\bar{\mathbb{D}}_+\bar{\mathbb{F}} - \bar{\epsilon}^-\mathbb{D}_-\mathbb{F}, \qquad (7.18)$$

where the transformations are to be seen as commutators. The transformations close to a supersymmetry (3.18) off-shell. The corresponding supersymmetry transformations for the gauge potentials can be derived using the expressions of the field strengths in terms of the potentials (7.8), together with the reality constraint (7.7),

$$\delta\mathbb{V}_\phi = -\epsilon^+\mathbb{D}_+\bar{\mathbb{V}}_\chi - \epsilon^-\mathbb{D}_-\mathbb{V}_\chi + \bar{\epsilon}^-\bar{\mathbb{D}}_-\mathbb{V}_\phi - \bar{\epsilon}^+\bar{\mathbb{D}}_+\mathbb{V}_\phi,$$
$$\delta\mathbb{V}_\chi = \epsilon^+\mathbb{D}_+\bar{\mathbb{V}}_\phi + \bar{\epsilon}^-\bar{\mathbb{D}}_-\mathbb{V}_\phi + \bar{\epsilon}^+\bar{\mathbb{D}}_+\mathbb{V}_\chi - \epsilon^-\mathbb{D}_-\mathbb{V}_\chi. \qquad (7.19)$$

From the definitions of the field strengths in (7.8), it is clear that any chiral term in the transformations for $\mathbb{V}_\phi$, and similarly any twisted chiral term in the transformations $\mathbb{V}_\chi$, will not appear in the physical spectra and are to be viewed as gauge transformations of the form (7.6). With this observation, the transformations for the gauge potentials close to a supersymmetry up to a gauge transformation,

$$[\delta(\epsilon_1), \delta(\epsilon_2)]\begin{pmatrix}\mathbb{V}_\phi\\\mathbb{V}_\chi\end{pmatrix} = \bar{\epsilon}^\pm_{[2}\epsilon^\pm_{1]}i\partial_{\pm\pm}\begin{pmatrix}\mathbb{V}_\phi\\\mathbb{V}_\chi\end{pmatrix} + \begin{pmatrix}\Lambda^r - \Lambda^\ell + 2\tilde{\alpha}\tilde{\mathbb{F}} + 2\bar{\tilde{\alpha}}\bar{\tilde{\mathbb{F}}}\\\bar{\Lambda}^{\bar{r}} - \Lambda^\ell + 2\alpha\mathbb{F} + 2\bar{\alpha}\bar{\mathbb{F}}\end{pmatrix}, \qquad (7.20)$$

where supersymmetry parameters are labeled collectively as $\alpha = i\bar{\epsilon}^+_{[2}\bar{\epsilon}^-_{1]}$ and $\tilde{\alpha} = i\bar{\epsilon}^+_{[2}\epsilon^-_{1]}$ and the gauge transformations are defined as

$$\begin{aligned}\Lambda^\ell &= \bar{\epsilon}^+_{[2}\epsilon^+_{1]}\bar{\mathbb{D}}_+\mathbb{D}_+(\bar{\mathbb{V}}_\chi + \mathbb{V}_\phi), \\ \Lambda^r &= \bar{\epsilon}^-_{[2}\epsilon^-_{1]}\bar{\mathbb{D}}_-\mathbb{D}_-(\mathbb{V}_\chi - \mathbb{V}_\phi).\end{aligned} \quad (7.21)$$

The supersymmetry transformations in (7.18) can be assumed to generalize for non-abelian field strengths,

$$\begin{aligned}\delta\mathbb{F} &= \epsilon^+[\bar{\nabla}_+, \bar{\tilde{\mathbb{F}}}] + \epsilon^-[\bar{\nabla}_-, \tilde{\mathbb{F}}], \\ \delta\tilde{\mathbb{F}} &= -\epsilon^+[\bar{\nabla}_+, \bar{\mathbb{F}}] - \bar{\epsilon}^-[\nabla_-, \mathbb{F}].\end{aligned} \quad (7.22)$$

In the abelian case, the gauge field strengths are gauge invariant. In the non-abelian case, however, the field strengths transform covariantly under gauge transformations, and the supersymmetry transformations close up to gauge transformations. Writing the gauge field strengths in a collective notation as $F = \{\mathbb{F}, \bar{\mathbb{F}}, \tilde{\mathbb{F}}, \bar{\tilde{\mathbb{F}}}\}$, two subsequent supersymmetry transformations commute to

$$[\delta(\epsilon_1), \delta(\epsilon_2)]F = i\bar{\epsilon}^+_{[2}\epsilon^+_{1]}[\nabla_{++}, F] + i\bar{\epsilon}^-_{[2}\epsilon^-_{1]}[\nabla_{=}, F] + [K(\epsilon_1, \epsilon_2), F], \quad (7.23)$$

where the non-abelian generalization of the anti-commutator of the supersymmetry derivatives is $\nabla_{\pm\pm} = \{\bar{\nabla}_\pm, \nabla_\pm\}$, $K$ is the gauge parameter

$$K = -\alpha\mathbb{F} + \bar{\alpha}\bar{\mathbb{F}} - \tilde{\alpha}\tilde{\mathbb{F}} + \bar{\tilde{\alpha}}\bar{\tilde{\mathbb{F}}} \quad (7.24)$$

and $\alpha$ is the collective notation for supersymmetry parameters defined above.

The supersymmetry transformations of the non-abelian gauge potentials are less straight-forward. In the real representation, the supersymmetry transformations of the covariant derivatives are

$$\begin{aligned}\delta\bar{\nabla}_+ &= i\epsilon^-\tilde{\mathbb{F}} - i\bar{\epsilon}^-\mathbb{F}, \\ \delta\bar{\nabla}_- &= i\epsilon^+\bar{\tilde{\mathbb{F}}} + i\bar{\epsilon}^+\mathbb{F},\end{aligned} \quad (7.25)$$

which can be derived either by making an ansatz for the covariant derivatives that is compatible with the chiralities and the supersymmetry transformations of the field strengths (7.22), or by starting from the transformations in the left representation and performing a gauge transformation to the real representation. Independent of the representation, a covariant derivative transforms under supersymmetry as

$$\delta\nabla = [\nabla, e^{-V}\delta(e^V)]. \quad (7.26)$$

The transformations on the gauge potentials can then be derived using (7.25)-(7.26) and take the form

$$\delta(e^{\mathbb{U}_L}) = \bar{\epsilon}^- e^{\mathbb{U}_R}\bar{\mathbb{D}}_- e^{-\mathbb{U}_R}e^{\mathbb{U}_L} - \epsilon^- e^{-\bar{\mathbb{U}}_R}\mathbb{D}_- e^{\bar{\mathbb{U}}_R}e^{\mathbb{U}_L},$$
$$\delta(e^{\mathbb{U}_R}) = \epsilon^+ e^{-\bar{\mathbb{U}}_L}\mathbb{D}_+ e^{\bar{\mathbb{U}}_L}e^{\mathbb{U}_R} - \bar{\epsilon}^+ e^{\mathbb{U}_L}\bar{\mathbb{D}}_+ e^{-\mathbb{U}_L}e^{\mathbb{U}_R},$$
$$\delta(e^{\bar{\mathbb{U}}_L}) = \bar{\epsilon}^- e^{\bar{\mathbb{U}}_L}e^{\mathbb{U}_R}\bar{\mathbb{D}}_- e^{-\mathbb{U}_R} - \epsilon^- e^{\bar{\mathbb{U}}_L}e^{-\bar{\mathbb{U}}_R}\mathbb{D}_- e^{\bar{\mathbb{U}}_R},$$
$$\delta(e^{\bar{\mathbb{U}}_R}) = \epsilon^+ e^{\bar{\mathbb{U}}_R}e^{-\bar{\mathbb{U}}_L}\mathbb{D}_+ e^{\bar{\mathbb{U}}_L} - \bar{\epsilon}^+ e^{\bar{\mathbb{U}}_R}e^{\mathbb{U}_L}\bar{\mathbb{D}}_+ e^{-\mathbb{U}_L}. \quad (7.27)$$

Finally, the supersymmetry transformations for the gauge potentials $e^{\mathbb{V}_\phi}$ and $e^{\mathbb{V}_\chi}$ can be derived using (7.14),

$$\delta(e^{\mathbb{V}_\phi}) = -\epsilon^+ e^{-\bar{\mathbb{V}}_\chi}\mathbb{D}_+ e^{\bar{\mathbb{V}}_\chi}e^{\mathbb{V}_\phi} - \epsilon^- e^{\mathbb{V}_\phi}e^{-\mathbb{V}_\chi}\mathbb{D}_- e^{\mathbb{V}_\chi} + \bar{\epsilon}^- \bar{\mathbb{D}}_- e^{\mathbb{V}_\phi} - \bar{\epsilon}^+ \bar{\mathbb{D}}_+ e^{\mathbb{V}_\phi},$$
$$\delta(e^{\mathbb{V}_\chi}) = \epsilon^+ e^{-\bar{\mathbb{V}}_\phi}\mathbb{D}_+ e^{\bar{\mathbb{V}}_\phi}e^{\mathbb{V}_\chi} + \bar{\epsilon}^- e^{\mathbb{V}_\chi}e^{-\mathbb{V}_\phi}\bar{\mathbb{D}}_- e^{\mathbb{V}_\phi} + \bar{\epsilon}^+ \bar{\mathbb{D}}_+ e^{\mathbb{V}_\chi} - \epsilon^- \mathbb{D}_- e^{\mathbb{V}_\chi}.$$
$$(7.28)$$

It serves as a consistency check that these transformations reduce to the transformations (7.19) in the abelian limit.

## 7.2 The large vector multiplet

The isometry corresponding to the *large vector multiplet* (LVM) is similar to (7.2), but mixes chiral and twisted chiral directions with the Killing vector

$$k_{\phi\chi} = i(\partial_\phi - \partial_{\bar{\phi}} - \partial_\chi + \partial_{\bar{\chi}}). \quad (7.29)$$

The T-duality corresponding to the large vector multiplet will be studied in great detail in chapter 8. In this section a notation will be used that differs slightly from the one in paper [III], but instead agrees with the notation used in the next chapter. A Lagrangian invariant along the direction defined by this Killing vector may be a function of any of the combinations

$$x = \phi + \bar{\phi}, \quad y = \chi + \bar{\chi}, \quad u = \phi + \chi, \quad v = \phi - \bar{\chi} \quad (7.30)$$

plus their complex conjugates. These are not independent, however, but related by the expressions

$$2\,\mathrm{Re}\,u = x+y, \quad 2\,\mathrm{Re}\,v = x-y, \quad z = -2\,\mathrm{Im}\,u = -2\,\mathrm{Im}\,v. \quad (7.31)$$

Hence, a Lagrangian invariant under the isometry (7.29) is a function of, e.g., the combinations $(x,y,z)$,

$$K = K(\phi + \bar{\phi}, \chi + \bar{\chi}, i(\phi - \bar{\phi} + \chi - \bar{\chi}), \mathbb{X}^\ell, \bar{\mathbb{X}}^{\bar{\ell}}, \mathbb{X}^r, \bar{\mathbb{X}}^{\bar{r}}). \quad (7.32)$$

Under the gauged isometry, the chiral and twisted chiral fields transform into chiral $\delta\phi = i\Lambda$ and twisted chiral $\delta\chi = i\tilde{\Lambda}$ parameters, respectively. As discussed in section 5.3, gauge potentials with suitable transformation properties,

$$\delta V_x = -i(\Lambda - \bar{\Lambda}),$$
$$\delta V_y = -i(\tilde{\Lambda} - \bar{\tilde{\Lambda}}),$$
$$\delta V_z = \Lambda + \bar{\Lambda} + \tilde{\Lambda} + \bar{\tilde{\Lambda}}, \quad (7.33)$$

are introduced to keep the action invariant under the gauged isometry. The set of gauge potentials $(V_x, V_y, V_z)$ is denoted the large vector multiplet [LRR+07, GM07]. The gauged action is

$$K = K\big(x + V_x,\, y + V_y,\, z + V_z\big). \quad (7.34)$$

Complex gauge potentials transforming into gauge parameters with definite chirality properties can be defined from the real gauge potentials in the large vector multiplet as

$$V_L = \tfrac{1}{2}[V_z + i(V_x + V_y)], \qquad \delta V_L = \Lambda + \tilde{\Lambda},$$
$$V_R = \tfrac{1}{2}[V_z + i(V_x - V_y)], \qquad \delta V_R = \Lambda + \bar{\tilde{\Lambda}}. \quad (7.35)$$

The potentials have the same real part, $V_L + \bar{V}_L = V_R + \bar{V}_R$, and are not independent. Using the complex gauge potentials, gauge invariant field strengths can be defined as

$$\mathbb{G}_+ = \bar{\mathbb{D}}_+ V_L, \quad \mathbb{G}_- = \bar{\mathbb{D}}_- V_R. \quad (7.36)$$

The field strengths satisfy left and right semichiral constraints, respectively. The field strengths $\mathbb{G}_\pm$ for the the large vector multiplet are not the only gauge invariant objects that can be constructed from the gauge potentials in (7.35). Higher-order field strengths involving two derivatives can be defined as [LRR+07]

$$W = -i\bar{\mathbb{D}}_+ \bar{\mathbb{D}}_- V_y, \qquad B = -\bar{\mathbb{D}}_+ \bar{\mathbb{D}}_- (V_z + iV_x),$$
$$\tilde{W} = -i\mathbb{D}_+ \bar{\mathbb{D}}_- V_x, \qquad \tilde{B} = -\mathbb{D}_+ \bar{\mathbb{D}}_- (V_z - iV_y). \quad (7.37)$$

As opposed to the field strengths $\mathbb{G}_\pm$, the higher-order field strengths $W, B$ and $\tilde{W}, \tilde{B}$ are chiral and twisted chiral, respectively.

In contrast to the semichiral vector multiplet, the large vector multiplet contains, when reduced to $N=(1,1)$ superspace, four extra gauge invariant spinor multiplets in addition to the $N=(1,1)$ gauge invariant field strength and three gauge invariant scalars [LRR+07]. The large number of gauge invariant components motivates the name *large* vector multiplet. An action for the large vector multiplet was constructed in [Ryb07], where the higher derivative terms resulting from the gauge invariant spinors were removed by a field redefinition and gauge invariants in the Kähler potential.

## 7.2.1 Non-abelian large vector multiplet

The non-abelian version of the large vector multiplet was developed in paper [LRR+09]. Similarly to the semichiral multiplet, the gauge transformations of the vector potentials (7.35) generalize straight-forwardly to the non-abelian case. But an important difference between the semichiral and the large vector multiplet is that two copies of covariant field strengths can be constructed for the large vector multiplet. For example, in the chiral representation, there are two covariant derivatives containing the supersymmetry derivative $\bar{\mathbb{D}}_-$ to the lowest order,

$$\bar{\nabla}_- = \bar{\mathbb{D}}_-, \quad \hat{\bar{\nabla}}_- = e^{iV_R}\bar{\mathbb{D}}_- e^{-iV_R}. \qquad (7.38)$$

The non-abelian generalization of the field strengths in (7.36) is the difference in these two sets of covariant derivatives,

$$\mathbb{G}_+ = i(\hat{\bar{\nabla}}_+ - \bar{\nabla}_+), \quad \mathbb{G}_- = i(\hat{\bar{\nabla}}_- - \bar{\nabla}_-). \qquad (7.39)$$

Being the difference of two covariant derivatives, the field strengths are indeed covariant tensors. The chirality constraints of the non-abelian field strengths are defined using both sets of covariant derivatives [LRR+09],

$$(\hat{\bar{\nabla}}_\pm + \bar{\nabla}_\pm)\mathbb{G}_\pm = 0. \qquad (7.40)$$

In addition to the gauge invariant tensors (7.39), two sets of additional field strengths can be constructed, that are the non-abelian generalizations of (7.37),

$$\begin{aligned} F &= \{\hat{\bar{\nabla}}_+, \bar{\nabla}_-\}, & \hat{F} &= \{\bar{\nabla}_+, \hat{\bar{\nabla}}_-\}, \\ \tilde{F} &= \{\hat{\bar{\nabla}}_+, \nabla_-\}, & \hat{\tilde{F}} &= \{\bar{\nabla}_+, \hat{\nabla}_-\}, \end{aligned} \qquad (7.41)$$

together with their complex conjugates. The approach of [Ryb07] to construct actions of the large vector multiplet was generalized in [LRR+09], where it was found that an action of the large vector multiplet can contain only combinations of the field strengths $(F, \bar{F}, \tilde{F}, \bar{\tilde{F}})$ *or* of the hatted field strengths $(\hat{F}, \hat{\bar{F}}, \hat{\tilde{F}}, \hat{\bar{\tilde{F}}})$. Mixed terms from the two sets of field strengths gives rise to higher derivative terms that cannot be eliminated by field redefinitions.

## 7.2.2 Additional supersymmetry

In the same way as in section 7.1, one can ask if and under what conditions the large vector multiplet allows for $N = (4,4)$ supersymmetry. Noting, however, that the two field strengths in (7.36) together with their complex conjugates are semichiral, and using the results from [I], [II] and [IV] saying that off-shell $N = (4,4)$ supersymmetry can only be imposed on a sigma model described by semichiral superfields if the target space is larger than four-dimensional, one can draw the conclusion that no ansatz on the field strengths $\mathbb{G}_\pm$ can close to a supersymmetry algebra. As in the semichiral sigma model in four dimensions discussed in section 6.2, there is also the possibility to impose twisted supersymmetry; linear transformations can be constructed acting on the field strengths that close to a pseudo-supersymmetry off-shell [III].

## 7.3 Results

The semichiral vector multiplet $(\mathbb{V}^\ell, \mathbb{V}^r, \mathbb{V}')$ with the corresponding gauge field strengths $(\mathbb{F}, \tilde{\mathbb{F}})$ allows for $N = (4,4)$ supersymmetry [III]. In the abelian case, linear supersymmetry transformations can be constructed for the gauge potentials and for the field strengths. The corresponding non-abelian transformations can also be constructed and close to a supersymmetry up to gauge transformations.

The large vector multiplet $(V_x, V_y, V_z)$, however, has gauge field strengths $\mathbb{G}_\pm$ with semichiral chirality properties. This is the important difference from the semichiral vector multiplet, whose field strengths satisfy chiral and twisted chiral chirality properties. This semichirality of the large vector multiplet obstructs the $N = (4,4)$ supersymmetry, as is expected from the results of semichiral fields and off-shell $N = (4,4)$ supersymmetry reviewed in chapter 6. As discussed in the same chapter, one could consider $N = (4,4)$ twisted supersymmetry, or only left/right-going supersymmetry. There is also a possibility that one could impose additional supersymmetry on-shell, analogously with the discussion in section 6.5. The on-shell discussion would be action dependent, and was not further investigated in paper [III].

From the result of paper [I, II, IV] and earlier work [GHR84], it is clear that a set of four chiral and twisted chiral fields allow off-shell $N = (4,4)$ supersymmetry, whereas a set of semichiral fields does not. The results of paper [III] is therefore expected and in agreement with previous results.

# 8. T-duality and N = (4,4) supersymmetry

> I never am really satisfied that I understand anything; because, understand it well as I may, my comprehension can only be an infinitesimal fraction of all I want to understand about the many connections and relations which occur to me, how the matter in question was first thought of or arrived at, etc., etc.
> *Ada Lovelace, computer programmer, mathematician (1815-1852)*

From chapter 6 it is clear that a semichiral model with four-dimensional target space does not admit off-shell $N=(4,4)$ supersymmetry [I]. The supersymmetry can only be obtained on-shell [IV] or in a larger target space [II].

However, from [GHR84], it is known that a model parametrized by a chiral and a twisted chiral field may be enhanced to to off-shell $N=(4,4)$ supersymmetry if and only if the generalized Kähler potential satisfies the Laplace equation. It is also known [IKR95, GMST99] that a chiral and twisted chiral model can be related to a semichiral model by T-duality, similar to the duality relating a chiral and a twisted chiral model as discussed in section 5.3.

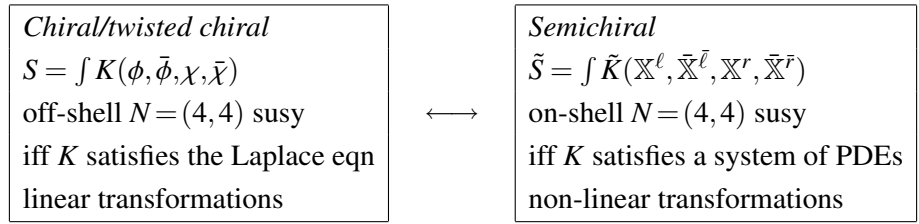

| *Chiral/twisted chiral* | *Semichiral* |
| --- | --- |
| $S = \int K(\phi,\bar{\phi},\chi,\bar{\chi})$ | $\tilde{S} = \int \tilde{K}(\mathbb{X}^\ell, \bar{\mathbb{X}}^{\bar{\ell}}, \mathbb{X}^r, \bar{\mathbb{X}}^{\bar{r}})$ |
| off-shell $N=(4,4)$ susy | on-shell $N=(4,4)$ susy |
| iff $K$ satisfies the Laplace eqn | iff $K$ satisfies a system of PDEs |
| linear transformations | non-linear transformations |

This situation raises several questions. The overall question and setup for paper [V] was: Starting with a chiral and twisted chiral sigma model with off-shell linear $N=(4,4)$ supersymmetry, and dualizing into a semichiral sigma model – what happens to the supersymmetry? In particular, the following questions were addressed.

RESEARCH QUESTIONS:
⋆ Is the $N=(4,4)$ supersymmetry of the chiral and twisted chiral model preserved along the isometry, or is the isometry incompatible with geometric data generated by the supersymmetry?

⋆ Can the non-linear supersymmetry transformations of the semichiral model be related to the linear transformations of the chiral and twisted chiral model? From where does the non-linearity originate?
⋆ How is the off-shell closure in the chiral and twisted chiral model and the on-shell closure in the semichiral model to be interpreted? Do the field equations of the latter model have any analogue in the former?
⋆ What does the Laplace equation correspond to in the dual semichiral model? Is the corresponding equation sufficient to make the semichiral action invariant under the supersymmetry transformations? If not, what are the additional constraints, and how can they be understood?
⋆ Does the T-duality provide a complementary and/or more insightful understanding of the on-shell $N=(4,4)$ supersymmetry of the semichiral model?

These questions were all investigated and answered in [V]. In the next sections, the details and methods of the paper will be discussed. The results of the paper and the answers to the questions will be summarized in section 8.6.

## 8.1 The N = 4 puzzle

The sigma model with manifest $N=(2,2)$ supersymmetry was introduced in sections 4.3-4.4 and can be parametrized by chiral $\phi$, twisted chiral $\chi$ and semichiral $\mathbb{X}^\ell$, $\mathbb{X}^r$ superfields. The chiral and the twisted chiral fields are constrained in two chiralities, $\bar{\mathbb{D}}_\pm \phi = 0$ and $\bar{\mathbb{D}}_+ \chi = \mathbb{D}_- \chi = 0$, whereas the semichiral fields are constrained only in one, $\bar{\mathbb{D}}_+ \mathbb{X}^\ell = 0$ and $\bar{\mathbb{D}}_- \mathbb{X}^r = 0$.

### 8.1.1 Linear off-shell supersymmetry in chiral/twisted model

The geometry of a sigma model parametrized by chiral and twisted chiral coordinates, $K(\phi,\bar\phi,\chi,\bar\chi)$, is bihermitian local product space (BiLP), and corresponds to the situation when the two complex structures enabling the extended $N=(2,2)$ supersymmetry commute, $[J^{(+)},J^{(-)}]=0$. A system of one chiral and one twisted chiral superfield and their complex conjugates admits transformations

$$\delta\phi = \bar\epsilon^+ \bar{\mathbb{D}}_+ \bar\chi + \bar\epsilon^- \bar{\mathbb{D}}_- \chi,$$
$$\delta\chi = -\bar\epsilon^+ \bar{\mathbb{D}}_+ \bar\phi - \epsilon^- \mathbb{D}_- \phi, \qquad (8.1)$$

that are linear and close to a $N=(4,4)$ supersymmetry off-shell. The action with Lagrangian $K(\phi,\bar\phi,\chi,\bar\chi)$ is invariant under the supersymmetry transformations if and only if the generalized Kähler potential satisfies the Laplace equation [GHR84],

$$K_{\phi\bar\phi} + K_{\chi\bar\chi} = 0. \qquad (8.2)$$

The additional supersymmetry implies the existence of two commuting integrable quaternionic structures (4.33),

$$J_i^{(\pm)} J_j^{(\pm)} = -\delta_{ij} + \varepsilon_{ijk} J_k^{(\pm)}, \quad i = 1,2,3$$

with $[J_i^{(+)}, J_j^{(-)}] = 0$, satisfying the compact algebra $SO(4) \simeq SU(2) \times SU(2)$. This implies that nine product structures can be formed from any two products of the six structures, $\Pi = J_i^{(\pm)} J_j^{(\mp)}$ [GHR84].

### 8.1.2 Non-linear on-shell supersymmetry in semichiral model

However, as shown in [I, II, IV] and reviewed in chapter 6, the analogous situation for a sigma model in semichiral coordinates is very different. If the target space is four-dimensional, the transformations for $N=(4,4)$ supersymmetry can only close on-shell, and are non-linear (6.56),

$$\delta \mathbb{X}^\ell = \bar{\epsilon}^+ \bar{\mathbb{D}}_+ f(\mathbb{X}^\ell, \bar{\mathbb{X}}^\ell, \mathbb{X}^r, \bar{\mathbb{X}}^r) + g(\mathbb{X}^\ell) \bar{\epsilon}^- \bar{\mathbb{D}}_- \mathbb{X}^\ell + h(\mathbb{X}^\ell) \epsilon^- \mathbb{D}_- \mathbb{X}^\ell,$$
$$\delta \mathbb{X}^r = \bar{\epsilon}^- \bar{\mathbb{D}}_- \tilde{f}(\mathbb{X}^\ell, \bar{\mathbb{X}}^\ell, \mathbb{X}^r, \bar{\mathbb{X}}^r) + \tilde{g}(\mathbb{X}^r) \bar{\epsilon}^+ \bar{\mathbb{D}}_+ \mathbb{X}^r + \tilde{h}(\mathbb{X}^r) \epsilon^+ \mathbb{D}_+ \mathbb{X}^r.$$

The action is invariant under the transformations if and only if the generalized Kähler potential $K(\mathbb{X}^\ell, \bar{\mathbb{X}}^\ell, \mathbb{X}^r, \bar{\mathbb{X}}^r)$ satisfies the system of non-linear partial differential equations (6.38)-(6.39), relating derivatives of the functions $f$, $\tilde{f}$ with the functions $g$, $\tilde{g}$, $h$, $\tilde{h}$ and second derivatives of the generalized Kähler potential.

Invariance of the action and on-shell closure of the supersymmetry algebra imply, together with the fact that the complex structures from the $N=(2,2)$ supersymmetry do not commute, that the target manifold is necessarily hyperkähler, as described in section 6.6.

## 8.2 Duality between the models

Recall from chapter 5 that the existence of isometries enable dualities between different sigma models. The vector multiplets needed to gauge an isometry mixing chiral and twisted chiral fields as described in section 7.2 is the large vector multiplet.

### 8.2.1 Duality transformations

Consider a sigma model parametrized by chiral $\phi$ and twisted chiral $\chi$ superfields, subject to an isometry defined by the translational Killing vector (7.29), $k = i(\partial_\phi - \partial_{\bar{\phi}} - \partial_\chi + \partial_{\bar{\chi}})$. Writing combinations of the fields as in section 7.2,

$$x = \phi + \bar{\phi}, \quad y = \chi + \bar{\chi}, \quad z = i(\phi - \bar{\phi} + \chi - \bar{\chi}), \tag{8.3}$$

together with the forth coordinate that parametrizes the direction of the isometry as $w = i(\phi - \bar{\phi} - \chi + \bar{\chi})$, an action respecting the isometry is given by the Lagrangian in (7.32), where here the light-cone coordinates of the world-sheet are written as $\xi^{\pm}$,

$$S = \int d^2\xi d^2\theta d^2\bar{\theta} K(x,y,z). \tag{8.4}$$

This action, parametrized by coordinates that are combinations of chiral and twisted chiral fields, will in this chapter be referred to as the *original* action. Under the gauged isometry transformation, the chiral and twisted chiral superfields transform into gauge parameters of the same chirality. The gauge potentials of the large vector multiplet $V^\mu = (V_x, V_y, V_z)$ transform as in (7.33). By choosing a gauge such that $x = y = z = 0$ and introducing semichiral field strengths as in (7.36),

$$\mathbb{G}_\pm = \tfrac{1}{2}\bar{\mathbb{D}}_\pm(V_z + iV_x \pm iV_y) \tag{8.5}$$

together with unconstrained spinorial Lagrange multipliers $X^\pm$, a first order action can be constructed as

$$S_{1st} = \int d^2\xi d^2\theta d^2\bar{\theta}\Big[K(V_x,V_y,V_z) - (X^+\mathbb{G}_+ + \bar{X}^+\bar{\mathbb{G}}_+ + X^-\mathbb{G}_- + \bar{X}^-\bar{\mathbb{G}}_-)\Big]. \tag{8.6}$$

Varying this action with respect to the Lagrange multipliers constrains the field strengths to vanish, $\mathbb{G}_\pm = 0$, which is solved by $V_x = x$, $V_y = y$ and $V_z = z$, reproducing the original chiral and twisted chiral model (8.4).

Instead integrating by parts and introducing semichiral fields as $\mathbb{X}^\ell = \bar{\mathbb{D}}_+ X^+$ and $\mathbb{X}^r = \bar{\mathbb{D}}_- X^-$, the first order action takes the form

$$S_{1st} = \int d^2\xi d^2\theta d^2\bar{\theta}\big[K(V_x,V_y,V_z) - \tilde{x}V_x - \tilde{y}V_y - \tilde{z}V_z\big], \tag{8.7}$$

where the coordinates $\tilde{x}^i = (\tilde{x},\tilde{y},\tilde{z})$ are combinations of semichiral superfields,

$$\begin{aligned}
\tilde{x} &= \tfrac{i}{2}(\mathbb{X}^\ell - \bar{\mathbb{X}}^\ell + \mathbb{X}^r - \bar{\mathbb{X}}^r),\\
\tilde{y} &= \tfrac{i}{2}(\mathbb{X}^\ell - \bar{\mathbb{X}}^\ell - \mathbb{X}^r + \bar{\mathbb{X}}^r),\\
\tilde{z} &= \tfrac{1}{2}(\mathbb{X}^\ell + \bar{\mathbb{X}}^\ell + \mathbb{X}^r + \bar{\mathbb{X}}^r).
\end{aligned} \tag{8.8}$$

Varying with respect to the gauge potentials implies that they are not independent, but rather functions of the combinations of semichiral fields,

$$\frac{\partial K}{\partial V_x} = \tilde{x}, \quad \frac{\partial K}{\partial V_y} = \tilde{y}, \quad \frac{\partial K}{\partial V_z} = \tilde{z}. \tag{8.9}$$

Inserting the relation $V^\mu = V^\mu(\tilde{x}^i)$ gives the dual semichiral action,

$$\tilde{S} = \int\Big[K\big(V_x(\tilde{x}^i),V_y(\tilde{x}^i),V_z(\tilde{x}^i)\big) - \tilde{x}V_x(\tilde{x}^i) - \tilde{y}V_y(\tilde{x}^i) - \tilde{z}V_z(\tilde{x}^i)\Big] = \int \tilde{K}(\tilde{x},\tilde{y},\tilde{z}), \tag{8.10}$$

where the integration measure is implicit. This semichiral action will hereafter be denoted the *dual* action. From the Legendre transformation (8.9)-(8.10), the following useful identities are obtained,

$$K_\mu = \delta_{\mu i}\tilde{x}^i, \quad \tilde{K}_i = -\delta_{i\mu}V^\mu. \tag{8.11}$$

Assuming that the Hessian $K_{\mu\nu}$ is invertible, these relations from the Legendre transform further imply that $\tilde{K}_{ij} = -\delta_{i\mu}(K^{-1})^{\mu\nu}\delta_{\nu j}$ and $\mathbb{D}x^\mu = (K^{-1})^{\mu\nu}\delta_{\nu i}\mathbb{D}\tilde{x}^i$.

### 8.2.2 Field equations and Bianchi identities

The Bianchi identities for the original model are obtained when varying the first order action (8.6) with respect to the Lagrange multipliers and obtaining the pure gauge condition,

$$\mathbb{G}_\pm = \tfrac{1}{2}\bar{\mathbb{D}}_\pm(V_z + iV_x \pm iV_y) = 0. \tag{8.12}$$

The Bianchi identities are automatically satisfied when the gauge potentials are identified as the combinations $x^\mu$ of the chiral and twisted chiral fields, due to their chirality constraints,

$$\begin{aligned}\tfrac{1}{2}\bar{\mathbb{D}}_+(z+ix+iy) &= \bar{\mathbb{D}}_+ x_L = i\bar{\mathbb{D}}_+(\phi+\chi) = 0, \\ \tfrac{1}{2}\bar{\mathbb{D}}_-(z+ix-iy) &= \bar{\mathbb{D}}_- x_R = i\bar{\mathbb{D}}_-(\phi-\bar{\chi}) = 0.\end{aligned} \tag{8.13}$$

The field equations for the dual semichiral model (8.10) are found by varying the dual action (8.10) with respect to the unconstrained $X^\pm$ in the semichiral fields, and take the form

$$\bar{\mathbb{D}}_\pm(\tilde{K}_{\tilde{z}} + i\tilde{K}_{\tilde{x}} \pm i\tilde{K}_{\tilde{y}}) = 0. \tag{8.14}$$

In the usual semichiral coordinates, (8.14) correspond to the field equations $\bar{\mathbb{D}}_+ K_\ell = \bar{\mathbb{D}}_- K_r = 0$ in (6.41). Using the identities (8.11) from the Legendre transform, the Bianchi identities (8.12) for the chiral and twisted chiral model are replaced by the field equations (8.14) in the dual semichiral model.

## 8.3 Supersymmetry and T-duality

### 8.3.1 Isometries preserving N = (4,4) supersymmetry

Along which isometries can the twisted chiral multiplet be dualized while still preserving the $N=(4,4)$ supersymmetry?

As described in section 8.1, the target space geometry of the $N=(4,4)$ supersymmetric chiral and twisted chiral model has six complex structures $J_i^{(\pm)}$ with $i = 1,2,3$, that are constant in the coordinates $(\phi, \bar{\phi}, \chi, \bar{\chi})$ [GHR84]. For an isometry defined by a Killing vector $k$ to maintain the supersymmetry, it must be holomorphic with respect to the all the complex structures, i.e., the complex structures must be preserved by the isometry,

$$\mathcal{L}_k J^\mu_{\ \nu} = k^\rho \partial_\rho J^\mu_{\ \nu} - \partial_\rho k^\mu J^\rho_{\ \nu} + \partial_\nu k^\rho J^\mu_{\ \rho} = 0. \tag{8.15}$$

Since all complex structures are constant, the first term vanishes and the vanishing of the remaining terms imply that the Killing vector is diagonal of the form

$$k = k^\phi(\phi)\partial_\phi + k^{\bar{\phi}}(\bar{\phi})\partial_{\bar{\phi}} + k^\chi(\chi)\partial_\chi + k^{\bar{\chi}}(\bar{\chi})\partial_{\bar{\chi}}, \tag{8.16}$$

with the derivatives of all coefficients equal; $k^\phi_{,\phi} = k^{\bar{\phi}}_{,\bar{\phi}} = k^\chi_{,\chi} = k^{\bar{\chi}}_{,\bar{\chi}}$. The solutions to these constraints are either that all coefficients are constants, so that the isometry represents a translation, e.g., $k = i(\partial_\phi - \partial_{\bar{\phi}} - \partial_\chi + \partial_{\bar{\chi}})$, or that they are linear with the same derivative, $k = \phi\partial_\phi + \bar{\phi}\partial_{\bar{\phi}} + \chi\partial_\chi + \bar{\chi}\partial_{\bar{\chi}}$, which represents a rescaling. The isometry considered here and in [V] is translational and should therefore preserve the additional supersymmetry.

### 8.3.2 Supersymmetry in the original model

In coordinates adapted to the duality transformation, the supersymmetry transformations (8.1) of the combinations of chiral and twisted chiral fields are

$$\begin{aligned}
\delta x &= \bar{\epsilon}^+ \bar{\mathbb{D}}_+ y + \bar{\epsilon}^- \bar{\mathbb{D}}_- y + \epsilon^+ \mathbb{D}_+ y + \epsilon^- \mathbb{D}_- y, \\
\delta y &= -\bar{\epsilon}^+ \bar{\mathbb{D}}_+ x - \bar{\epsilon}^- \bar{\mathbb{D}}_- x - \epsilon^+ \mathbb{D}_+ x - \epsilon^- \mathbb{D}_- x, \\
\delta z &= i\bar{\epsilon}^+ \bar{\mathbb{D}}_+ (y-x) + i\bar{\epsilon}^- \bar{\mathbb{D}}_- (y+x) - i\epsilon^+ \mathbb{D}_+ (y-x) - i\epsilon^- \mathbb{D}_- (y+x),
\end{aligned} \tag{8.17}$$

and the Laplace equation that follows from invariance of the action is

$$K_{xx} + K_{yy} + 2K_{zz} = 0. \tag{8.18}$$

However, if the chirality constraints for the chiral and twisted chiral fields had not been used, the action would not be invariant under the transformations; non-linear terms would have to be added, and the new terms would have to satisfy certain differential equations.

### 8.3.3 Adding non-linear terms

As just seen, the transformations of the chiral and twisted chiral fields are linear, whereas the semichiral analogues are non-linear. The underlying reason for the linearity of the former model is the first order differential constraint in the Bianchi identities for the original model, i.e. the chiralities of the chiral and twisted chiral fields.

Consider a model of unconstrained gauge potentials, $K(V_x, V_y, V_z)$ and add non-linear terms that are proportional to the Bianchi identities,

$$\delta V_x = \bar{\epsilon}^+ \left[ \bar{\mathbb{D}}_+ V_y + \tfrac{i}{2}\alpha \bar{\mathbb{D}}_+(V_z + iV_x + iV_y) \right] + \ldots$$
$$\delta V_y = \bar{\epsilon}^+ \left[ -\bar{\mathbb{D}}_+ V_x + \tfrac{i}{2}\gamma \bar{\mathbb{D}}_+(V_z + iV_x + iV_y) \right] + \ldots$$
$$\delta V_z = i\bar{\epsilon}^+ \left[ \bar{\mathbb{D}}_+(V_y - V_x) + \tfrac{i}{2}\varepsilon \bar{\mathbb{D}}_+(V_z + iV_x + iV_y) \right] + \ldots \quad (8.19)$$

where $\alpha$, $\gamma$, $\varepsilon$ are arbitrary functions. When the Bianchi identities (8.12) are evoked, these transformations on the unconstrained fields $V^\mu$ reduce to the linear transformations (8.17) on the chiral and twisted chiral coordinates. It will be useful to write the transformations in the same compact notation used in chapter 6,

$$\delta V^\mu = \bar{\epsilon}^\pm U^{(\pm)\mu}{}_\nu \bar{\mathbb{D}}_\pm V^\nu + \epsilon^\pm V^{(\pm)\mu}{}_\nu \mathbb{D}_\pm V^\nu, \quad (8.20)$$

where the notation for the unconstrained fields $V^\mu$ should not be confused with the transformation matrices $V^{(\pm)}$. The transformation matrices are

$$U^{(+)} = \frac{1}{2} \begin{pmatrix} -\alpha & (2-\alpha) & i\alpha \\ -(2+\gamma) & -\gamma & i\gamma \\ -i(2+\varepsilon) & i(2-\varepsilon) & -\varepsilon \end{pmatrix},$$

$$U^{(-)} = \frac{1}{2} \begin{pmatrix} -\beta & (2+\beta) & i\beta \\ -(2+\delta) & \delta & i\delta \\ i(2-\kappa) & i(2+\kappa) & -\kappa \end{pmatrix}, \quad (8.21)$$

and $V^{(\pm)}$ are the complex conjugates of $U^{(\pm)}$.

Now consider invariance of the action. If the Bianchi identities (8.12) hold, the action with Lagrangian $K(V^\mu)$ is supersymmetry invariant if and only if the Laplace equation is satisfied. But for unconstrained fields $V^\mu$, invariance of the action implies $K_{\mu[\nu} U^{(\pm)\mu}{}_{\rho]} = 0$. This system of partial differential equations can be solved if the non-linear terms in the transformations are certain functions of second derivatives of $K$ and first order derivatives of a function $f$,

$$\alpha = -i(K^{-1})^{1\mu} f_\mu + 1, \qquad \beta = -i(K^{-1})^{1\mu} \hat{f}_\mu - 1,$$
$$\gamma = -i(K^{-1})^{2\mu} f_\mu - 1, \qquad \delta = -i(K^{-1})^{2\mu} \hat{f}_\mu - 1,$$
$$\varepsilon = -(K^{-1})^{3\mu} f_\mu, \qquad \kappa = -(K^{-1})^{3\mu} \hat{f}_\mu, \quad (8.22)$$

where the functions $f$ and $\hat{f}$ satisfy the partial differential equations

$$f_z + if_x = -(K_x + K_y)_y - i(K_x + K_y)_z,$$
$$f_z + if_y = (K_x + K_y)_x + i(K_x + K_y)_z,$$
$$\hat{f}_z + i\hat{f}_x = -(K_x - K_y)_y + i(K_x - K_y)_z,$$
$$\hat{f}_z - i\hat{f}_y = -(K_x - K_y)_x - i(K_x - K_y)_z. \quad (8.23)$$

This solution for the transformation parameters may seem ad-hoc, but is chosen to agree with the transformations found in [IV] and the notation used in chapter 6. The important observation that should be stressed is the following: in the first order action (8.7), the $V^\mu$ are unconstrained and the $\tilde{x}^i$ are combinations of semichiral fields. The action with Lagrangian $K(V^\mu)$ is no longer invariant provided that the Laplace equation is satisfied, instead invariance of the action implies the equations $K_{\mu[\nu}U^{(\pm)\mu}_{\rho]} = 0$. Transformations for $V^\mu$ can be defined in terms of certain functions (8.22), such that these constraints are satisfied. Using the Bianchi identities for the original model, the unconstrained fields convert to combinations of chiral and twisted chiral fields, $V^\mu = x^\mu$, and the transformations reduce to the known transformations (8.17) for the original chiral and twisted chiral model.

### 8.3.4 Supersymmetry in the dual model

With the added non-linear terms in $\delta V^\mu$ together with the Legendre identities (8.11) from the T-duality, the transformations for the dual semichiral coordinates can easily be derived,

$$\begin{aligned}
\delta \tilde{x}^i &= \delta^{i\mu} K_{\mu\nu} \delta V^\nu \\
&= \delta^{i\mu} K_{\mu\nu} \left( \bar{\epsilon}^\alpha U^{(\alpha)\nu}_{\rho} \bar{\mathbb{D}}_\alpha V^\rho + \epsilon^\alpha V^{(\alpha)\nu}_{\rho} \mathbb{D}_\alpha V^\rho \right) \\
&= \bar{\epsilon}^\alpha \left( \delta^{i\mu} K_{\mu\nu} U^{(\alpha)\nu}_{\rho} (K^{-1})^{\rho\sigma} \delta_{\sigma j} \right) \bar{\mathbb{D}}_\alpha \tilde{x}^j + \epsilon^\alpha \left( \delta^{i\mu} K_{\mu\nu} V^{(\alpha)\nu}_{\rho} (K^{-1})^{\rho\sigma} \delta_{\sigma j} \right) \mathbb{D}_\alpha \tilde{x}^j \\
&= \bar{\epsilon}^\alpha \tilde{U}^{(\alpha)i}{}_j \bar{\mathbb{D}}_\alpha \tilde{x}^j + \epsilon^\alpha \tilde{V}^{(\alpha)i}{}_j \mathbb{D}_\alpha \tilde{x}^j,
\end{aligned} \qquad (8.24)$$

where the chirality indices $\alpha = +, -$ are summed over. From this expression, the transformation matrices of the semichiral coordinates can be read off in terms of the transformations of the chiral and twisted coordinates as

$$\begin{aligned}
\tilde{U}^{(\alpha)i}{}_j &= \delta^{i\mu} K_{\mu\nu} U^{(\alpha)\nu}_{\rho} (K^{-1})^{\rho\sigma} \delta_{\sigma j}, \\
\tilde{V}^{(\alpha)i}{}_j &= \delta^{i\mu} K_{\mu\nu} V^{(\alpha)\nu}_{\rho} (K^{-1})^{\rho\sigma} \delta_{\sigma j}.
\end{aligned} \qquad (8.25)$$

Inserting the explicit expressions for $\alpha, \beta, \ldots$ from (8.22) and again using Legendre identities, the dual transformations read

$$\tilde{U}^{(+)} = \frac{1}{2} \begin{pmatrix} if_{\tilde{x}} - 1 & if_{\tilde{y}} - 1 & i(f_{\tilde{z}} - 2) \\ if_{\tilde{x}} + 1 & if_{\tilde{y}} + 1 & i(f_{\tilde{z}} + 2) \\ f_{\tilde{x}} + i & f_{\tilde{y}} - i & f_{\tilde{z}} \end{pmatrix},$$

$$\tilde{U}^{(-)} = \frac{1}{2} \begin{pmatrix} i\hat{f}_{\tilde{x}} + 1 & i\hat{f}_{\tilde{y}} - 1 & i(\hat{f}_{\tilde{z}} + 2) \\ -i\hat{f}_{\tilde{x}} + 1 & -(i\hat{f}_{\tilde{y}} + 1) & -i(\hat{f}_{\tilde{z}} - 2) \\ \hat{f}_{\tilde{x}} - i & \hat{f}_{\tilde{y}} - i & \hat{f}_{\tilde{z}} \end{pmatrix}. \qquad (8.26)$$

Explicitly, the combinations of the semichiral fields $\tilde{x}^i$ then transform under the $N=(4,4)$ supersymmetry as

$$\delta\tilde{x} = \tfrac{i}{2}\bar{\epsilon}^+\left[\bar{\mathbb{D}}_+ f - \bar{\mathbb{D}}_+(2\tilde{z} - i\tilde{x} - i\tilde{y})\right] + \tfrac{i}{2}\bar{\epsilon}^-\left[\bar{\mathbb{D}}_- \hat{f} + \bar{\mathbb{D}}_-(2\tilde{z} - i\tilde{x} + i\tilde{y})\right],$$
$$\delta\tilde{y} = \tfrac{i}{2}\bar{\epsilon}^+\left[\bar{\mathbb{D}}_+ f + \bar{\mathbb{D}}_+(2\tilde{z} - i\tilde{x} - i\tilde{y})\right] - \tfrac{i}{2}\bar{\epsilon}^-\left[\bar{\mathbb{D}}_- \hat{f} - \bar{\mathbb{D}}_-(2\tilde{z} - i\tilde{x} + i\tilde{y})\right],$$
$$\delta\tilde{z} = \tfrac{1}{2}\bar{\epsilon}^+\left[\bar{\mathbb{D}}_+ f + i\bar{\mathbb{D}}_+(\tilde{x} - \tilde{y})\right] + \tfrac{1}{2}\bar{\epsilon}^-\left[\bar{\mathbb{D}}_- \hat{f} - i\bar{\mathbb{D}}_-(\tilde{x} + \tilde{y})\right] \quad (8.27)$$

plus complex conjugate parts. These agree with the obtained transformations for semichiral fields (6.56) with $g = 1$ and $\tilde{g} = -1$.[†]

An illuminating consequence of this procedure is that the on-shell condition for algebra closure now becomes evident. The algebra closure in the original model follows due to the Bianchi constraints. From the T-duality, these equations are replaced by field equations in the dual semichiral model, hence, the transformations on the semichiral fields are expected to close only on-shell.

### 8.3.5 *Duality between BiLP and hyperkähler geometry*

As discussed in section 4.4, the target space parametrized by chiral and twisted chiral fields describes the section of bihermitian geometry where the two complex structures of the $N=(2,2)$ supersymmetry commute, so called BiLP geometry, whereas the semichiral fields parametrize the region where they don't. With additional supersymmetry, the target space of the chiral and twisted chiral model is bihyperhermitian with two commuting quaternionic structures [GHR84], whereas the semichiral model is hyperkähler [IV].

From the previous section it is clear that translations and rescalings preserve the $N=(4,4)$ supersymmetry of the twisted chiral multiplet. When dualized along a translational isometry with equal amounts in the chiral and the twisted chiral directions, the Laplace equation transforms into an analogue of the Monge-Ampère equation $\{J^{(+)}, J^{(-)}\} = 2c$ with vanishing $c$, implying that the semichiral model is hyperkähler [BSvdLVG99]. In the coordinates $x^\mu$, the Laplace equation is the linear equation (8.18), whereas the Monge-Ampère analogue (4.78) with $c = 0$ in the dual coordinates is the non-linear relation

$$(\tilde{K}_{\tilde{x}\tilde{x}} + \tilde{K}_{\tilde{y}\tilde{y}})\tilde{K}_{\tilde{z}\tilde{z}} + 2\tilde{K}_{\tilde{x}\tilde{x}}\tilde{K}_{\tilde{y}\tilde{y}} - 2\tilde{K}_{\tilde{x}\tilde{y}}^2 - \tilde{K}_{\tilde{x}\tilde{z}}^2 - \tilde{K}_{\tilde{y}\tilde{z}}^2 = 0. \quad (8.28)$$

One could also consider a translational isometry with unequal amounts in the chiral and twisted chiral directions, where now $c$ is non-vanishing but still constant, such that the torsion vanishes and the geometry is hyperkähler [Cri12].

Another option is dualizing along a rescaling isometry. This gives a non-constant $c$ [Cri12], hence the dual geometry has non-trivial torsion and is not hyperkähler, while the $N=(4,4)$ supersymmetry is still preserved. Investigating this route would be an interesting research project for the future.

---

[†] In paper [IV], the obtained result was that $g$ and $\tilde{g}$ are phases and $gh = \tilde{g}\tilde{h} = -1$. For simplification, the special case $g = -\tilde{g} = 1$ was considered here.

## 8.4 Reducing to (1,1) superspace

Reducing the original chiral and twisted chiral model (8.4) to a sigma model in $N=(1,1)$ superspace (4.25) reveals that the geometric structures $E = g+b$ are independent of the forth coordinate that parametrizes the direction of the isometry, $w| = i(\phi - \bar\phi - \chi + \bar\chi)$. The same is true for the semichiral dual model; when reduced to $N=(1,1)$ superspace, after eliminating the auxiliary fields, the action is

$$S = \int d^2\xi d^4\theta \tilde{K}(\tilde{x}^i) \xrightarrow[(1,1)]{} S = -\frac{1}{4}\int d^2\xi d^2\theta D_+\tilde{X}^a \tilde{E}_{ab} D_-\tilde{X}^b, \quad (8.29)$$

where the coordinates are $\tilde{X}^a = (\mathbb{X}^L, \mathbb{X}^R)| = (X^\ell, \bar{X}^{\bar\ell}, X^r, \bar{X}^{\bar r})$, and $\tilde{E}_{ab}$ depends only on derivatives of the potential $\tilde{K}$. In other words, the coordinate functions of the metric and the $b$-field are independent of the coordinate that is excluded by the isometry, $\tilde{w}| = \frac{1}{2}(X^\ell + \bar{X}^{\bar\ell} - X^r - \bar{X}^{\bar r})$,

$$g \sim g_{\ell\ell}(\tilde{x},\tilde{y},\tilde{z})dX^\ell dX^\ell + g_{\ell\bar\ell}(\tilde{x},\tilde{y},\tilde{z})dX^\ell d\bar{X}^{\bar\ell} + \ldots, \quad (8.30)$$

which does not alter the fact that the metric is non-degenerate.

## 8.5 Examples

### 8.5.1 Flat space

To illustrate the results reviewed in this chapter, consider the special case of a quadratic Lagrangian. A potential for flat space that satisfies the Laplace equation is, in terms of the combinations of chiral and twisted chiral coordinates,

$$K = \frac{1}{2}(x^2 - y^2) + \frac{r}{4}\left(z^2 - (x-y)^2\right), \quad (8.31)$$

where $r$ is an arbitrary real constant. The dual semichiral model is obtained by the T-duality procedure described in section 8.2 and is

$$\tilde{K} = -\frac{1}{2}(\tilde{x}^2 - \tilde{y}^2) - \frac{r}{4}\left((\tilde{x}+\tilde{y})^2 + \frac{4\tilde{z}^2}{r^2}\right). \quad (8.32)$$

For any value of $r$, this potential satisfies (8.28) with $c=0$, i.e., the Kähler potential will satisfy the Monge-Ampère equation and there is no $b$-field, which is expected, since the dualization was performed along a translational isometry by equal amounts on $\phi$ and $\chi$.

For the supersymmetry transformations, the analogue of (8.31) with unconstrained fields $V^\mu$ is considered. Functions that satisfy the partial differential equations in (8.23) can easily be found,

$$f = sV_L - i(V_x + V_y),$$
$$\hat{f} = tV_R + ir(V_x + V_y) + i(V_x - V_y), \quad (8.33)$$

where $s$ and $t$ are two arbitrary constants and $V_{L,R}$ are the combinations of the chiral and twisted chiral fields defined in (7.35) that vanish when the Bianchi identities are evoked, (8.12). The terms multiplying the integration constants $s$ and $t$ in $f$ and $\hat{f}$ will vanish due to the Bianchi identities for the original model, and on-shell in the dual model. This holds in general; a term $s \cdot g(V_L)$ in $f$ will transform the fields as

$$\delta\tilde{x} = \bar{\epsilon}^+ \bar{\mathbb{D}}_+ (sg(V_L)) = \bar{\epsilon}^+ sg'(V_L)\bar{\mathbb{D}}_+ V_L = -\bar{\epsilon}^+ \frac{s}{2} g'(V_L)\bar{\mathbb{D}}_+(\tilde{K}_{\tilde{z}} + i\tilde{K}_{\tilde{x}} + i\tilde{K}_{\tilde{y}}), \quad (8.34)$$

and the last expression vanishes due to the field equation (8.14). The same is valid for a term $t \cdot h(V_R)$ in $\hat{f}$. The parameters in $U^{(\pm)}$ are defined in (8.22) and take the constant expressions

$$\alpha = \tfrac{1}{2}(-2r + rs + s), \qquad \beta = \tfrac{1}{2}(2r(r+1) + t),$$
$$\gamma = \tfrac{1}{2}(-2r + rs - s), \qquad \delta = \tfrac{1}{2}(2r(r-1) + t),$$
$$\varepsilon = -\tfrac{s}{r}, \qquad \kappa = -\tfrac{t}{r}. \quad (8.35)$$

The dual transformation matrices can then be derived by Legendre transform and take the form

$$\tilde{U}^{(+)} = \frac{1}{4}\begin{pmatrix} 2r - s - rs & -4 + 2r - rs + s & 2i(-2 + \tfrac{s}{r}) \\ 4 + 2r - rs - s & 2r - rs + s & 2i(2 + \tfrac{s}{r}) \\ i(-2r + rs + s) & i(-2r + rs - s) & 2\tfrac{s}{r} \end{pmatrix}, \quad (8.36)$$

and

$$\tilde{U}^{(-)} = \frac{1}{4}\begin{pmatrix} -2r(r+1) - t & -4 + 2r(1-r) - t & 2i(2 + \tfrac{t}{r}) \\ 4 + 2r(r+1) + t & 2r(r-1) + t & 2i(2 - \tfrac{t}{r}) \\ i(2r(r+1) + t) & i(2r(r-1) + t) & 2\tfrac{t}{r} \end{pmatrix}. \quad (8.37)$$

The semichiral action (8.32) is invariant under these transformations. The field equations in flat space defined by the generalized Kähler potential in (8.32) are

$$(r+1)\bar{\mathbb{D}}_+\tilde{x} + (r-1)\bar{\mathbb{D}}_+\tilde{y} - \tfrac{2i}{r}\bar{\mathbb{D}}_+\tilde{z} = 0,$$
$$\bar{\mathbb{D}}_-\tilde{x} + \bar{\mathbb{D}}_-\tilde{y} - \tfrac{2i}{r}\bar{\mathbb{D}}_-\tilde{z} = 0. \quad (8.38)$$

The integration constants $s$ and $t$ in the transformations multiply field equations and vanish when (8.38) are used. Using the field equations, one can then check explicitly that the transformation defined by the matrices in (8.36)-(8.37) close to a supersymmetry on-shell.

### 8.5.2 Non-quadratic potential

Non-flat generalized Kähler potentials can also be constructed. One example, inspired by similar examples in [BSvdLVG99], is

$$K(x,y,z) = z \cdot \big(F(x+iy) + \bar{F}(x-iy)\big). \tag{8.39}$$

The potential satisfies the Laplace equation (8.18), hence the original chiral and twisted chiral sigma model has $N=(4,4)$ supersymmetry off-shell. As the functions $F$, $\bar{F}$ one can consider, for example, $F = \frac{1}{2}(x+iy)^2$, so that the original Lagrangian takes the cubic form

$$K(x,y,z) = z \cdot (x^2 - y^2). \tag{8.40}$$

Gauging the potential by introducing the unconstrained gauge potentials $V^\mu$, the Legendre transform gives the identities corresponding to (8.11),

$$\begin{aligned} K_{V_x} &= 2V_z \cdot V_x = \tilde{x}, \\ K_{V_y} &= -2V_z \cdot V_y = \tilde{y}, \\ K_{V_z} &= V_x^2 - V_y^2 = \tilde{z}, \end{aligned} \tag{8.41}$$

Solving for $V^\mu$ and inserting back into the first order action, the dual generalized Kähler potential takes the form

$$\tilde{K}(\tilde{x},\tilde{y},\tilde{z}) = -\sqrt{\tilde{z}(\tilde{x}^2 - \tilde{y}^2)}. \tag{8.42}$$

This potential satisfies (8.28), which is equivalent to the Monge-Ampère equation, hence the dual model describes hyperkähler geometry. The dual potential in semichiral coordinates reads

$$\tilde{K}(\mathbb{X}^\ell, \bar{\mathbb{X}}^{\bar{\ell}}, \mathbb{X}^r, \bar{\mathbb{X}}^{\bar{r}}) = -\sqrt{\tfrac{1}{2}(\mathbb{X}^\ell + \bar{\mathbb{X}}^{\bar{\ell}} + \mathbb{X}^r + \bar{\mathbb{X}}^{\bar{r}})i(\mathbb{X}^\ell - \bar{\mathbb{X}}^{\bar{\ell}})i(\mathbb{X}^r - \bar{\mathbb{X}}^{\bar{r}})}. \tag{8.43}$$

The determinant of the Hessian corresponding to this Lagrangian is well-defined as long as the potential (8.42) is non-vanishing.

To find the supersymmetry transformations, functions satisfying (8.23) are required. Making an ansatz for $f$ and $\hat{f}$ to be quadratic, in order to satisfy the partial differential equations, they must be of the form

$$\begin{aligned} f &= 2\big(sV_L^2 + 2V_x V_y - i(V_x + V_y)V_z\big), \\ \hat{f} &= 2\big(tV_R^2 + 2V_x V_y + i(V_x - V_y)V_z\big), \end{aligned} \tag{8.44}$$

where $s$ and $t$ are arbitrary integration constants. Again, the terms multiplying $s$ and $t$ in the transformations will vanish when Bianchi identities are used, or equivalently, on-shell for the transformations of the semichiral fields. With these functions, the transformation parameters take the form in (8.22). For clarity, only the on-shell part of the transformations are displayed here and are

$$\begin{aligned} \alpha &= 2iV_y \frac{V_x^2+V_y^2}{(V_x^2-V_y^2)V_z}, & \beta &= 2iV_y \frac{V_x^2+V_y^2}{(V_x^2-V_y^2)V_z}, \\ \gamma &= 2iV_x \frac{V_x^2+V_y^2}{(V_x^2-V_y^2)V_z}, & \delta &= 2iV_x \frac{V_x^2+V_y^2}{(V_x^2-V_y^2)V_z}, \\ \varepsilon &= -\frac{4V_xV_y}{V_x^2-V_y^2}, & \kappa &= -\frac{4V_xV_y}{V_x^2-V_y^2}. \end{aligned} \tag{8.45}$$

The supersymmetry transformations for the semichiral model can now be derived by the Legendre transform $\tilde{U} = (K)(U)(K^{-1})$. The on-shell content of the transformation matrices is

$$\tilde{U}^{(+)} = \begin{pmatrix} \frac{2i\tilde{y}(\tilde{x}^2+\tilde{y}^2)\bar{z}}{(\tilde{x}^2-\tilde{y}^2)^2} & -1 - \frac{2i\tilde{x}(\tilde{x}^2+\tilde{y}^2)\bar{z}}{(\tilde{x}^2-\tilde{y}^2)^2} & -i - \frac{2i\tilde{x}\tilde{y}}{\tilde{x}^2-\tilde{y}^2} \\ 1 + \frac{2i\tilde{y}(\tilde{x}^2+\tilde{y}^2)\bar{z}}{(\tilde{x}^2-\tilde{y}^2)^2} & -\frac{2i\tilde{x}(\tilde{x}^2+\tilde{y}^2)\bar{z}}{(\tilde{x}^2-\tilde{y}^2)^2} & i - \frac{2i\tilde{x}\tilde{y}}{\tilde{x}^2-\tilde{y}^2} \\ \frac{2\tilde{y}(\tilde{x}^2+\tilde{y}^2)\bar{z}}{(\tilde{x}^2-\tilde{y}^2)^2} & -\frac{2\tilde{x}(\tilde{x}^2+\tilde{y}^2)\bar{z}}{(\tilde{x}^2-\tilde{y}^2)^2} & -\frac{2\tilde{x}\tilde{y}}{\tilde{x}^2-\tilde{y}^2} \end{pmatrix} \tag{8.46}$$

and similar for $\tilde{U}^{(-)}$. One can explicitly check that, for arbitrary values of the integration constants $s$ and $t$ (not displayed in (8.46) since this is the on-shell part only), $\bar{\mathbb{D}}_\pm(\bar{\delta}^{(\pm)}\tilde{x}^i) = 0$ and the partial differential equations $\tilde{K}_{i[j}\tilde{U}^{(\pm)i}{}_{k]} = 0$ are satisfied, hence the semichiral action with generalized Kähler potential (8.42) is invariant under these supersymmetry transformations.

## 8.6 Results

The questions posted at the beginning of this chapter have now been answered and are summarized in table 8.1.

Translational and rescaling isometries are compatible with the additional supersymmetry, hence the $N=(4,4)$ supersymmetry of the chiral and twisted chiral model is preserved. To relate the non-linear transformations of the semichiral model to the linear ones for the chiral and twisted chiral model, new non-linear terms must be added to the transformations of the chiral and twisted chiral fields. These terms vanish when the chiral and twisted chiral Bianchi identities are imposed, and equivalently when the semichiral field equations are used, but they are crucial for the invariance of the semichiral action.

On the chiral and twisted chiral side, the $N=(4,4)$ supersymmetry algebra closes due to the Bianchi identities. These equations are T-dual to the field equations for the semichiral model, and hence, from the point of view of T-duality, it is clear that the algebra closes on-shell for the semichiral fields. This is related to the approach in [LR83], where additional supersymmetry is constructed for a first order action and then derived for the two dual actions.

The chiral/twisted chiral action is invariant under the supersymmetry transformations if the potential satisfies the Laplace equation, which by T-duality corresponds to the Monge-Ampère equation on the semichiral side. This implies that the target space of the semichiral model is hyperkähler but does not, however, imply that the semichiral action is invariant under the $N=(4,4)$ supersymmetry.

| *Chiral/twisted model* | *Semichiral model* |
|---|---|
| $S = \int K(x^\mu) = \int K(x,y,z)$ | $\tilde{S} = \int \tilde{K}(\tilde{x}^i) = \int K(\tilde{x},\tilde{y},\tilde{z})$ |
| Bianchi identities: | Bianchi identities: |
| $\bar{\mathbb{D}}_+(z+ix+iy)=0$ | $\bar{\mathbb{D}}_+\bar{\mathbb{D}}_-(\tilde{z}-i\tilde{x})=0$ |
| $\bar{\mathbb{D}}_-(z+ix-iy)=0$ | $\bar{\mathbb{D}}_+\mathbb{D}_-(\tilde{z}-i\tilde{y})=0$ |
| Field equations: | Field equations: |
| $\bar{\mathbb{D}}_+\bar{\mathbb{D}}_-(K_z-iK_x)=0$ | $\bar{\mathbb{D}}_+(\tilde{K}_{\tilde{z}}+i\tilde{K}_{\tilde{x}}+i\tilde{K}_{\tilde{y}})=0$ |
| $\bar{\mathbb{D}}_+\mathbb{D}_-(K_z-iK_y)=0$ | $\bar{\mathbb{D}}_-(\tilde{K}_{\tilde{z}}+i\tilde{K}_{\tilde{x}}-i\tilde{K}_{\tilde{y}})=0$ |
| Supersymmetry: | Supersymmetry: |
| $\delta x^\mu = \bar{\epsilon}^\alpha U^{(\alpha)\mu}{}_\nu \bar{\mathbb{D}}_\alpha x^\nu + c.c.$ | $\delta \tilde{x}^i = \bar{\epsilon}^\alpha \tilde{U}^{(\alpha)i}{}_j \bar{\mathbb{D}}_\alpha \tilde{x}^j + c.c.$ |
| $U^{(\alpha)}$ constant $3\times 3$ matrices | $\tilde{U} = (\partial\partial K) U (\partial\partial K)^{-1}$ |
| susy algebra closes off-shell | susy algebra closes on-shell |
| (using Bianchi identities) | (using field equations) |

*Figure 8.1:* The $N=(4,4)$ supersymmetry transformations for the chiral and twisted chiral model and the semichiral model can be understood from T-duality.

The generalized Kähler potential $\tilde{K}$ must satisfy an additional system of partial differential equations in order for the semichiral action to be supersymmetry invariant. That no additional constraints appear for the chiral and twisted chiral Lagrangian is again due to the Bianchi identities; but the equivalent equations of motion on the semichiral side cannot be used in the invariance of the action, hence the constraints remain for the semichiral Lagrangian. These differential equations are more transparent in the original chiral and twisted chiral coordinates, hence, finding transformations for $N=(4,4)$ supersymmetry is more straight-forward in the T-duality setting.

# 9. Summary

> Science makes people reach selflessly for truth and objectivity; it teaches people to accept reality, with wonder and admiration, not to mention the deep awe and joy that the natural order of things brings to the true scientist.
> *Lise Meitner, physicist (1878–1968)*

The research presented in this thesis concerns non-linear sigma models, supersymmetry and geometry. Non-linear sigma models with extended supersymmetry have constrained target space geometries, and can serve as tools for investigating and constructing new geometries. Since sigma models are fundamental objects in string theory, analyzing the geometrical and topological properties of sigma models is necessary for a full understanding of the underlying structures of string theory.

The most general two-dimensional sigma model with two manifest supersymmetries in each chirality can be described by a Lagrangian that is a scalar function of three kinds of superfields: chiral, twisted chiral and semichiral superfields, together with their complex conjugates, $K(\phi,\bar{\phi},\chi,\bar{\chi},\mathbb{X}^\ell,\bar{\mathbb{X}}^{\bar{\ell}},\mathbb{X}^r,\bar{\mathbb{X}}^{\bar{r}})$. The choice of superfields, i.e., the choice of supersymmetry representation, determines the geometry of the target space. A sigma model parametrized by only chiral and anti-chiral superfields $\phi,\bar{\phi}$, or by twisted and anti-twisted superfields $\chi,\bar{\chi}$, has a target space which is necessarily Kähler. Geometry with torsion can be described if both chiral and twisted chiral superfields are used as coordinates, implying a bihermitian target space with two commuting complex structures, so called BiLP geometry. The section of the target space where the two complex structures do not commute is described by a semichiral sigma model, parametrized by left and right semichiral superfields,

$$
\begin{aligned}
&K(\phi,\bar{\phi}) \text{ and } K(\chi,\bar{\chi}) &&\text{Kähler,} \\
&K(\phi,\bar{\phi},\chi,\bar{\chi}) &&\text{bihermitian with } [J^{(+)},J^{(-)}]=0, \\
&K(\mathbb{X}^\ell,\bar{\mathbb{X}}^{\bar{\ell}},\mathbb{X}^r,\bar{\mathbb{X}}^{\bar{r}}) &&\text{bihermitian with } [J^{(+)},J^{(-)}]\neq 0.
\end{aligned} \quad (9.1)
$$

Since bihermitian geometry is equivalent to generalized Kähler geometry, the $N=(2,2)$ supersymmetric sigma model gives a local description of general-ized Kähler geometry, with the Lagrangian $K$ being the generalized Kähler potential.

The sigma model parametrized by one (anti-) chiral and one (anti-) twisted chiral superfield can be enhanced with additional linear supersymmetry if the generalized Kähler potential satisfies the linear Laplace equation. The target space geometry becomes bihyperhermitian with two commuting quaternionic structures. The corresponding situation for a semichiral sigma model was previously not known.

In [I], it was found that a semichiral sigma model with four-dimensional target space, i.e., parametrized by only one set of left and right semichiral fields, cannot incorporate additional supersymmetry off-shell. However, linear transformations can be constructed that close to a *pseudo*-supersymmetry, resulting in a sigma model with $N=(4,4)$ twisted supersymmetry. The action is invariant under the twisted supersymmetry provided that the Lagrangian satisfies a system of linear partial differential equations. Non-trivial solutions can be found, and the corresponding target space geometry is neutral hyperkähler, with vanishing torsion and a metric with indefinite signature.

Off-shell $N=(4,4)$ supersymmetry can be imposed if the target space dimension is increased, as was described in paper [II]. The supersymmetry transformations leave the action invariant if and only if the generalized Kähler potential satisfies a system of partial differential equations. The geometrical structures can be encoded in Yano $f$-structures, a generalization of complex structures allowing for degeneracies in the transformation matrices.

Using the field equations, the $N=(4,4)$ supersymmetry can close *on-shell* in arbitrary $4n$-dimensional space. This was first observed in paper [II] and further investigated in paper [IV]. The transformations are necessarily non-linear. For the case of a four-dimensional target space, the non-linear partial differential equations resulting from the invariance of the action simplify, and solutions can be found. The geometrical constraints from the additional on-shell supersymmetry imply that the target space geometry is hyperkähler.

Hence, the semichiral sigma model with four-dimensional target space differs from the chiral and twisted chiral sigma model in several ways. Whereas linear off-shell $N=(4,4)$ supersymmetry can be imposed on the latter model, the semichiral model has non-linear transformations that close only on-shell.

Different sigma models can be related to each other by T-duality. The duality process involves gauging isometries using certain gauge potentials. The $N=2$ vector multiplet that gauges isometries mixing chiral and twisted chiral fields is called the large vector multiplet, and the vector multiplet gauging isometries mixing semichiral fields is the semichiral multiplet. In paper [III], it was found that the semichiral vector multiplet allows off-shell $N=(4,4)$ supersymmetry, and the explicit transformations were constructed for the gauge potentials and the field strengths in both the abelian and non-abelian case.

No transformations can close to additional supersymmetry on the large vector multiplet, however. This is in agreement with the results of the previous papers, since the large vector multiplet has semichiral field strengths; from [I] and [II] it is known that one set of semichiral fields does not allow for additional off-shell supersymmetry.

In paper [V], the discrepancy between the semichiral model and the chiral and twisted chiral model was investigated. Starting from the latter model with linear off-shell $N=(4,4)$ supersymmetry, the semichiral analogue was constructed by T-duality. Non-linear terms have to be added to the transformations in order to arrive at the semichiral sigma model obtained in paper [IV]. These non-linear terms vanish due to Bianchi identities and never appear for the chiral and twisted chiral fields. The Bianchi identities are dual to field equations for the semichiral model, hence by T-duality, the supersymmetry transformations are expected to close only on-shell on the semichiral fields. The T-duality replaces the Laplace equation by the Monge-Ampère equation on the semichiral side, implying that the target space parametrized by semichiral coordinates must be hyperkähler, in agreement with the results of paper [IV]. As shown in paper [II] and [IV], the generalized Kähler potential $K(\mathbb{X}^\ell, \bar{\mathbb{X}}^{\bar{\ell}}, \mathbb{X}^r, \bar{\mathbb{X}}^{\bar{r}})$ has to satisfy a system of non-linear partial differential equations for the action to be invariant under the $N=(4,4)$ supersymmetry. Using T-duality, these equations take a more transparent form, and solutions are easier obtained.

As a summary, the papers [I-V] give a conclusive understanding of the semichiral sigma models and $N=(4,4)$ supersymmetry. Off-shell supersymmetry can be obtained if the sigma model is parametrized by at least two sets of semichiral fields. For one set of left and right semichiral fields, the additional supersymmetry transformations are necessarily non-linear and close only on-shell, which is to be expected from T-duality.

# Acknowledgements

It is a pleasure to take this opportunity to thank the people who have supported me and made the writing of this thesis possible.

First and foremost, my deepest gratitude is to my supervisor, Ulf Lindström. He has truly been an invaluable support, not only during the writing of this thesis, but throughout my whole PhD studies. By presenting interesting problems, giving me helpful feed-back and working side by side with me on different projects, he has guided me through achievements and struggles and patiently encouraged me to become an independent researcher. Again, thank you, Ulf!

Also, I am extremely thankful to my assistant supervisor Maxim Zabzine, who read the manuscript and gave me many constructive comments for improvement, and to Johan Alm, Johan Blåbäck and Tove Fraurud who also read parts of the manuscript and helped me improving it.

I have been fortunate to have several brilliant collaborators and want to thank my co-authors, Martin Roček and Itai Ryb, for insightful discussions, it has been a privilege to work with you. Also, huge thanks to Marcos Crichigno, for resolving my confusion on several issues, and to Malte Dyckmanns and Gabriele Tartaglino-Mazzucchelli, for new perspectives and ideas.

Many inspiring colleagues make the Division of Theoretical Physics and the Department of Physics and Astronomy a good working environment. Sincere thanks to all my fellow graduate student colleagues: Anton, for company during late night working-hours; Itai, for the fighting spirit; Jacob, for taking over the fika list; Joel, for opening interesting discussions on all kinds of various topics; Johan B, for being a great friend and colleague, and for sharing my path since undergraduate studies; Johan K, for pedagocally sharing sharp insights and always helping out; Kasper, for promoting a strong passion for classical music; Magdalena, for motivation and encouragement; Martin, for being a good room-mate; Niklas, for strongly contributing to a more cheerful atmosphere at the division; Olof, for excellent evenings at the pub; Raul, for continuing the representation in the PhD council; Sergey, for the happy attitude; Shuang-Wei, for trying to maintain my level of Chinese and for brilliant cooking; Valentina, for your warm, generous heart and the lovely trip to Perugia; and Xubiao, for tasty dumplings. Lisa, Thomas, Joe, Rolf, Ulf D, Staffan, all postdocs, and all my other colleagues; thank you for the nice and intellectual atmosphere, for the great work you carry out every day and for the pleasant (and delicious) Friday-fikas.

A few colleagues in other groups at the department have made my time as a PhD student extra valuable. In particular I want to thank Karin Schönning, continuously working to improve equality in academia, and Lena Heijkenskjöld, for walks, laughs and for bringing me cakes.

During my time as a graduate student I have participated in the organizing of several large events, and I am grateful to all the co-organizers who made the Strings conference 2011 and the Diversities in the Cultures of Physics summer school in Berlin and Uppsala 2012 so successful.

The administrators and IT personnel of the department all deserve acknowledgement for their enduring support in all sorts of questions; I especially want to thank Inger, Marja and Bertil.

And, although most of them might never read this, I also want to thank my students, for the positive spirit during class, for taking the studies seriously and for helping me to improve my teaching.

Part of the research presented in this thesis was carried out when I was at Stony Brook, NY. I would like to thank Stony Brook University for hospitality, and the equality grant from the Dept. of Physics and Astronomy, Uppsala University, for funding the stay.

I believe that many PhD students sometimes despair or have the overwhelming feeling that the world is coming to an end because a problem seems to be unsolvable, or an annoying sign in the calculation is wrong. In order to come back to reality and regain the motivation, I have relied on my friends and family; none of this would have been possible without their unlimited love and support.

My wonderful friends – you know who you are! – thank you for friendship, parties, discussions, never-ending breakfasts and dinners, travels and lazy hang-arounds, for support during hard times, and shared joy in the good.

I want to thank my parents for endless care and love; my mother, for always believing in me, no matter what I do, and my father, for triggering my interest in research and theoretical physics by sending me packages with interesting books and popular articles about new research, no matter where in the world my letter box was at the time. And my brother, for providing new perspectives and unconditioned solidarity and love. Also, my enlarged family – thank you for reviving holidays and for being there, you are an important part of my life.

Saving the best for last, with all my heart, I want to thank Johan. For always listening and understanding, for providing me with lunch and dinner packages when I was working late, for sharing with me both setbacks and successes. But most of all, for making life – Monday mornings in February as well as relaxed summer days – so wonderful to live.118

# Svensk sammanfattning

Jag tror att det finns en inneboende nyfikenhet i människan. Vi vill veta vad som finns på andra sidan havet, vad åska är, om universum har ett slut. Genom historien har människor vänt sig till religioner och skapelseberättelser för att försöka förstå svårförklarliga fenomen. En annan väg, som jag följer, är den vetenskapliga. Metodiskt och logiskt undersöker forskaren världen genom observationer, experiment och teoretiska modeller. Ibland kommer observationen först, ibland föregås den av teori. Men trots att den vetenskapliga principen följer vissa lagar och strukturer, så är forskning en väldigt kreativ process. Utifrån ett vitt ark skapas något nytt, resultat som tidigare varit okända.

Under 1900-talet har vår förståelse av fysikens lagar helt revolutionerats. Einsteins allmänna relativitetsteori har gett en helt ny förståelse av gravitation som en krökning i rumtiden. Kvantfysiken som utvecklades ungefär samtidigt var en nyskapande förklaring till fysik på atomnivå, och ligger till grund för den standardmodell som beskriver partikelfysik, hur atomer är uppbyggda och samverkar.

Men det finns fortfarande frågor inom fysiken vi ännu inte har svaret på. Varför finns det tre generationer partiklar? Hur kan hierarkin mellan olika krafter tolkas, eller värdet på alla olika parametrar? Hur förklaras neutrinons massa? Vad är mörk materia och mörk energi? Hur ska universums tidiga inflation och nuvarande homogenitet förstås? Varför observerar vi mer materia än antimateria?

Strängteori är ett försök att formulera en teori som beskriver både kvantfysik och gravitation. Det grundläggande antagandet i strängteori är att de minsta beståndsdelarna i världen inte är punktformiga partiklar, utan endimensionella strängar. Tanken är att olika vibrationer hos strängarna ska ge upphov till de partiklar som vi observerar, på samma sätt som en gitarrsträng ger upphov till olika toner. Sedan strängteori började utvecklas på 1970-talet så har teorin rönt stor uppmärksamhet och inspirerat många världsledande fysiker. Experimentella data som skulle kunna bekräfta eller dementera teorin har dock hittills lyst med sin frånvaro. Men skulle det visa sig att strängteori inte är en korrekt beskrivning av vår fysikaliska verklighet, så har teorin ändå lett till många viktiga resultat inom andra delar av fysik, samt inom matematik.

## Bakgrund

De fem artiklarna som den här doktorsavhandlingen baserar sig på handlar om särskilda sigmamodeller med utökad supersymmetri, och om deras korrelation till rummet de rör sig i. Intuitivt kan man tänka sig en sigmamodell som en sträng, en gummisnodd, som rör sig i något rum. Det har visat sig att när man kräver att en sigmamodell har en viss symmetri så händer det saker med rummet som sigmamodellen befinner sig i: det kröks, vrids eller får andra strukturer.

Eftersom de supersymmetriska sigmamodellerna är en viktig beståndsdel i strängteori, så kan man förstå den matematiska strukturen hos strängteori genom att studera sigmamodeller med supersymmetri. Omvänt så kan nya spännande geometrier och topologier förstås och modelleras med hjälp av supersymmetriska sigmamodeller. För att förstå bakgrunden till artiklarna introduceras här i tur och ordning de tre koncepten: geometri, supersymmetri och sigmamodeller.

### *Geometri*

*Geometri* handlar om hur en yta eller en volym ser ut. En badboll är rund, och, även om vi inte ser det från jordytan, så vet vi att vår planet är (approximativt) rund. Detta beskrivs matematiskt som att ytan har positiv krökning. En hästsadel, å andra sidan, har negativ krökning. Metriken *g* mäter avstånd och beskriver en ytas krökning lokalt, torsionen *H* beskriver istället hur ett rum är vridet. *Topologi* handlar om en ytas globala egenskaper. Även om en kaffekopp och en munk ser olika ut så har de båda ett hål. Om munken var gjord av lera så skulle man kunna omforma den till en kaffekopp utan att skapa eller fylla igen några hål: de har samma topologi.

I denna avhandling studeras i huvudsak olika aspekter av komplex geometri. En *mångfald* är ett rum (av godtycklig dimension) där varje punkt beskrivs av reella koordinater $(x^1,\ldots,x^n)$, exempelvis det reella talplanet $\mathbb{R}^2$, som beskrivs av två koordinater $(x,y)$. För att alla punkter i mångfalden ska tilldelas koordinater på ett konsistent sätt så krävs att det finns deriverbara funktioner mellan koordinatsystemen. På samma sätt är en *komplex mångfald* ett rum där varje punkt beskrivs av komplexa koordinater $(z^1,\ldots z^n)$, där $z = x+iy$. För att övergången mellan två koordinatsystem ska ske konsistent så måste koordinatbytesfunktionerna vara *holomorfa*. Detta är ekvivalent med existensen av en *komplex struktur*, en integrabel struktur som kvadrerar till minus ett, $J^2 = -1$. Intuitivt kan man tänka sig den komplexa strukturen som en multiplikation med det komplexa talet *i* eller $-i$.

En *Kählermångfald* är ett specialfall av en komplex mångfald, där den så kallade Kählerformen $\omega = gJ$ är sluten, $d\omega = 0$. Detta innebär att metriken lokalt kan skrivas som derivator på en *Kählerpotential*, $g_{\mu\bar{\nu}} = \partial_\mu \partial_{\bar{\nu}} K$.

En *hyperkomplex* mångfald har inte en, utan tre komplexa strukturer $(I, J, K)$ som alla kvadrerar till minus ett, är integrabla och dessutom uppfyller kvaternionalgebran, $IJK = -1$. På samma sätt har en mångfald som är *hyperkähler* tre Kählerformer som alla är slutna.

## Supersymmetri

De fysikaliska lagar som bygger upp den fysikaliska världsbilden bygger i stort på symmetrier. En symmetri är en transformation under vilken systemet är invariant. Exempelvis så förändras inte klotet om man roterar på det, eller triangeln om man vrider den en tredjedels varv. Triangeln har en global diskret symmetri, medan klotet är ett exempel på ett system med en global kontinuerlig (rotations-)symmetri. Einsteins allmänna relativitetsteori bygger istället på en *lokal* symmetri, nämligen invarians under koordinatbyten av rumtidskoordinaterna. Lokala symmetrier kallas också gaugeteorier.

Naturen är uppbyggd av två fundamentalt skilda sorters partiklar: bosoner och fermioner. Bosoner har ett inre spinn som alltid är ett heltal. De är kraftpartiklar, exempelvis fotoner, som utgör ljus och förmedlar den elektromagnetiska kraften. Fermioner däremot har ett inre spinn som alltid är ett halvtal. De utgör materia, exempelvis elektroner och kvarkar, som bildar atomer. *Supersymmetri* relaterar dessa två sorters partiklar och stoppar ihop dem i en och samma *supermultiplet*. Matematiskt är supersymmetrialgebran en utvidgning av Poincarésymmetri, som är grunden för speciell relativitetsteori, med udda supersymmetrigeneratorer $Q$. Om $X$ betecknar ett bosoniskt fält och $\psi$ ett fermioniskt, så kan ett *superfält* definieras som en kombination av dessa två fält, $\phi = X + \theta\psi$, där $\theta$ är en så kallad Grassmannkoordinat.

## Supersymmetriska sigmamodeller

Intuitivt kan man föreställa sig en sigmamodell som en sträng som rör sig i något rum. Det är också från denna bild som man kan härleda deras *verkan*, som beskriver deras dynamik och rörelse. När strängen rör sig i tiden så skapas en *världsyta*, som beskrivs av en koordinat längs strängen och en tidskoordinat. Verkan är en integral, som i ett särskilt val av koordinater på världsytan tar formen

$$S = \int d^2x \partial_{++} X (g+b) \partial_= X, \tag{9.2}$$

där återigen $g$ betecknar metriken och $b$-fältet ger upphov till torsion $H = db$, och $X$ är en boson. När metriken är en funktion av fälten, $g = g(X)$ så kallas sigmamodellen *icke-linjär*. Mer generellt är en sigmamodell en avbildning från en världsyta (som inte behöver vara två-dimensionell) till ett målrum, tillsammans med en verkan. I en supersymmetrisk sigmamodell är det bosoniska fältet *X* utbytt mot ett superfält som även innehåller fermionska fält, och verkan är en integral även över de udda Grassmannkoordinaterna.

Transformationer som formar en utökad supersymmetri kan konstrueras om målrummet är utrustat med två komplexa strukturer och metriken *g* uppfyller vissa geometriska krav. I ett utökat superrum är denna $N = 2$ supersymmetri manifest och den mest generella modellen beskrivs av särskilda koordinater som kallas *kirala φ*, *tvistat kirala χ* respektive *semikirala* $\mathbb{X}^\ell, \mathbb{X}^r$,

$$S = \int d^2x d^2\theta d^2\bar{\theta} K(\phi, \bar{\phi}, \chi, \bar{\chi}, \mathbb{X}^\ell, \bar{\mathbb{X}}^{\bar{\ell}} \mathbb{X}^r, \bar{\mathbb{X}}^{\bar{r}}). \qquad (9.3)$$

Geometrin som beskrivs av denna modell kallas bihermitsk, eller generaliserad Kähler, och funktionen *K* kallas den *generaliserade Kählerpotentialen*.

## Frågeställningar och resultat

Som har beskrivits ovan, så finns ett intimt samband mellan supersymmetriska sigmamodeller och geometri. Att addera supersymmetri till olika sigmamodeller och analysera vilka geometriska krav som blir följden har varit en effektiv metod dels för att undersöka och konstruera nya geometrier, och dels för att förstå de matematiska strukturerna i strängteori.

### *Semikirala sigmamodeller och N = 4 supersymmetri*

En allmän tvådimensionell sigmamodell med två manifesta supersymmetrier kan skrivas i termer av kirala, tvistat kirala och semikirala superfält. Den modellen som beskrivs av kirala och tvistat kirala superfält kan ha ytterligare supersymmetri om den generaliserade Kählerpotentialen uppfyller en viss slags partiell differentialekvation, Laplaceekvationen. Motsvarande situation för en semikiral modell, som skrivs i termer av semikirala koordinater, var tidigare inte känd. I artiklarna [I], [II] och [IV] studerade vi därför frågorna:

* *Kan en semikiral sigmamodell ha fyra supersymmetrier, på samma sätt som den kirala och tvistat kirala modellen? Påverkas situationen av målrummets dimension? Kan fältekvationer användas för att sluta supersymmetrialgebran på skalet? Vad blir den resulterande geometrin?*

Det visade sig att en sigmamodell som beskrivs av ett vänster- och ett högersemikiralt fält inte kan ha ytterligare supersymmetri av skalet. Däremot så kan transformationer som sluter till en pseudo-supersymmetri konstrueras [I], vilket innebär att geometrin är neutralt hyperkähler, med en metrik av obestämd signatur.

Förutsättningarna för extra supersymmetrier ändras helt om målrummets dimension utökas. Med två uppsättningar semikirala fält så är det möjligt att sluta den utökade supersymmetrialgebran (utan att använda sig av fältekvationerna) [II]. Situationen ändras även om fältekvationerna används. Då kan även en semikiral modell med fyradimensionellt målrum ha extra supersymmetri.

I fyra dimensioner förenklas ekvationerna och algebran sluter sig på skalet med endast ett ytterligare villkor. Den resulterande geometrin är hyperkähler, och konkreta exempel för denna geometri kan konstrueras [IV].

### *Vektormultipleter*

Olika sigmamodeller relaterar till varandra genom så kallad *T-dualitet*, som genomförs med hjälp av olika *vektormultipleter*. I artikel [III] studerades:

∗ *Kan den semikirala och/eller den stora vektormultipleten ha extra supersymmetrier?*

Resultatet blev att den semikirala vektormultipleten kan ha extra supersymmetri, och de explicita linjära transformationerna kunde konstrueras. Däremot visade det sig att den stora vektormultipleten inte kan inkorporera extra supersymmetri [III]. Detta går att förstå och står i överensstämmelse med resultaten beskrivna ovan. Den semikirala vektormultipleten liknar den kirala och tvistat kirala sigmamodellen och kan således ha extra supersymmetri. Den stora vektormultipleten däremot liknar den semikirala sigmamodellen. Från resultaten i artikel [I] och [II] vet vi, att en modell med fyra semikirala fält inte kan ha extra supersymmetri om man inte använder sig av fältekvationerna.

### *T-dualitet och N = 4 supersymmetri*

Sigmamodellen med kirala och tvistat kirala fält kan relateras till den semikirala modellen som studerades i [I, II, IV] med hjälp av T-dualitet. Detta inspirerade till fortsatta studier för att förstå hur den semikirala sigmamodellen förhåller sig till andra modeller. I artikel [V] undersöktes slutligen:

∗ *Hur relaterar de extra supersymmetrierna i de semikirala och (tvistat) kirala modellerna, finns det samband genom T-dualitet? Varför krävs fältekvationerna för att algebran ska slutas, och vad är den semikirala motsvarigheten till Laplaceekvationen?*

Resultatet var intressant och gav en djupare förståelse för supersymmetriska sigmamodeller och T-dualitet. Genom T-dualiteten är det väntat att supersymmetrialgebran i den semikirala modellen kan slutas endast med hjälp av fältekvationer, med andra ord på skalet. Laplaceekvationen motsvaras av Monge-Ampèreekvationen, vilket innebär att geometrin hos den semikirala modellen är hyperkähler, i överensstämmelse med resultaten i artikel [IV].